\documentclass[11pt]{scrarticle}
\usepackage[T1]{fontenc}
\usepackage[UKenglish]{babel}
\usepackage{amsmath, amsfonts, amsthm, amssymb, mathtools, stmaryrd}

\usepackage[margin=1in,letterpaper]{geometry}

\usepackage[pdfusetitle,hidelinks,hypertexnames=false]{hyperref}
\usepackage[capitalise,noabbrev]{cleveref}
\hypersetup{
    hidelinks,
    colorlinks,
    linkcolor=[rgb]{0.3,0.3,0.6},
    citecolor=[rgb]{0.2,0.6,0.2},
    urlcolor=[rgb]{0.6,0.2,0.2}
}

\usepackage{csquotes}
\usepackage[osf,sc]{mathpazo}
\addtokomafont{disposition}{\normalfont\bfseries} 
\usepackage{relsize}
\usepackage{microtype}
\usepackage{thmtools, thm-restate}
\makeatletter
\renewcommand\thmt@autorefsetup{%
  \@xa\def\csname\thmt@envname autorefname\@xa\endcsname\@xa{\thmt@thmname}%
}
\newcommand{\restatethm}[2]{%
  \begingroup
  \edef\restatethm@ctr{\the\c@definition}%
  \renewcommand{\thedefinition}{\ref*{#1}}%
  \renewcommand{\theHdefinition}{restated.#1}%
  #2*%
  \global\c@definition=\restatethm@ctr\relax
  \endgroup}
\makeatother
\usepackage{subcaption}
\usepackage{enumitem}
\usepackage{comment}

\usepackage[
    backend=biber,
    style=alphabetic,
    url=false,
    doi=false,
    isbn=false,
    sorting=nyt,
    backref=true,
    maxbibnames=10,
    maxcitenames=10,
    mincitenames=1,
]{biblatex}
\DefineBibliographyExtras{UKenglish}{}
\AtEveryBibitem{\clearlist{language}}
\AtEveryBibitem{\clearfield{issn}}

\usepackage{orcidlink}
\usepackage{booktabs}

\usepackage{comment}

\theoremstyle{definition}
\newtheorem{definition}{Definition}[section]

\newtheorem{remark}[definition]{Remark}

\theoremstyle{plain}
\newtheorem{example}[definition]{Example}
\newtheorem{lemma}[definition]{Lemma}
\newtheorem{corollary}[definition]{Corollary}
\newtheorem{theorem}[definition]{Theorem}

\newtheorem{proposition}[definition]{Proposition}
\newtheorem{observation}[definition]{Observation}
\crefname{observation}{Observation}{Observations}
\Crefname{observation}{Observation}{Observations}

\Crefname{fact}{Fact}{Facts}

\theoremstyle{remark}
\newtheorem{claim}{Claim}[definition]

\Crefname{claim}{Claim}{Claims}
\newenvironment{claimproof}[1][Proof of Claim]{\begin{proof}[#1] }{ \end{proof}}
\setlist[enumerate, 1]{font=\upshape, itemsep=0.05em}
\setlist[enumerate, 2]{font=\upshape, itemsep=0.05em, nolistsep}
\setlist[itemize, 1]{noitemsep, nolistsep,font=\upshape}
\setlist[itemize, 2]{noitemsep, nolistsep,font=\upshape}

\usepackage{todonotes}
\usepackage{tikz}
\usetikzlibrary{calc,positioning}

\usepackage{multicol}

\usepackage{abstract}

\KOMAoptions{captions=nooneline}   
\setcapindent{0pt}                 
\addtokomafont{captionlabel}{\bfseries}

\DefineBibliographyStrings{english}{%
  backrefpage  = {cited on page},
  backrefpages = {cited on pages}
}

\newcommand{\doiorurl}{%
  \iffieldundef{doi}
    {\iffieldundef{url}
       {}
       {\strfield{url}}}
    {http://dx.doi.org/\strfield{doi}}%
}

\newcommand{\myhref}[1]{%
  \ifboolexpr{%
    test {\ifhyperref}
    and
    not test {\iftoggle{bbx:url}}
    and
    not test {\iftoggle{bbx:doi}}
  }
    {\href{\doiorurl}{#1}}
    {#1}%
}

\DeclareFieldFormat{title}{\myhref{\mkbibemph{#1}}}
\DeclareFieldFormat
  [article,inbook,incollection,inproceedings,patent,thesis,unpublished]
  {title}{\myhref{\mkbibquote{#1\isdot}}}

\newif\ifcomments
\commentstrue    
\ifcomments
  \renewcommand{\comment}[2]{\marginpar{\tiny{\textbf{#1: }\textit{#2}}}}
  \newcommand{\important}[1]{{\color{red}#1}}
  \newcommand{\note}[1]{{\color{blue}#1}}
\else
  \renewcommand{\comment}[2]{}
  \newcommand{\important}[1]{}
  \newcommand{\note}[1]{}
\fi

\declaretheoremstyle[bodyfont=\normalfont \itshape]{thmstyle} 

\Crefname{mainthm}{Theorem}{Theorems}

  {\vspace{1ex}\noindent{\em Proof.}\hspace{0.5em}\def\myproof@name{#1}}%
  {\hfill{\tiny \qed\ (\myproof@name)}\vspace{1ex}}
\newenvironment{proof-sketch}{\medskip\noindent{\em Proof Sketch.}}{\qed\bigskip}
\newenvironment{proof-attempt}{\medskip\noindent{\em Proof attempt.}}{\bigskip}

\newcommand{\K}{\mathbb{K}}

\newcommand{\poly}{\operatorname{poly}}

\newcommand{\maxorb}{\operatorname{maxOrb}}
\newcommand{\maxsup}{\operatorname{maxSup}}
\DeclareMathOperator{\tw}{tw}
\DeclareMathOperator{\pw}{pw}

\DeclareMathOperator{\perm}{perm}

\DeclareMathOperator{\tr}{tr}

\DeclareMathOperator{\id}{id}
\DeclareMathOperator{\Aut}{\mathbf{Aut}}
\DeclareMathOperator{\Sym}{\mathbf{Sym}}

\DeclareMathOperator{\Stab}{\mathbf{Stab}}
\DeclareMathOperator{\StabP}{\mathbf{Stab}^{\bullet}}

\usepackage{xspace}

\newcommand{\symVP}{\ensuremath{\mathsf{symVP}}\xspace}
\newcommand{\symVNP}{\ensuremath{\mathsf{symVNP}}\xspace}
\newcommand{\symVBP}{\ensuremath{\mathsf{symVBP}}\xspace}
\newcommand{\symVS}{\ensuremath{\mathsf{symVS}}\xspace}
\newcommand{\symVF}{\ensuremath{\mathsf{symVF}}\xspace}
\newcommand{\symVAC}{\ensuremath{\mathsf{symVAC}}}

\newcommand{\VAC}{\ensuremath{\mathsf{VAC}}\xspace}
\newcommand{\VP}{\ensuremath{\mathsf{VP}}\xspace}

\newcommand{\VNP}{\ensuremath{\mathsf{VNP}}\xspace}

\newcommand{\VBP}{\ensuremath{\mathsf{VBP}}\xspace}

\newcommand{\VF}{\ensuremath{\mathsf{VF}}\xspace}

\renewcommand{\epsilon}{\varepsilon}

\renewcommand{\phi}{\varphi}

\renewcommand{\epsilon}{\varepsilon}
\newcommand{\ignore}[1]{}

\DeclareMathOperator{\td}{td}

\let\sup\relax
\DeclareMathOperator{\sup}{sup}

\newcommand{\bbN}{\mathbb{N}}
\newcommand{\bbQ}{\mathbb{K}}

\newcommand{\Xx}{\mathcal{X}}
\newcommand{\Yy}{\mathcal{Y}}

\DeclareMathOperator{\fw}{fw}

\newcommand{\leanbaseurl}{https://laxarchive.org/lax-461425/Lax461425.}
\definecolor{leangreen}{rgb}{0.0,0.4,0.1}
\newcommand{\lean}[1]{%
  \marginpar{\raggedright\footnotesize\href{\leanbaseurl #1.html}{Lean formalisation on \textsc{LaxLeanArchive}}}%
}
\DeclareMathOperator{\VS}{VS}
\title{Depth Hierarchies for \\Symmetric Algebraic Circuits\footnote{A Lean formalisation of the main results is available at \url{https://laxarchive.org/lax-461425/}.}}

\newcommand{\authoremail}[1]{\href{mailto:#1}{\nolinkurl{#1}}}

\author{%
    \begin{tabular}{@{}c@{\hspace{0.75em}}c@{\hspace{0.75em}}c@{\hspace{0.75em}}c@{}}
        Anuj Dawar\thanks{University of Cambridge, United Kingdom. \texttt{\{\href{mailto:anuj.dawar@cl.cam.ac.uk}{anuj.dawar}, \href{mailto:benedikt.pago@cl.cam.ac.uk}{benedikt.pago}\}@cl.cam.ac.uk}.\protect\label{aff:cambridge}}\enspace\orcidlink{0000-0003-4014-8248}%
        &
        Prateek Dwivedi\thanks{Universität des Saarlandes, Germany, and Université Savoie Mont Blanc, France. \protect\authoremail{dwivedi@cs.uni-saarland.de}.}\enspace\orcidlink{0000-0002-0572-3721}%
        &
        Benedikt Pago\footref{aff:cambridge}\enspace\orcidlink{0000-0001-6377-1230}
        &
        Tim Seppelt\thanks{IT-Universitetet i København, Denmark. \protect\authoremail{tise@itu.dk}.}\enspace\orcidlink{0000-0002-6447-0568}%
    \end{tabular}
}
\hypersetup{pdfauthor={Anuj Dawar, Prateek Dwivedi, Benedikt Pago, Tim Seppelt}}

\begin{document}
	\pagenumbering{roman}
	\maketitle

    \begin{abstract}
        Algebraic complexity studies the cost of computing families of polynomials by algebraic circuits, with the central goal of proving super-polynomial circuit lower bounds and separations between different models of algebraic computation. Since such lower bounds remain elusive for general circuits, a common strategy is to focus on restricted models. 
        A recent line of work (ICALP 2020, ITCS \& STOC 2026) introduces and studies symmetry as a syntactic circuit restriction, showing exponential symmetric lower bounds for the permanent, as well as separations between symmetric analogues of algebraic formulas ($\VF$), skew circuits ($\VS$), and circuits ($\VP$). 
        
        Motivated by the progress of this research programme, we investigate the role of \emph{depth} as a complexity parameter in symmetric circuits.
        For every fixed $\Delta \geq 1$, we exhibit a family of polynomials computable by symmetric circuits of polynomial size and depth $2\Delta+3$, but requiring size $n^{\Omega(\log n)}$ at depth $\Delta$. For a parameter called \emph{support depth}, which captures the semantically relevant concept of depth for symmetric circuits, this complexity gap arises already between $\Delta$ and $\Delta+1$.

        We also define symmetric versions of the bounded-depth algebraic complexity classes $\VAC^k$, for all $k \in \bbN$, that we call $\symVAC^k$, and establish two strict hierarchies: Firstly, a hierarchy within $\symVAC^0$, corresponding to constant-depth symmetric computation, and secondly, a hierarchy of the form $\symVAC^k \subsetneq \symVAC^{k+1}$ for all $k$. The latter shows that symmetric circuits of depth $(\log n)^{k+1}$ are strictly more powerful than depth-$(\log n)^{k}$ circuits. 
        Our results are obtained via a graph-theoretic characterisation of the symmetric complexity classes in terms of homomorphism polynomials. This also allows us to place the bounded-depth classes within the landscape of symmetric algebraic complexity developed by Dwivedi, Pago and Seppelt (STOC 2026).
        Finally, we complement this picture by introducing $\symVNP$, a symmetric analogue of $\VNP$, and establishing $\symVNP$-completeness of the permanent under symmetric projections.   
    \end{abstract}





    \newpage
    \tableofcontents
    \newpage
    \pagenumbering{arabic}

\section{Introduction}

Algebraic complexity theory studies families of polynomials in terms of the algebraic circuit size that is required to represent them. 
An algebraic circuit is a directed acyclic graph
whose internal vertices correspond to addition and multiplication gates and whose input gates are
labelled with variables or constants from a fixed underlying field. It computes a polynomial in the input variables in the natural way, and the circuit size is defined as the total number of gates and wires it contains. 
In 1979, Valiant postulated the conjecture that the \emph{permanent} polynomial, i.e., the determinant without signs, does not admit polynomial-size algebraic circuits \cite{valiant_completeness_1979}. 
A central part of the algebraic complexity programme is thus to develop techniques for proving circuit size lower bounds. 
The common approach is to begin with restricted circuit models, where lower bounds are easier to establish. 
A substantial line of work has focussed on proving lower bounds for example on monotone or multilinear circuits and formulas \cite{agrawal_selection_2014}. 
Arguably, these results tell us more about the weakness of the restricted models than about the true complexity of the permanent: In fact, many of them, like the exponential lower bound for multilinear formulas computing the permanent \cite{Raz04}, apply equally to the determinant, even though the determinant is known to admit polynomial-size circuits. 

In recent years, a new circuit restriction has been investigated \cite{DW2022,dawar_symmetric_2020,dps_itcs,dwivedi_lower_2026}, in which determinant and permanent provably do exhibit a polynomial versus exponential complexity gap, in line with Valiant's conjecture. This model is that of \emph{symmetric circuits}. 
Not only has the separation between determinant and permanent \cite{dawar_symmetric_2020} been established for this model, but also the separations of symmetric analogues of the major classes in algebraic complexity, that is, $\VF$, $\VBP/\VS$, and $\VP$ \cite{dwivedi_lower_2026} (corresponding to algebraic formulas, branching programs or skew circuits, and general circuits, respectively).
Briefly, the notion of symmetry means the following: Whenever a polynomial $p \in \K[X]$ is invariant under a permutation group $\Gamma$ acting on the variables $X$, then it makes sense to require $\Gamma$-symmetry also of a circuit $C$ that computes $p$. A circuit $C$ is \emph{$\Gamma$-symmetric} if, for every permutation $\pi \in \Gamma$ that acts on the input variables, there also exists a permutation $\sigma$ of the gates of $C$ that is consistent with $\pi$ on the input gates, and preserves the wire structure of the circuit (that is, is an automorphism of $C$). For a formal definition, see \cref{sec:matrixSymmetry}. In this paper and in most of the works referenced above, the group $\Gamma$ is always $\Sym_n \times \Sym_n$, and it acts via row and column permutations on variables $X_n \coloneqq \{ x_{ij} \mid i,j \in [n] \}$ that are arranged in an $n \times n$ matrix. 
This notion is called \emph{matrix symmetry}, and it applies to matrix polynomials such as the permanent and many other natural polynomials\footnote{The determinant is symmetric, too, but only under the \emph{simultaneous} action of $\Sym_n$ on the rows and columns, not under all independent row and column permutations.} from algebraic complexity.

The recent lower bounds and separation results for symmetric circuits establish this model as an important and promising concept that merits further investigation.
In this paper, we focus on the interplay of symmetry and \emph{depth} in algebraic circuits. Bounded-depth circuit classes are a natural notion that has been studied extensively both in the algebraic and the Boolean setting. 
We prove \emph{depth hierarchies} for symmetric circuits and place them in the existing symmetric algebraic circuit complexity landscape.
A depth hierarchy is a statement along these lines: For every $\Delta \in \bbN$, there exist families of polynomials that are computable by depth-$\Delta$ circuits of polynomial size but require super-polynomial size to be computed by a circuit of depth $< \Delta$.
Such depth hierarchy results were first shown for restricted circuit models, such as constant-depth multilinear formulas \cite{RazY09, chillara_near-optimal_2018}, and recently for general constant-depth algebraic circuits in the breakthrough by Limaye, Srinivasan and Tavenas \cite{LST2021}.
Other known results in this style are the separation between multilinear $\mathsf{NC}^1$ and $\mathsf{NC}^2$ circuits \cite{Raz06}, and depth hierarchies of various strengths in the Boolean setting, that is, for $\mathsf{AC}^0$ circuits \cite{Sipser83, HastadCircuitBounds, HastadRST17}.

\subsection{Results}

Our main results establish depth hierarchies for symmetric algebraic circuits, both within the constant-depth regime between $\Delta$ and $\Delta+1$, as well as between depths $(\log n)^k$ and $(\log n)^{k+1}$ for all constants~$k$. The former is a symmetric analogue of the depth hierarchy inside $\VAC^0$, which is the class of polynomials computable by constant-depth polynomial-size circuits, and the latter shows a hierarchy between symmetric analogues of the classes $\VAC^{k}$ and $\VAC^{k+1}$ for all $k$. 

These results are not implied by the hierarchies that are known in the non-symmetric setting, and are in some sense incomparable:
While the symmetric model is syntactically restricted, its complexity measure is more lenient than for standard circuits. Namely, the natural size parameter of symmetric circuits is their \emph{orbit size}, which is the number of gates in its largest orbit under the action of the symmetry group. Thus, a circuit can be considered efficient under symmetry aspects even when it contains a super-polynomial number of gates, as long as for each gate, there is at most a polynomial number of gates that are symmetric to it.
A circuit with super-polynomial orbit size has in particular super-polynomial size, so lower bound results for symmetric circuits apply to the standard notion of size (i.e., total number of gates), too. 
Our depth hierarchy holds both with respect to the orbit size and standard circuit size measure.

\begin{theorem}[label=thm:depth-hierarchyInformal,restate=thmDepthHierarchyInformal]
    \lean{DepthHierarchy}
Fix any $\Delta \geq 1$. There is a family $(p_n)_{n \in \bbN}$ of matrix-symmetric polynomials such that
    \begin{enumerate}
        \item Every matrix-symmetric circuit of depth $\leq \Delta$ computing
         $p_n$ has size at least \(  n^{\Omega(\log_2 n)} \).
        \item There is a matrix-symmetric circuit of depth $2\Delta+3$ and size $O(n^{\Delta +2})$ computing $p_n$.
    \end{enumerate}
\end{theorem}

Stated in terms of the notion of \emph{support depth}, which is the semantically relevant depth parameter for symmetric circuits (see \cref{def:support-depth}), the separation is in fact between $\Delta$ and $\Delta+1$, see \cref{thm:depth-hierarchy} for the formal statement.
This is similar to other depth hierarchy results where \emph{product depth}, rather than depth, is the relevant measure. 
Moreover, support depth and depth agree up to a factor of~$2$ at polynomial orbit size, as we will see in \cref{cor:depth-vs-scd}.

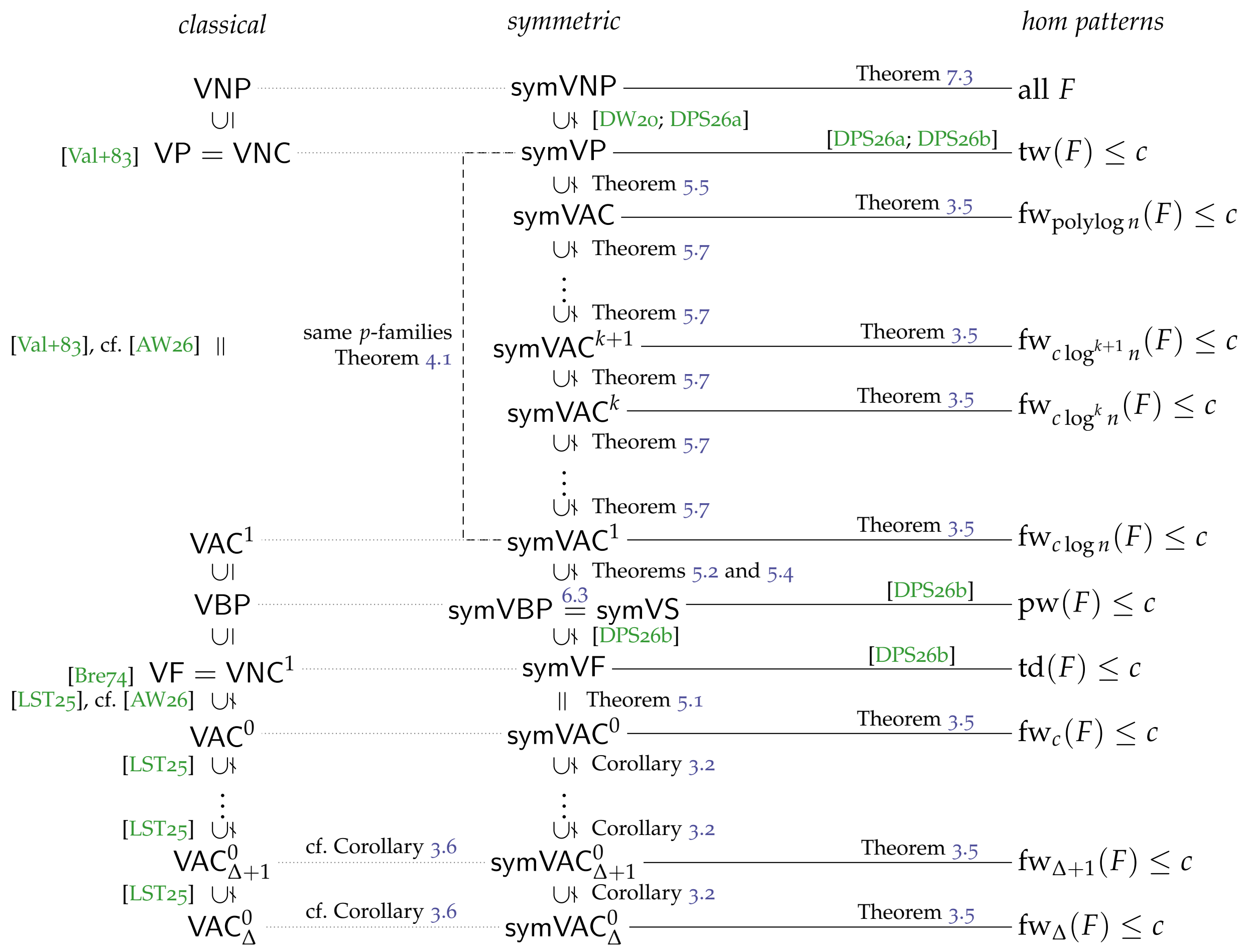
\begin{figure}
    \centering
    \begin{tikzpicture}[
        x=1cm, y=0.85cm,
        cls/.style={inner sep=2pt},
        rel/.style={font=\small, inner sep=1pt},
        ref/.style={font=\scriptsize, inner sep=1pt, anchor=west},
        corr/.style={densely dotted, gray},
        head/.style={font=\small\itshape},
    ]
        \newcommand{\strict}{\rotatebox{90}{$\subsetneq$}}
        \newcommand{\incl}{\rotatebox{90}{$\subseteq$}}
        \newcommand{\eqq}{\rotatebox{90}{$=$}}

        \node[head] at (0,14) {classical};
        \node[head] at (4.5,14) {symmetric};
        \node[head, anchor=west] at (10.4,14) {hom patterns};

        \node[cls] (sD)  at (4.5,0) {$\symVAC^0_\Delta$};
        \node[cls] (sD1) at (4.5,1) {$\symVAC^0_{\Delta+1}$};
        \node[cls] (s0d) at (4.5,2) {$\vdots$};
        \node[cls] (s0)  at (4.5,3) {$\symVAC^0$};
        \node[cls] (sF)  at (4.5,4) {$\symVF$};
        \node[cls] (sBP) at (4.5,5) {$\symVBP \overset{\ref{thm:symVBP-symVS}}{=} \symVS$};
        \node[cls] (s1)  at (4.5,6) {$\symVAC^1$};
        \node[cls] (s1d) at (4.5,7) {$\vdots$};
        \node[cls] (sk)  at (4.5,8) {$\symVAC^k$};
        \node[cls] (sk1) at (4.5,9) {$\symVAC^{k+1}$};
        \node[cls] (skd) at (4.5,10) {$\vdots$};
        \node[cls] (s)   at (4.5,11) {$\symVAC$};
        \node[cls] (sP)  at (4.5,12) {$\symVP$};
        \node[cls] (sNP) at (4.5,13) {$\symVNP$};
        \foreach \a/\b/\r/\c in {
            sD/sD1/\strict/{\cref{cor:symVAC0-hierarchy}},
            sD1/s0d/\strict/{\cref{cor:symVAC0-hierarchy}},
            s0d/s0/\strict/{\cref{cor:symVAC0-hierarchy}},
            s0/sF/\eqq/{\cref{thm:symVAC-symVF}},
            sF/sBP/\strict/{\cite{dwivedi_lower_2026}},
            sBP/s1/\strict/{\cref{cor:symVBP-symVAC1,cor:symVBP-symVAC1-strict}},
            s1/s1d/\strict/{\cref{thm:symVACk-hierarchy}},
            s1d/sk/\strict/{\cref{thm:symVACk-hierarchy}},
            sk/sk1/\strict/{\cref{thm:symVACk-hierarchy}},
            sk1/skd/\strict/{\cref{thm:symVACk-hierarchy}},
            skd/s/\strict/{\cref{thm:symVACk-hierarchy}},
            s/sP/\strict/{\cref{thm:symVAC1-separation}},
            sP/sNP/\strict/{\cite{dawar_symmetric_2020,dps_itcs}}%
        } {
            \node[rel] (r-\a) at ($(\a)!0.5!(\b)$) {\r};
            \node[ref] at ([xshift=4pt]r-\a.east) {\c};
        }
        \draw[densely dashed] (s1.west) -- ++(-0.5,0) |- node[ref, pos=0.25, anchor=east, align=right,xshift=-.1cm] {same $p$-families\\\cref{thm:log-depth-reduction}} (sP.west);

        \foreach \a/\t/\c in {
            sD/{$\fw_\Delta(F) \leq c$}/{\cref{thm:bounded-depth-characterisation}},
            sD1/{$\fw_{\Delta+1}(F) \leq c$}/{\cref{thm:bounded-depth-characterisation}},
            s0/{$\fw_c(F) \leq c$}/{\cref{thm:bounded-depth-characterisation}},
            sF/{$\td(F) \leq c$}/{\cite{dwivedi_lower_2026}},
            sBP/{$\pw(F) \leq c$}/{\cite{dwivedi_lower_2026}},
            s1/{$\fw_{c \log n}(F) \leq c$}/{\cref{thm:bounded-depth-characterisation}},
            sk/{$\fw_{c \log^k n}(F) \leq c$}/{\cref{thm:bounded-depth-characterisation}},
            sk1/{$\fw_{c \log^{k+1} n}(F) \leq c$}/{\cref{thm:bounded-depth-characterisation}},
            s/{$\fw_{\mathrm{polylog}\, n}(F) \leq c$}/{\cref{thm:bounded-depth-characterisation}},
            sP/{$\tw(F) \leq c$}/{\cite{dps_itcs,dwivedi_lower_2026}},
            sNP/{all $F$}/{\cref{thm:all-matrix-symmetric-symVNP}}%
        } {
            \node[cls, anchor=west] (g-\a) at (\a -| 10.4,0) {\t};
            \draw[corr, solid, black] (\a.east) -- node[ref, text=black, pos=0.75, anchor=south] {\c} (g-\a.west);
        }

        \node[cls] (cD)  at (0,0) {$\mathsf{VAC}^0_\Delta$};
        \node[cls] (cD1) at (0,1) {$\mathsf{VAC}^0_{\Delta+1}$};
        \node[cls] (c0d) at (0,2) {$\vdots$};
        \node[cls] (c0)  at (0,3) {$\mathsf{VAC}^0$};
        \node[cls] (cF)  at (0,4) {$\VF = \mathsf{VNC}^1$};
        \node[cls] (cBP) at (0,5) {$\VBP$};
        \node[cls] (cA1) at (0,6) {$\mathsf{VAC}^1$};
        \node[cls] (cP)  at (0,12) {$\VP = \mathsf{VNC}$};
        \node[cls] (cNP) at (0,13) {$\VNP$};
        \foreach \a/\b/\r/\c in {
            cD/cD1/\strict/{\cite{LST2021}},
            cD1/c0d/\strict/{\cite{LST2021}},
            c0d/c0/\strict/{\cite{LST2021}},
            c0/cF/\strict/{\cite{LST2021}, cf.\ \cite{andrews_constant-depth_2026}},
            cF/cBP/\incl/{},
            cBP/cA1/\incl/{},
            cA1/cP/\eqq/{\cite{valiant_fast_1983}, cf.\ \cite{andrews_constant-depth_2026}},
            cP/cNP/\incl/{}%
        } {
            \node[rel] (r-\a) at ($(\a)!0.5!(\b)$) {\r};
            \node[ref, anchor=east] at ([xshift=-4pt]r-\a.west) {\c};
        }
        \node[ref, anchor=base east] at ([xshift=-2pt]cF.base west) {\cite{brent_parallel_1974}};
        \node[ref, anchor=base east] at ([xshift=-2pt]cP.base west) {\cite{valiant_fast_1983}};

        \foreach \a/\b/\c in {sD/cD/{cf.\ \cref{cor:depth-vs-scd}}, sD1/cD1/{cf.\ \cref{cor:depth-vs-scd}}, s0/c0/{}, sF/cF/{}, sBP/cBP/{}, s1/cA1/{}, sP/cP/{}, sNP/cNP/{}} {
            \draw[corr] (\a.west) -- node[ref, text=black, anchor=south] {\c} (\b.east);
        }
    \end{tikzpicture}
    \caption{The classical landscape of algebraic complexity classes (left) and the symmetric landscape developed here (middle).
    Dotted lines connect symmetric classes to their classical counterparts (by identifying size with orbit size, depth with support depth).
    On the classical side, all classes consist of $p$-families, that is, families of polynomials with polynomially bounded degree: $\VP$ and $\VNP$ by definition, $\mathsf{VAC}^0$, $\VF = \mathsf{VNC}^1$, and $\VBP$ automatically, and $\mathsf{VAC}^1$ and $\mathsf{VNC}$ are restricted to $p$-families.
    Each symmetric class consists precisely of the linear combinations of homomorphism polynomials $\hom_{F,n}$ for patterns $F$ satisfying the condition on the right, for a constant $c$. 
    Restricted to $p$-families, $\symVP$ and $\symVAC^1$ are equal.
    }
    \label{fig:landscape}
\end{figure}

\cref{thm:depth-hierarchyInformal} motivates introducing bounded-depth symmetric algebraic complexity classes and locating them in the symmetric algebraic complexity hierarchy 
\[
\symVF \subsetneq \symVS \subsetneq \symVP
\]
that was established in \cite{dwivedi_lower_2026}.
Using sophisticated machinery from finite model theory \cite{grohe_iteration_2025} and the theory of homomorphism indistinguishability \cite{roberson_oddomorphisms_2022}, 
we place the bounded-depth classes in the existing picture, and unconditionally establish the separations depicted in \cref{fig:landscape}.

\begin{definition}\label{def:symVAC}
    For $\Delta \geq 0$, let $\symVAC^0_\Delta$ denote the class of families $(p_n)$ of matrix-symmetric polynomials for which there is a constant $c$ such that, for all sufficiently large $n$, the polynomial $p_n$ is computed by a matrix-symmetric circuit of orbit size $\leq n^c$ and support depth $\leq \Delta$.
    Let $\symVAC^0 \coloneqq \bigcup_{\Delta \geq 0} \symVAC^0_\Delta$.

    For $k \geq 1$, let $\symVAC^k$ denote the class of families $(p_n)$ of matrix-symmetric polynomials for which there is a constant $c$ such that, for all sufficiently large $n$, the polynomial $p_n$ is computed by a matrix-symmetric circuit of orbit size $\leq n^c$ and support depth $\leq c \log^k n$.
    Let $\symVAC \coloneqq \bigcup_{k \geq 1} \symVAC^k$.
\end{definition}

\paragraph{Symmetric algebraic complexity classes and homomorphism polynomials.}
The classes $\symVF$, $\symVS$, and $\symVP$ comprise matrix-symmetric polynomials that are computable by matrix-symmetric formulas, skew circuits, and circuits of polynomial orbit size, respectively. 
The main technical contribution of \cite{dawar_symmetric_2020, dwivedi_lower_2026} was a characterisation of these syntactic circuit classes in terms of a graph-theoretic semantics: Each of these classes consists precisely of linear combinations of \emph{homomorphism polynomials} of patterns of bounded treedepth, bounded pathwidth and bounded treewidth, respectively, cf.\ \cref{thm:symv-to-hom}. 
A homomorphism polynomial \[ \hom_{F,n}(X)=\sum_{h \colon V(F) \to [n]} \prod_{ab\in E(F)}x_{h(a),h(b)} \] is parametrised by a pattern graph $F$ and the size $n$ of the target graph. Its variables can be instantiated with the adjacency matrix of an $n$-vertex graph $G$, and $\hom_{F,n}(G)$ evaluates to the number of homomorphisms from $F$ to $G$. 
The separation of the symmetric algebraic complexity classes then follows from the separation of the graph distinguishing power of the corresponding classes of homomorphism polynomials.

The right column of \cref{fig:landscape} describes the structural graph parameter of the homomorphism polynomials whose linear combinations capture the respective symmetric algebraic complexity class. 
For example, the correspondence between $\symVP$ and $\tw(F) \leq c$ of \cite{dps_itcs} is to be read as follows: 
A family $(p_n)_{n \in \bbN}$ is in $\symVP$ if and only if there is a fixed constant $c \in \bbN$ such that for every $n \in \bbN$
\begin{equation}\label{eq:intro-symvp}
p_n(X) = \sum_{F \text{ with } \tw(F) \leq c} \alpha_{F,n} \cdot \hom_{F,n}(X),
\end{equation}
for some choice of coefficients $\alpha_{F,n} \in \bbQ$.

In \cref{thm:bounded-depth-characterisation}, we prove a characterisation of the polynomials computed by matrix-symmetric circuits of a given orbit size and support depth.
The graph parameter arising here is \emph{$\Delta$\nobreakdash-forestwidth} $\fw_\Delta$, a graph parameter inspired by $\Delta$\nobreakdash-treewidth of \textcite{bhargav_monotone_2025}:
For $\Delta \geq 0$,
a graph $F$ has \emph{$\Delta$\nobreakdash-treewidth} $\tw_\Delta(F) \leq k$
if it admits a tree decomposition $(T, \beta)$ of width~$\leq k$ such that the decomposition tree is of depth at most $\Delta$.
For example, $\tw_0(F) = |V(F)|-1$, as $\Delta = 0$ forces the tree decomposition to consist of a single vertex, and $\tw_1(F)$ is closely related to the vertex-cover number of~$F$, see \cite{bhargav_monotone_2025}.
Bounded-depth tree decompositions are studied in finite model theory \cite{adler_going_2026} and also arise as divide-and-conquer schemes for iterated matrix multiplication \cite{LST2021}.
$\Delta$-forestwidth differs from $\Delta$-treewidth only in the additional technical requirement that the root bag be empty.

In order to describe the bounded-depth setting, our characterisation \cref{thm:bounded-depth-characterisation} needs to be more precise than previous results \cite{dwivedi_lower_2026,dps_itcs}, which it fully subsumes.
In particular, \cref{thm:bounded-depth-characterisation} does not only hold asymptotically but at every fixed host size~$n$, see \cref{sec:depth-hierarchy} for a discussion.
This allows us to overcome the limitation of \cite{dwivedi_lower_2026,dps_itcs} to circuits of polynomial orbit size.


\paragraph*{Comparing classical and symmetric hierarchies.}
The depth hierarchy within $\symVAC^0$ parallels the one within $\VAC^0$, though it follows from completely different arguments. As noted earlier, it is not just a special case of the result for $\VAC^0$, among other things because of the distinction of size versus orbit size. 
The separations between $\symVAC^k$ and $\symVAC^{k+1}$ have an essentially uninteresting analogue in the non-symmetric setting: Since the definition of $\VAC^k$ bounds total circuit size rather than orbit size, there is an upper bound of $2^{O(\log^{k+1} n)}$ on the degree of the polynomials that $\VAC^k$ can compute---therefore, $\VAC^{k+1}$ contains polynomials that are not in $\VAC^k$ simply because their degree is too large. The definition of $\symVAC^k$, however, puts no upper bound on the total circuit size, but rather on the orbit size. This does not entail an upper bound on the degree, so the separation between $\symVAC^k$ and $\symVAC^{k+1}$ is a genuine result, not an immediate consequence of the definition. 
It builds on quite sophisticated lower bounds on the iteration number of the Weisfeiler--Leman algorithm for graph isomorphism \cite{grohe_iteration_2025}. 
This iteration number translates into symmetric circuit depth via our homomorphism polynomial characterisation.

A new separation that we establish in the symmetric setting and which is not known to hold for non-symmetric circuits is $\symVBP = \symVS \subsetneq \symVAC^1$. Even the inclusion, not just the separation, is surprising, given that no degree restrictions are imposed in these classes. For example, $\symVS$ contains the homomorphism polynomials of paths with $2^n$ vertices, whose natural representation in $\symVS$ has depth $2^{\Omega(n)}$. It is not at all straightforward to show that these polynomials are also in
$\symVAC^1$, i.e., require only logarithmic depth. For $p$-families, this follows from \cref{thm:log-depth-reduction}, but for families of higher degree, this requires substantially different techniques from invariant theory, as our graph-theoretic criterion alone does not yield this.

\paragraph*{Symmetric VBP and VNP.}
As depicted in the figure, we also provide definitions of the symmetric classes $\symVNP$ and $\symVBP$, for which no natural definitions were known so far.
While $\symVS$ was defined in terms of symmetric skew circuits, $\symVBP$ comprises the families of polynomials computable by symmetric algebraic branching programs (ABPs) of polynomial orbit size. In the general non-symmetric setting, skew circuits and ABPs are polynomially equivalent. As we prove in \cref{thm:symVBP-symVS}, via the homomorphism polynomial characterisation, this carries over to the symmetric setting, so that we also have $\symVS = \symVBP$. 
In order to define the class $\symVNP$, we place a natural symmetry condition on the definition of $\VNP$ in terms of hypercube sums over $\VP$-polynomials, see \cref{def:symVNP}. This definition turns out to yield the correct symmetric analogue of $\VNP$ in the sense that it satisfies the following neat properties: Firstly, $\symVNP$ is equal to the class of \emph{all} matrix-symmetric polynomials, see \cref{thm:all-matrix-symmetric-symVNP}. Secondly, the permanent is complete for $\symVNP$ under \emph{symmetric} projections (\cref{thm:permanent-completeness})---this is in analogy to the $\VNP$-completeness of the permanent under not necessarily symmetric projections.
Symmetric formulas are strictly weaker than symmetric circuits~\cite[Theorem~A]{dwivedi_lower_2026}. Nevertheless, \cref{cor:symvnp-formula-witnesses} shows that their Boolean sums define the same class. Thus, as in classical $\VNP$, restricting witnesses to formulas does not reduce expressive power.

\subsection{Techniques} 
On a high level, the proof of \cref{thm:depth-hierarchyInformal} follows the strategy established in \cite{dwivedi_lower_2026,dps_itcs} for separating symmetric algebraic complexity classes.
Its two ingredients are a \emph{characterisation} of a symmetric algebraic complexity class in terms of homomorphism polynomials for patterns from some graph classes of restricted structure and a \emph{separation} result asserting that a certain polynomial cannot be written as a linear combination of homomorphism polynomials for patterns from some graph class.
However, the ideas in \cite{dwivedi_lower_2026,dps_itcs} are insufficient for \cref{thm:depth-hierarchyInformal} and need to be substantially refined.

\paragraph{Characterisations.}
The characterisations are insufficient as they only apply to circuits of polynomial orbit size. Our lower bounds require considering circuits of orbit size $n^{\Omega(\log n)}$ and thus a tighter analysis.
Moreover, the most technical ingredient of \cite{dwivedi_lower_2026,dps_itcs}, a lemma describing how support-changing gates affect the inductively constructed linear combination of homomorphism polynomials, needs to be replaced entirely.
For details, see \cref{sec:extended-differences}.

\paragraph{Separation.}
Even though every matrix-symmetric polynomial can be written as a linear combination of homomorphism polynomials, this linear combination is by no means unique; see \cite[Example~3.1]{dwivedi_lower_2026} and \cite{Cur2022} for examples.
\Cref{fig:landscape} also illustrates this:
By \cref{cor:symVBP-symVAC1,thm:bounded-depth-characterisation,thm:symv-to-hom}, it holds that $\symVBP \subseteq \symVAC^1$, i.e., every polynomial that can be written as a linear combination of homomorphism polynomials for patterns of pathwidth~$O(1)$ can also be written as such a combination for patterns of $O(\log n)$-forestwidth $O(1)$.
However, there are families of graphs of constant pathwidth and super-constant $O(\log n)$-forestwidth.
Nonetheless, their homomorphism polynomials lie in $\symVAC^1$.

\textcite{dps_itcs} introduced the \emph{non-uniform descriptive complexity monotonicity} for ruling out such failures of homomorphism polynomials of distinct patterns to be linearly independent.
For example, in order to show that a matrix-symmetric polynomial $p_n$ is not in $\symVP$, one may (recalling \eqref{eq:intro-symvp}) construct $n$-vertex graphs $G$ and $H$ that are \emph{homomorphism indistinguishable} over all graphs of treewidth $\leq c$, i.e., admit the same number of homomorphisms from every such graph, 
but differ in the value of $p_n(G) \neq p_n(H)$.
The construction of $G$ and $H$ in \cite{dps_itcs}
crucially relies on the fact that the bound on the treewidth $\leq c$ is constant while the host size~$n$ is unbounded.
It therefore cannot yield lower bounds as in \cref{thm:depth-hierarchyInformal} or lower bounds against $\symVAC^k$ where \enquote{treewidth $\leq c$} is replaced by \enquote{$c\log^k n$-forestwidth $\leq c$}, a condition depending on the host size~$n$.
In \cref{thm:non-uniform-non-uniform}, we overcome this issue using techniques from structural graph theory and homomorphism indistinguishability \cite{roberson_oddomorphisms_2022}.
Our upgraded non-uniform descriptive complexity monotonicity supersedes \cite{dwivedi_lower_2026,dps_itcs} and has a more direct proof.

\subsection{Future work}

We highlight the following three directions for future work that may build on the results we have obtained.
Firstly, our logarithmic depth reduction for $p$-families (\cref{thm:log-depth-reduction}) raises the question of whether symmetric circuits can be reduced further, to depth $3$ or $4$, and at what cost in orbit size.
In the classical setting, depth reduction began with balancing circuits~\cite{hyafil_parallel_1979,valiant_fast_1983} and formulas~\cite{brent_parallel_1974,BCE1995}. Later work gave sharper bounds for homogeneous formulas~\cite{FLMST2026} and syntactically multilinear circuits~\cite{KOS2019}.
The depth-four~\cite{agrawal_arithmetic_2008,Koiran2012,Tavenas15} and depth-three reductions over the rationals~\cite{KKS16,Tavenas15} allow sufficiently strong lower bounds for shallow circuits to be lifted to general circuits. This is the motivation for seeking analogous reductions that preserve symmetry.

Another promising direction is to port the symmetry framework to \emph{algebraic proof complexity}. In \cite{DawarGKP25}, a symmetric version of the \emph{Ideal Proof System} (IPS), which is based on algebraic circuits, was introduced. Can symmetries be exploited in order to prove strong lower bounds against this proof system? 
An active recent line of work has established lower bounds for very restricted fragments of IPS, for example ro-ABP or constant depth multilinear IPS proofs \cite{ForbesSTW21, govindasamy2022simple, HakoniemiLT24, ElbazGLT26}.
A natural next step in this direction would be to attempt lower bounds for bounded depth symmetric IPS proofs. The machinery we have developed in this paper may serve as a tool to approach this problem.

Finally, we believe that our definition of the class $\symVBP$ is a first step towards combining the symmetry-based lower bound approach with an ongoing programme on stronger (multilinear read-once oblivious) ABP lower bounds, e.g., \cite{ChatterjeeKSS24, BhargavDS25, FabrisLSY26}. Our definition and homomorphism polynomial characterisation of symmetric ABPs may be used in combination with other techniques to make new progress in this line of research, too.

\section{Preliminaries}

We work over a field $\K$ of characteristic zero. Write $[n]=\{1,\ldots,n\}$ and let $\Sym_n$ be its permutation group. We follow the circuit and graph-polynomial conventions of \cite{dps_itcs,dwivedi_lower_2026}.

\subsection{Circuits and polynomial families}

An \emph{algebraic circuit} is a finite directed acyclic graph whose input gates are labelled by variables or field constants and whose internal gates are labelled by $+$ or $\times$. Input gates with the same label are identified. We allow unbounded fan-in and parallel wires, counted with multiplicity in each operation. There is a unique output gate, and every gate lies on a path to it. The \emph{size} is the number of gates and wires, counting multiplicities. The \emph{depth} is the maximum number of wires on an input-to-output path. Likewise, the depth of a rooted tree or forest counts edges, so a single node has depth~$0$.

A \emph{formula} has no parallel wires and its internal gates induce a tree; input gates may be shared. A circuit is \emph{skew} if each multiplication gate has at most one incoming wire from an internal gate.

A function is \emph{$p$-bounded} if it is bounded by a polynomial in its arguments. A \emph{$p$-family} is a polynomial family whose number of variables and degree are $p$-bounded. The classes $\VP$ and $\VF$ consist of $p$-families computed by circuits and formulas, respectively, of polynomial total size. A $p$-family $(f_n)$ belongs to $\VNP$ if there are a $p$-family $(g_n)\in\VP$ and a $p$-bounded function $q$ such that
\[
    f_n(X_n)=\sum_{\boldsymbol{e}\in\{0,1\}^{q(n)}}g_n(X_n,\boldsymbol{e}).
\]

\subsection{Matrix symmetry and orbit size}\label{sec:matrix-symmetry-orbits}
\label{sec:matrixSymmetry}
For $n\geq 1$, put
\(
    \Xx_n=\{x_{i,j}:i,j\in[n]\}.
\)
The group $\Gamma_n=\Sym_n\times\Sym_n$ acts on $\Xx_n$ by $(\sigma,\tau)x_{i,j}=x_{\sigma(i),\tau(j)}$. A polynomial is \emph{matrix-symmetric} if it is invariant under this action. For a group $\Gamma$ acting on a set $\Omega$ and $S\subseteq\Omega$, write $\Stab_\Gamma(S)$ for the setwise stabiliser and $\StabP_\Gamma(S)$ for the pointwise stabiliser. We omit $\Gamma$ when it is clear.

Let $\Gamma$ act on the variables of a circuit $C$, fixing field constants. Write $\lambda(v)$ for the label of a gate $v$. A circuit automorphism preserves internal operation labels and the directions and multiplicities of wires. Such an automorphism $\sigma$ \emph{extends} $\gamma\in\Gamma$ if
$ \lambda(\sigma(v))=\gamma(\lambda(v))$ 
for every input gate $v$.

\begin{definition}[Symmetric circuits]\label{def:symmetric-ckts}
    \lean{MatrixSymmetry}
    A circuit is \emph{$\Gamma$-symmetric} if every $\gamma\in\Gamma$ extends to a circuit automorphism. For $\Gamma=\Gamma_n$ acting on $\Xx_n$, we call it matrix-symmetric.
\end{definition}

Let $\Aut_\Gamma(C)$ be the group of all automorphisms extending some element of $\Gamma$. The \emph{orbit size} is
\[
    \maxorb(C)=\max_{v\in V(C)}
        \bigl|\{\sigma(v):\sigma\in\Aut_\Gamma(C)\}\bigr|.
\]

\subsection{Rigid circuits, supports, and support depth}\label{sec:supports}

The group $\Gamma_n$ acts on $[n] \uplus [n]$ and on $\Xx_n$.

\begin{definition}\label{def:rigid-reduced}
    A matrix-symmetric circuit $C$ is \emph{rigid} if every element of $\Gamma_n$ extends to a unique automorphism of $C$.
    It is \emph{reduced} if no two distinct input gates have the same label and no two distinct internal gates have the same label and the same multiset of children.
\end{definition}

If $C$ is rigid, then $\Gamma_n$ acts on the gates of $C$, and the orbit size of $C$ is the maximal index of a stabiliser $\Stab(g) \leq \Gamma_n$ of a gate~$g$.

\begin{lemma}[label=lem:reduced-rigid,restate=lemReducedRigid]
    \lean{ReducedRigid}
    Every reduced matrix-symmetric circuit is rigid.
\end{lemma}

\begin{lemma}[name=Rigidification,label=lem:reduction,restate=lemReduction]
    For every matrix-symmetric circuit $C$, there is a reduced matrix-symmetric circuit $C'$ computing the same polynomial whose size, depth, and orbit size are at most those of $C$.
    \lean{Rigidification}
\end{lemma}

The proofs are given in \cref{app:supports}. 
We call the circuit $C'$ constructed there the \emph{rigidification} of $C$. The rigidification lemma here has appeared in similar forms in the literature on symmetric circuits before, e.g., in \cite{dps_itcs, dwivedi_lower_2026}. In all these instances, the lemma actually yields not only a rigid, but also a \emph{reduced} circuit (which was not always made explicit). Reducedness is in fact a stronger condition than rigidity, and the support theorems for symmetric circuits, as presented below, are actually based on this stronger assumption.

Let $C$ be rigid.
A set $S \subseteq [n] \uplus [n]$ \emph{supports} a gate $g$ if the pointwise stabiliser $\StabP(S)$ fixes $g$.
We write $S_L$ and $S_R$ for the two sides of $S$.
If $g$ has a support $S$ with $|S_L|, |S_R| < n/2$, then $g$ has a unique minimal support $\sup(g) \subseteq S$, which is contained in every support of $g$,
and $\sup(\pi g) = \pi \sup(g)$ for all $\pi \in \Gamma_n$ \cite[Lemmas~2.1 and~2.2]{dps_itcs}.
The \emph{support size} $\maxsup(C)$ of $C$ is the least $k$ such that every gate of $C$ has a support of size $\leq k$.
If $\maxsup(C) < n/2$, then every gate $g$ of $C$ has a minimal support and $\maxsup(C) = \max_g |\sup(g)|$.
In this case, a wire from $h$ to $g$ is \emph{support-changing} if $\sup(h) \not\subseteq \sup(g)$, and
the \emph{support depth}\label{def:support-depth} of $C$ is the maximum number of support-changing wires on a path from the output gate to an input gate.
If $\maxsup(C) \geq n/2$, we let the support depth of $C$ be its depth.
For an arbitrary matrix-symmetric circuit, we define its support size and support depth as those of its rigidification.
In any case, the support depth of a circuit is at most its depth.

\begin{observation}\label{obs:support-orbit}
    Every rigid matrix-symmetric circuit of support size $\leq k$ has orbit size $\leq n^k$.
\end{observation}
\begin{proof}
    Let $S$ be a support of size $\leq k$ of a gate $g$.
    Since $\StabP(S)$ fixes $g$, the orbit of $g$ has size at most the number of images of $S$ under $\Gamma_n$, which is $\leq n^{|S|}$.
\end{proof}

Conversely, small orbit size implies small support size.
In contrast to \cite{dps_itcs,dwivedi_lower_2026}, we allow circuits of superpolynomial orbit size.
For a function $f \colon \mathbb{N} \to \mathbb{N}$ bounding the orbit size, let
\begin{equation}
    \kappa_f(n) \coloneqq \min\left\{ r\in \mathbb{N} : \binom{n}{r} > f(n) \right\}.
\end{equation}
If $f(n) \in n^{O(1)}$, then $\kappa_f(n) \in O(1)$, i.e.\ polynomial orbit size corresponds to constant support size, as in \cite{dps_itcs,dwivedi_lower_2026}.
The proof of the following theorem  is given in \cref{app:supports}.

\begin{theorem}[name=Support theorem,label=thm:support,restate=thmSupport]
    \lean{SupportTheorem}
    Let $f \colon \mathbb{N} \to \mathbb{N}$.
    Let $n \geq 20$ such that $3\kappa_f(n) \leq n+2$.
    Then every gate of a reduced matrix-symmetric circuit of orbit size $\leq f(n)$ has a minimal support of size $\leq \kappa_f(n) - 1$.
\end{theorem}

\subsection{Homomorphism polynomials and graph parameters}

A \emph{pattern} is a finite bipartite multigraph $F$ with a fixed bipartition $A\uplus B$. For $n \geq 1$, its \emph{$n$-th homomorphism polynomial} is
\[
    \hom_{F,n}(X)=
    \sum_{\substack{h_A \colon A\to[n]\\h_B \colon B\to[n]}} \,
        \prod_{ab\in E(F)}x_{h_A(a),h_B(b)}.
\]
The product counts edges with multiplicity. Evaluated on the bi-adjacency matrix of a bipartite graph, this polynomial counts bipartition-respecting homomorphisms.

\begin{observation}\label{obs:no-isolated-vertices}
    If $F + K_1$ is obtained from $F$ by adding an isolated vertex to the left part of the bipartition, 
    then $\hom_{F+ K_1, n} = n \hom_{F, n}$ for all $n \geq 1$. 
    By adjusting coefficients, we may assume that the patterns have no isolated vertices.
\end{observation}

A \emph{tree decomposition} of a graph $F$ is a tree $T$ with bags $\beta(t)\subseteq V(F)$ covering $V(F)$. Each edge has both endpoints in some bag, and the bags containing any fixed vertex induce a connected subtree. Its width is $\max_t|\beta(t)|-1$. Minimising this width gives the \emph{treewidth} $\tw(F)$; requiring $T$ to be a path gives the \emph{pathwidth} $\pw(F)$. A forest decomposition uses a forest in place of $T$.

An \emph{elimination forest} of $F$ is a rooted forest on $V(F)$ in which the endpoints of every edge are comparable in the ancestor order. The \emph{treedepth} $\td(F)$ is the minimum possible maximum number of vertices on a root-leaf path in such a forest. Thus $\td(K_1)=1$, while the depth of a one-node rooted tree is $0$. These graph parameters ignore edge multiplicities.

\subsection{Symmetric algebraic complexity classes}\label{sec:symmetric-complexity-classes}

We recall the complexity classes introduced by \textcite{dwivedi_lower_2026}.

\begin{definition}[\cite{dwivedi_lower_2026}]\label{def:symVP} \lean{SymmetricClasses}
    The classes $\symVP$, $\symVS$, and $\symVF$ contain all families $(p_n)$ of matrix-symmetric polynomials computed by matrix-symmetric circuits, skew circuits, and formulas, respectively, of orbit size $n^{O(1)}$.
\end{definition}

These classes impose no degree or total-size bound. The class $\symVBP$ of symmetric algebraic branching programs is defined in \cref{sec:symmetric-abps}.
\Cref{thm:symVBP-symVS} shows that $\symVBP = \symVS$.

\begin{theorem}[\cite{dwivedi_lower_2026,dps_itcs}]\label{thm:symv-to-hom}
    Let $(p_n)$ be a family of matrix-symmetric polynomials.
    \begin{enumerate}
        \item \lean{SymmetricVPCharacterisation} $(p_n) \in \symVP$ if, and only if, there is a constant $k$ such that, for all sufficiently large $n$, $p_n$ is a linear combination of homomorphism polynomials for patterns of treewidth $\leq k$.
        \item \lean{SkewCircuitCharacterisation} $(p_n) \in \symVS$ if, and only if, there is a constant $k$ such that, for all sufficiently large $n$, $p_n$ is a linear combination of homomorphism polynomials for patterns of pathwidth $\leq k$.
        \item \lean{SymmetricVFCharacterisation} $(p_n) \in \symVF$ if, and only if, there is a constant $k$ such that, for all sufficiently large $n$, $p_n$ is a linear combination of homomorphism polynomials for patterns of treedepth $\leq k$. 
    \end{enumerate}
\end{theorem}

\section{Symmetric constant-depth hierarchy}
\label{sec:depth-hierarchy}

In this section, we prove the following depth hierarchy theorem for matrix-symmetric circuits, which makes \cref{thm:depth-hierarchyInformal} precise.

\begin{theorem}[label=thm:depth-hierarchy,restate=thmDepthHierarchy]
    \lean{DepthHierarchy}
    Let $\Delta \geq 1$ and $ n \geq 20$. Let $T_{\Delta,n}$ denote the full $n$-ary tree of depth~$\Delta$.
    \begin{enumerate}
        \item Every matrix-symmetric circuit of support depth $\leq \Delta$ computing
         $\hom_{T_{\Delta,n},n}$ has orbit size, and hence size, at least
        \[
            n^{\log_2 n - \Delta\log_2\log_2 n - 3}.
        \]
        \item There is a matrix-symmetric circuit of
        support depth $\Delta+1$, depth $2\Delta+3$, orbit size $\leq n^2$, and size $\leq 13n^{\Delta+2}$ computing $\hom_{T_{\Delta,n},n}$.
    \end{enumerate}
\end{theorem}

Thus, for fixed $\Delta$, the size gap between support depth $\Delta$ and $\Delta+1$ for these polynomials is $n^{(1-o(1))\log_2 n}$. \cref{thm:depth-hierarchy} has the following immediate corollary.

\begin{corollary}\label{cor:symVAC0-hierarchy}
    For every $\Delta \geq 0$,
    it holds that $\symVAC^0_\Delta \subsetneq \symVAC^0_{\Delta+1}$.
\end{corollary}

The second assertion of \cref{thm:depth-hierarchy} is standard and proven in \cref{thm:converse}.
For the first assertion, we substantially refine the machinery of \cite{dps_itcs,dwivedi_lower_2026}.
The proof is conducted in two steps:
In \cref{sec:characterisations}, we give a translation between matrix-symmetric algebraic circuits and homomorphism polynomials such that support size and support depth of the former correspond to width and depth of tree decompositions of the patterns of the latter.
In \cref{sec:separation},
we show that homomorphism polynomials for $T_{\Delta, n}$ cannot be written as linear combinations of homomorphism polynomials of patterns of  $\Delta$-forestwidth $o(\log n)$.

\subsection{Characterisations}
\label{sec:characterisations}

The first step is to characterise the polynomials computed by bounded-depth matrix-symmetric circuits in terms of homomorphism polynomials.
The graph-theoretic parameter of the patterns arising here is a modification of the \emph{$\Delta$-treewidth}\footnote{In the definition of $\Delta$-treewidth of \cite{bhargav_monotone_2025}, the forest decomposition is taken to be a tree decomposition.} of \textcite{bhargav_monotone_2025}.

\begin{definition}
    \lean{Forestwidth}\label{def:forestwidth}
    Let $\Delta \geq 1$.
    The \emph{$\Delta$-forestwidth} $\fw_\Delta(F)$ of $F$ is the least $k \geq 0$ such that $F$ admits a tree decomposition $(T, \beta)$ with root $r$ such that $T$ has depth $\leq \Delta$, $\beta(r) = \emptyset$, and $|\beta(t)| \leq k+1$ for all $t \in V(T)$.
    Equivalently, deleting $r$ yields a forest decomposition of $F$ in which every root-leaf path has at most $\Delta$ nodes.
\end{definition}

Our \cref{thm:depth-hierarchy} follows from the following observation.

\begin{example}[\protect{\cite[Theorem~17]{bhargav_monotone_2025}}]\label{example:bccd17}
    Let $\Delta, k \geq 1$.
    Let $T_{\Delta, k}$ denote the full $k$-ary tree of depth~$\Delta$.
    Then $\fw_{\Delta+1}(T_{\Delta, k}) = 1$ whereas $\fw_{\Delta}(T_{\Delta, k}) \geq k-1$.
\end{example}

The main result facilitating all separations in \cref{fig:landscape}
is the following characterisation of the polynomials computed by bounded-depth matrix-symmetric circuits.
The characterisation is exact when the circuits are measured by their supports rather than their orbit size, see \cref{sec:supports}.
\begin{theorem}
    \lean{BoundedDepthCharacterisation}
    \label{thm:bounded-depth-characterisation}
    Let $\Delta \geq 1$, $k \geq 2$, $n \geq 2k+1$.
    For a polynomial $p \in \mathbb{K}[\Xx_n]$, the following are equivalent:
    \begin{enumerate}
        \item $p$ is computed by a reduced matrix-symmetric circuit of support size $\leq k$ and support depth $\leq \Delta$.
        \item $p$ is a linear combination of homomorphism polynomials $\hom_{F,n}$ for patterns $F$ with $\fw_\Delta(F) \leq k-1$.
    \end{enumerate}
\end{theorem}

\Cref{thm:bounded-depth-characterisation} can be translated to a statement on orbit size via \cref{thm:support}.
Indeed, if $n \geq 20$, $n^2 \leq f(n)$, and $3\kappa_f(n) \leq n+2$, then every polynomial computed by a matrix-symmetric circuit of orbit size $\leq f(n)$ and support depth $\leq \Delta$ is a linear combination of homomorphism polynomials for patterns $F$ with $\fw_\Delta(F) \leq \kappa_f(n) - 2$.
Conversely, for $k \geq 1$ and $n \geq 2k+3$, linear combinations of homomorphism polynomials for patterns with $\fw_\Delta(F) \leq k$ admit matrix-symmetric circuits of orbit size $\leq n^{k+1}$ and support depth $\leq \Delta$.
The two directions do not match in terms of orbit size: for instance, for a fixed pattern $F$ of treewidth $t$ and all sufficiently large $n$, every matrix-symmetric circuit computing $\hom_{F,n}$ has orbit size $\geq \binom{n}{t+1}$.

As a corollary, we observe that support depth and depth agree up to a factor of two, at the expense of the circuit size.
Clearly, every matrix-symmetric circuit of depth $\leq \Delta$ has support depth $\leq \Delta$.
\Cref{cor:depth-vs-scd} gives the converse:

\begin{corollary}\label{cor:depth-vs-scd}
    Let $n \geq 20$, $\Delta \geq 1$, and $n^2 \leq s$ such that $\kappa \coloneqq \kappa_s(n)$ satisfies $3\kappa \leq n+2$.
    Every polynomial computed by a matrix-symmetric circuit of orbit size $\leq s$ and support depth $\leq \Delta$ is computed by a reduced matrix-symmetric circuit of depth $\leq 2\Delta + 2$, orbit size $\leq n^{\kappa-1}$, support size $\leq \kappa - 1$, and support depth $\leq \Delta$.
\end{corollary}
\begin{proof}
    Let $p$ be computed by such a circuit.
    By \cref{lem:reduction}, we may suppose that it is reduced.
    By \cref{thm:support}, it has support size $\leq \kappa - 1$, where $\kappa \geq 3$ as $s \geq n^2 > \binom{n}{2}$, and $n \geq 2(\kappa - 1) + 1$ as $3\kappa \leq n+2$.
    By \cref{thm:bounded-depth-characterisation}, $p$ is a linear combination of $\hom_{F,n}$ for patterns $F$ with $\fw_\Delta(F) \leq \kappa - 2$.
    The circuit constructed for this linear combination in \cref{app:converse} has the claimed properties by \cref{thm:converse,obs:support-orbit}, where the output sum gate adds one to the depth.
\end{proof}

\subsection{Separation}
\label{sec:separation}

Towards \cref{thm:depth-hierarchy},
it remains to argue that the homomorphism polynomials for $T_{\Delta, n}$ cannot be written as linear combinations of homomorphism polynomials of patterns of $\Delta$-forestwidth $o(\log n)$.
This will be implied by the following general theorem.

\begin{theorem}[restate=thmNonUnfirom,label=thm:non-uniform-non-uniform]
    \lean{NonUniformClosure}
    Let $\mathcal{F}$ be a class of simple graphs. Suppose that
    $\mathcal{F}$ contains the $1$-vertex graph $K_1$,
    that
    $\mathcal{F}$ is minor-closed, and that $\mathcal{F}$ is closed under weak oddomorphism.
    Let $G \not\in \mathcal{F}$ be a connected graph.
    Let $F$ be a bipartite multigraph containing $G$ as a minor.
    Let $N \geq \gamma(G)  \coloneqq \sum_{v \in V(G)} 2^{\deg_G(v) - 1}$.
    Then $\hom_{F, N} \not\in \mathfrak{H}_N(\mathcal{F})$.
\end{theorem}

Here, for a graph class $\mathcal{F}$ and $n \geq 1$, we write $\mathfrak{H}_n(\mathcal{F})$ for the $\mathbb{K}$-linear span of the homomorphism polynomials $\hom_{F, n}$ for all bipartite multigraphs $F$ whose underlying simple graph is in $\mathcal{F}$.

The statement playing the same role as \cref{thm:non-uniform-non-uniform} in \cite{dps_itcs} differs in that it only permits $G$ to be an induced minor of $F$, and its size bounds are less tight and thus insufficient for our needs. Moreover, the proof of \cref{thm:non-uniform-non-uniform} is more direct.

The assumption in \cref{thm:non-uniform-non-uniform} that $\mathcal{F}$ is closed under weak oddomorphism requires an explanation:
A \emph{weak oddomorphism} \cite{roberson_oddomorphisms_2022} is a homomorphism between two graphs satisfying certain parity constraints. 
These constraints arise from analysing the CFI construction \cite{cai_optimal_1992}, the most versatile method for constructing highly indistinguishable graphs. 
A graph class $\mathcal{F}$ is \emph{closed under weak oddomorphisms} if, for every $F \in \mathcal{F}$ and every weak oddomorphism $F \to G$, it holds that $G \in \mathcal{F}$.

\textcite{roberson_oddomorphisms_2022} showed that graph classes $\mathcal{F}$ that are closed under weak oddomorphisms (and under disjoint unions and taking summands) are \emph{homomorphism distinguishing closed}, i.e., for every $F' \not\in \mathcal{F}$, there exist $G$ and $H$ that admit the same number of homomorphisms $\hom(F, G) = \hom(F, H)$ from every $F\in \mathcal{F}$ but $\hom(F', G) \neq \hom(F', H)$.
In this case, $\hom_{F', N} \not\in \mathfrak{H}_N(\mathcal{F})$
for $N \geq |V(G)|, |V(H)|$.
Our \cref{thm:non-uniform-non-uniform} combines this result with a compact reduction to graph minors inspired by \cite{seppelt_logical_2024}.

The classes of graphs of treewidth~$\leq k$, pathwidth~$\leq k$, and treedepth~$\leq k$ are closed under weak oddomorphisms \cite{neuen_homomorphism-distinguishing_2024,seppelt_homomorphism_2024,adler_going_2026}.
These results underpin the $\symVF \subsetneq \symVS \subsetneq \symVP$ hierarchy of \textcite{dwivedi_lower_2026}.
We prove the analogous statement (in an approximate sense) for the class of graphs of $\Delta$-forestwidth $\leq k$ in \cref{cor:non-rich-class}.
This requires explicitly describing the forbidden minors of graphs of $\Delta$-forestwidth $\leq k$.
Combining \cref{thm:non-uniform-non-uniform,sec:characterisations} yields \cref{thm:depth-hierarchy} in \cref{proof:depth-hierarchy}.

\section{Symmetric depth reduction}\label{sec:depth-reduction}

\textcite{hyafil_parallel_1979} showed that polynomials of degree $d$ computed by circuits of size $s$ have circuits of depth $O(\log s \log d)$ and size $s^{O(\log d)}$.
\textcite{valiant_fast_1983} reduced the size to $\poly(s,d)$, with depth $O(\log d \log(sd))$ and fan-in two; their proof yields depth $O(\log d)$ if sum gates may have unbounded fan-in, see e.g.\ \cite{andrews_constant-depth_2026}.
That is, $\VP \subseteq \mathsf{VAC}^1$, noting that by definition $\VP$ contains only $p$-families.
We prove the analogue for matrix-symmetric circuits, where orbit size takes the role of size.

\begin{theorem}\label{thm:log-depth-reduction}
    $\symVP$ and $\symVAC^1$ contain the same $p$-families.
\end{theorem}

The proof of \cref{thm:log-depth-reduction} differs from \cite{valiant_fast_1983}: it combines \cref{thm:bounded-depth-characterisation} with the balanced tree decompositions of \cref{thm:bodlander}.

\begin{theorem}[\protect{\textcite[Theorem~4.2 and proof of Theorem~4.1]{goos_nc-algorithms_1989}}]\label{thm:bodlander}
    Let $n, k \geq 1$.
    Every $n$-vertex graph $F$ of treewidth $\leq k$ admits a tree decomposition with bags of size $\leq 3(k+1)$ whose tree is a rooted binary tree of depth $\leq 2 \lceil \log_{\frac54}(2n) \rceil$.
\end{theorem}

\begin{proof}[Proof of \cref{thm:log-depth-reduction}]
    By definition, $\symVAC^1 \subseteq \symVP$, so we only need to show the other inclusion.
    Let $(p_n) \in \symVP$ be a $p$-family.
    By \cref{thm:symv-to-hom}, there exists a constant $k \geq 1$ such that, for all sufficiently large $n$, $p_n$ can be written as a linear combination of homomorphism polynomials
    \[
        p_n = \sum_{F} \alpha_{F, n} \hom_{F,n}
    \]
    for bipartite multigraphs $F$ of treewidth $\tw(F) \leq k$.
    The degree of $\hom_{F, n}$ is equal to the number of edges of $F$.
    Since $|V(F)| \leq 2 |E(F)|$ as $F$ contains wlog no isolated vertices by \cref{obs:no-isolated-vertices},
    we have that the number of vertices in $F$ is bounded by a polynomial.
    By \cref{thm:bodlander},
    the patterns $F$ admit a tree decomposition of width $\leq 3(k+1)$ and depth $O(\log n)$.
    By the easy direction of \cref{thm:bounded-depth-characterisation}, i.e., \cref{thm:converse},
    it follows that $(p_n) \in \symVAC^1$.
\end{proof}

Following the same idea as in the above proof but making the parameter dependencies more precise, we show the following symmetric analogue of \cite{valiant_fast_1983}.
In contrast to \cref{thm:log-depth-reduction}, \cref{cor:vsbr} also applies to circuits of superpolynomial orbit size.
The theorem is proven in \cref{app:vsbr}.

\begin{theorem}[label=cor:vsbr,restate=thmVSBR]
    Let $\varepsilon > 0$.
    Let $(p_n)$ be a family of matrix-symmetric polynomials of degree $\leq d(n)$ computed by matrix-symmetric circuits of orbit size $\leq s(n) \leq 2^{n^{1-\varepsilon}}$.
    Then $(p_n)$ is computed by matrix-symmetric circuits of depth $O(\log d(n))$ and orbit size $\leq \max\{s(n), n^2\}^{3/\varepsilon}$ for all sufficiently large $n$.
\end{theorem}

We finally note that 
the depth $O(\log d(n))$ in \cref{cor:vsbr} cannot be improved, even if the orbit size may grow polynomially.
Let $P_m$ denote the $m$-vertex path.

\begin{theorem}[label=cor:log-depth-optimal,restate=thmDepthLogOptimal]
    Let $a \geq 1$ and $d \colon \mathbb{N} \to \mathbb{N}$ with $2 \leq d(n) \leq n^a$.
    \begin{enumerate}
        \item The polynomials $\hom_{P_{d(n)},n}$ have degree $< d(n)$ and are computed by matrix-symmetric circuits of orbit size $\leq n^3$ and depth $O(\log d(n))$.
        \item For every $c$, every matrix-symmetric circuit of orbit size $\leq n^c$ computing $\hom_{P_{d(n)},n}$ has support depth, and hence depth, $\Omega_{c,a}(\log d(n))$.
    \end{enumerate}
\end{theorem}

\section{Symmetric polylogarithmic-depth hierarchy}

We  explore the relationship between the bounded-depth complexity classes $\symVAC^k$ for $k \geq 0$ and $\symVF$, $\symVBP$, and $\symVP$.
Using graph-theoretic arguments, we show that $\symVAC^0 = \symVF$ in \cref{thm:symVAC-symVF}.
Using invariant-theoretic techniques, we show that $\symVBP \subseteq \symVAC^1$ in \cref{cor:symVBP-symVAC1}.
Using the separation techniques from \cref{sec:separation}, we show that this inclusion is proper, i.e., $\symVBP \subsetneq \symVAC^1$, in \cref{cor:symVBP-symVAC1-strict}.
Finally, we show strictness of the $\symVAC^k$-hierarchy and that $\symVAC = \bigcup_{k\geq 1} \symVAC^k$ is a proper subclass of \symVP.
Here, we employ iteration number lower bounds for the Weisfeiler--Leman algorithm by \textcite{grohe_iteration_2025}.

\subsection{Constant-depth reduction for formulas}

We first prove a symmetric analogue of a result of \textcite{brent_parallel_1974}
asserting that every formula of size $s$ can be rewritten to a formula of depth $O(\log s)$ and size $\poly(s)$.

\begin{theorem}\label{thm:symVAC-symVF}
    $\symVAC^0 = \symVF$.
\end{theorem}
\begin{proof}
    By transforming tree decompositions into elimination forests and vice versa, cf.\ \cite{adler_going_2026}, every non-empty graph $F$ and every $\Delta \geq 1$ satisfy
    \begin{equation}\label{eq:fw-td}
        \fw_{\td(F)}(F) \leq \td(F) - 1 \qquad \text{and} \qquad
        \td(F) \leq (\fw_{\Delta}(F) + 1)\Delta.
    \end{equation}
    Moreover, $\fw_{\Delta}(F) \leq \fw_{\Delta'}(F)$ for $\Delta \geq \Delta'$.

    Let $(p_n) \in \symVF$.
    By \cref{thm:symv-to-hom}, for some $\Delta \geq 2$ and all sufficiently large $n$, $p_n$ is a linear combination of homomorphism polynomials for patterns of treedepth $\leq \Delta$.
    By \eqref{eq:fw-td}, these patterns satisfy $\fw_\Delta(F) \leq \fw_{\td(F)}(F) \leq \Delta - 1$.
    By \cref{thm:bounded-depth-characterisation,obs:support-orbit}, $p_n$ is computed by matrix-symmetric circuits of support depth $\leq \Delta$ and orbit size $\leq n^\Delta$ for $n \geq 2\Delta + 1$.
    Hence, $(p_n) \in \symVAC^0_\Delta$.

    Conversely, let $(p_n) \in \symVAC^0_\Delta$ be computed by matrix-symmetric circuits of orbit size $\leq n^c$ and support depth $\leq \Delta$, where $c \geq 2$ is an integer and $\Delta \geq 1$.
    For all sufficiently large $n$, we have $\kappa_{n^c}(n) \leq c+1$ and $3(c+1) \leq n+2$, so the rigidification has support size $\leq c$ by \cref{thm:support}.
    By \cref{thm:bounded-depth-characterisation}, $p_n$ is a linear combination of homomorphism polynomials for patterns with $\fw_\Delta(F) \leq c - 1$, and thus $\td(F) \leq c\Delta$ by \eqref{eq:fw-td}.
    Hence, $(p_n) \in \symVF$ by \cref{thm:symv-to-hom}.
\end{proof}

\subsection{Bounded pathwidth in logarithmic support depth}
\label{sec:depth-reduction2}
We now turn to the relationship between $\symVAC^1$ and $\symVBP$.
By \cref{thm:symv-to-hom} and \cref{thm:bounded-depth-characterisation},
the complexity classes $\symVAC^1$ and $\symVBP$ are characterised as the polynomials that can be written, for some constant $c$, as linear combinations of homomorphism polynomials for patterns $F_n$ of $\fw_{c \log n}(F_n) \leq c$ and $\pw(F_n) \leq c$, respectively.
We show the following.
\begin{theorem}[label=cor:symVBP-symVAC1, restate=corVBPAC]
    $\symVBP \subseteq \symVAC^1$.
\end{theorem}
This is notable since the exponentially long paths~$P_{2^n}$ have pathwidth~$1$, but for every constant~$c$ and all sufficiently large~$n$, it holds that $\fw_{c \log n}(P_{2^n}) \geq c + 1$ by \cref{ex:path-rich,lem:rich-fw}.
The proof of \cref{cor:symVBP-symVAC1} is therefore not combinatorial but invariant-theoretic.
The key ingredient is the following consequence of the Nagata--Higman theorem \cite{higman_conjecture_1956}
with Razmyslov's bound \cite{razmyslov_trace_1974} on the nilpotency index.

\begin{theorem}[\protect{\cite[Theorems~3.2--3.4]{procesi_invariant_1976}, \cite{razmyslov_trace_1974}}]\label{lem:trace-reduction}
    Suppose that $\mathbb{K}$ has characteristic zero.
    Let $N, m \geq 1$.
    Let $Q$ be a commutative $\mathbb{K}$-algebra and $A_1, \dots, A_m \in Q^{N \times N}$.
    For a word $w = w_1 \cdots w_\ell$ over $[m]$, write $A_w \coloneqq A_{w_1} \cdots A_{w_\ell}$.
    Then every $A_w$ is a linear combination of matrices $A_u$ with $|u| < N^2$ whose coefficients are polynomials over $\mathbb{K}$ in the traces $\tr(A_v)$ with $|v| \leq N^2$.
\end{theorem}

We encode patterns of bounded pathwidth as series compositions of bilabelled graphs, cf.\ \cite{mancinska_quantum_2020,grohe_homomorphism_2025,seppelt_homomorphism_2024}.
To a bilabelled graph, we may associate a homomorphism matrix counting homomorphisms with prescribed label images. 
We apply \cref{lem:trace-reduction} to these matrices.
The theorem then yields that every homomorphism polynomial of a bounded pathwidth pattern is a polynomial in traces of words in homomorphism matrices of polynomial length. 
Applying traces amounts to identifying in- and out-labels of bilabelled graphs and thus preserving bounded pathwidth.
Finally, we apply \cref{thm:bodlander} to reduce to logarithmic depth.
The full proof of \cref{cor:symVBP-symVAC1} is conducted in \cref{app:vbp-vac1}.
Finally, we observe that the inclusion of \cref{cor:symVBP-symVAC1} is strict.

\begin{theorem}\label{cor:symVBP-symVAC1-strict}
    $\symVBP \subsetneq \symVAC^1$.
\end{theorem}
\begin{proof}
    Let $B_n$ be the complete binary tree with $n$ leaves, as in \cite{dwivedi_lower_2026}.
    Since $\pw(B_n) \in \Theta(\log n)$ is unbounded, $(\hom_{B_n,n}) \notin \symVBP$ by \cite{dwivedi_lower_2026}.
    The bags $\{v, \mathrm{parent}(v)\}$ together with an empty root bag show that $\fw_{\Delta}(B_n) \leq 1$ for $\Delta \in O(\log n)$.
    By \cref{thm:bounded-depth-characterisation,obs:support-orbit}, $\hom_{B_n,n}$ is computed by matrix-symmetric circuits of orbit size $\leq n^2$ and support depth $O(\log n)$.
\end{proof}

\subsection{Separating \texorpdfstring{$\symVAC^1$}{symVAC1} from \texorpdfstring{$\symVP$}{symVP}}
\label{sec:depth-reduction3}

Using the robust lower bounds on the iteration number of the Weisfeiler--Leman algorithm of \textcite{grohe_iteration_2025}, we show that some families in $\symVP$ require superpolylogarithmic support depth.
By \cref{thm:log-depth-reduction}, the witnesses of this separation are not $p$-families.

\begin{theorem}\label{thm:symVAC1-separation}
    $\symVAC \subsetneq \symVP$.
\end{theorem}

\Cref{thm:symVAC1-separation} is inferred from the following result.

\begin{theorem}[\protect{\cite[Theorem~1.2]{grohe_iteration_2025}}]\label{thm:gln}
    There are absolute constants $k_0 \in \mathbb{N}$ and $\alpha, \varepsilon > 0$ such that, for every $d \geq k \geq k_0$ and every $n \geq \alpha d^8 k^6$, there is a pair of $k$-ary $n$-element relational structures $\mathfrak{A}$ and $\mathfrak{B}$ that are distinguished by $k$-WL, but $d$-WL does not distinguish $\mathfrak{A}$ and $\mathfrak{B}$ after $n^{\varepsilon k}$ refinement rounds.
\end{theorem}

We fix $k \coloneqq k_0$, choose $d \approx n^{1/8}$, and transfer $\mathfrak{A}$ and $\mathfrak{B}$ to a pair of $N \times N$ matrices with $N \coloneqq (n+1)^k$.
By \cref{thm:bounded-depth-characterisation}, circuits of orbit size $2^{N^\eta}$ and support depth $\Delta$ compute linear combinations of homomorphism polynomials for patterns $F$ with $\fw_\Delta(F) < N^\eta$.
For small $\eta > 0$, we have $kN^\eta \leq d$, so these polynomials cannot distinguish the two matrices unless $kN^\eta\Delta > n^{\varepsilon k}$, since $d$-WL does not distinguish $\mathfrak{A}$ and $\mathfrak{B}$ within $n^{\varepsilon k}$ refinement rounds.
Hence, $\Delta \geq N^\theta$ for some $\theta > 0$.
The proof is given in \cref{app:symvac1-separation}.
A version of the lower bounds of \textcite{grohe_iteration_2025} with a tunable number of rounds separates the levels of $\symVAC$.

\begin{theorem}\label{thm:symVACk-hierarchy}
    For every $k \geq 1$, $\symVAC^k \subsetneq \symVAC^{k+1}$.
\end{theorem}



\section{Symmetric algebraic branching programs}\label{sec:symmetric-abps}

We introduce symmetric algebraic branching programs, the symmetric analogue of algebraic branching programs. We then show that the resulting class $\symVBP$ coincides with $\symVS$.

We state the results for square variable matrices $\Xx_n$ and polynomial families $(f_n)$. All symmetric classes in this section refer to this setting, with $\Gamma_n=\Sym_n\times\Sym_n$. The same arguments apply to rectangular variable matrices, with orbit size polynomial in the sum of the two dimensions.

An \emph{algebraic branching program} (ABP) in variables $X$ is a finite directed acyclic graph $G=(V,E,\ell)$ with a unique source $s$ and a unique sink $t$. Each edge $e$ is labelled by an affine linear form $\ell(e)$ over $\K$ in $X$. The polynomial computed by $G$ is the sum, over all directed paths from $s$ to $t$, of the product of their edge labels.

\begin{definition}[Symmetric ABPs]
\label{def:symmetricABP}
\lean{SymmetricABPs}
Let $G=(V,E,\ell)$ be an ABP in variables $X$ and let $\Gamma\leq\Sym(X)$. We call $G$ \emph{$\Gamma$-symmetric} if every $\pi\in\Gamma$ extends to an automorphism $\sigma$ of $(V,E)$ fixing $s$ and $t$ such that
\[
    \ell(\sigma(e))=\pi(\ell(e)) \qquad\text{for every }e\in E.
\]
Here $\pi$ permutes the variables and fixes field constants.
Let $\Aut_\Gamma(G)$ be the group of all automorphisms extending some $\pi\in\Gamma$. The \emph{orbit size} of $G$ is
\[
    \maxorb(G)=\max_{v\in V}
        \bigl|\{\sigma(v):\sigma\in\Aut_\Gamma(G)\}\bigr|.
\]
\end{definition}



\begin{definition}[$\symVBP$]
\lean{SymmetricABPs}
The class $\symVBP$ contains all families of matrix-symmetric polynomials $(p_n)_{n \in \bbN}$ in variables $\Xx_n$ that can be represented by $\Gamma_n$-symmetric ABPs whose orbit size is polynomial in $n$.
\end{definition}

Recall from \cref{sec:symmetric-complexity-classes} that $\symVS$ consists of matrix-symmetric polynomial families computed by symmetric skew circuits of polynomial orbit size. In classical algebraic complexity, $\VBP=\mathsf{VS}$: polynomial-size ABPs and skew circuits compute the same polynomial families~\cite{mahajan_algebraic_2013}. The following theorem extends this equivalence to the symmetric setting.
\begin{theorem}\label{thm:symVBP-symVS}
\lean{SymmetricABPCharacterisation}
\(
\symVBP = \symVS.
\)
\end{theorem}

The following two lemmas prove both directions.

\begin{lemma}
\lean{SkewFromABP}
$\symVBP \subseteq \symVS$.
\end{lemma}
\begin{proof}
We use the standard conversion from ABPs to skew circuits~\cite[Section~3, p.~11]{mahajan_algebraic_2013} and verify that it preserves symmetry and polynomial orbit size.
Let $G$ be a symmetric ABP of orbit size $B$, and write $\ell(u,v)=a_{uv,0}+\sum_{x\in X}a_{uv,x}x$.
For each vertex $v$, introduce a sum gate $P_v$. Give $P_s$ the single input $1$, and, for $v\neq s$, set
\[
    P_v=\sum_{(u,v)\in E}
        \left(a_{uv,0}P_u+\sum_{x\in X}a_{uv,x}xP_u\right).
\]
Each term is a separate product gate feeding directly into $P_v$. Retain terms with zero coefficients. Every product gate has just one internal child, so the circuit is skew. Induction over $G$ shows that $P_v$ computes the path sum from $s$ to $v$. Choose $P_t$ as the output gate.

Every extending automorphism of $G$ acts on this circuit by permuting the vertex and variable indices.
To bound orbit size, we must consider all extending circuit automorphisms. Any such automorphism permutes the sum gates $P_v$, fixes $P_s$ as the only sum gate with an input child, and fixes the output $P_t$. A product gate has parent $P_v$ and unique internal child $P_u$, which identify the endpoints of its ABP edge. Its input labels identify the coefficient and the variable, if present. Since we retained all terms, including those with zero coefficients, this recovers every edge and its affine label. Thus every extending circuit automorphism induces an extending ABP automorphism, and sum-gate orbits have size at most $B$. Each product gate is determined by an ABP edge $(u,v)$ and the choice of the constant term or a variable $x\in X$. The induced ABP automorphism gives at most $B$ possible images for each of $u$ and $v$, and each resulting edge has $|X|+1$ product gates. Thus every product-gate orbit has size at most $B^2(|X|+1)$. With input gates shared by label, variable inputs have orbits of size at most $|X|$, and constant inputs are fixed. Hence the circuit has polynomial orbit size.
\end{proof}

\begin{lemma}
\lean{SymmetricABPCharacterisation}
$\symVS \subseteq \symVBP$.
\end{lemma}
\begin{proof}
Let $(f_n)\in\symVS$. By \cref{thm:symv-to-hom}(2), there is a constant $k$ such that, for all sufficiently large $n$, $f_n$ is a finite linear combination of homomorphism polynomials $\hom_{F,n}$ with $\pw(F)\leq k$. For each remaining $n$, a homomorphism expansion uses patterns with at most $n$ vertices on each side, hence pathwidth less than $2n$. Enlarging $k$ covers these finitely many cases.

Fix $n$ and one such pattern $F$. We recall the standard path-decomposition construction underlying the upper bound in \cite[Theorem~17]{KomarathPR22}. The ABP chooses the images of the pattern vertices along a path decomposition. A state remembers only the images of vertices in the current bag.

Take a path decomposition with bags of size at most $k+1$, starting and ending with the empty bag. Between consecutive bags, forget vertices that leave before introducing those that enter. Process each edge occurrence once, at a bag containing both endpoints, by repeating that bag. This gives a sequence of introduction, forgetting, and edge-processing steps without increasing the bag size.

For each bag, create a layer with one vertex for every assignment of its left and right vertices to the left and right copies of $[n]$, respectively. Connect consecutive layers according to the step:
\begin{itemize}
    \item To introduce a vertex, connect each assignment to all its extensions, with edge label $1$.
    \item To forget a vertex, connect each assignment to its restriction, with edge label $1$.
    \item To process an edge $ab$, connect each assignment $h$ to its copy in the next layer, with edge label $x_{h(a),h(b)}$.
\end{itemize}
The empty endpoint bags give a unique source and sink. Each path chooses one image for every pattern vertex. An image is retained until that vertex is forgotten, after all its incident edges have been processed. Thus paths correspond bijectively to the maps defining $\hom_{F,n}$, with the same products of edge labels. This also accounts for isolated vertices and parallel edge occurrences. Denote this ABP by $G_F$.

\begin{claim}
\lean{ABPsForPatterns}
The ABP $G_F$ is matrix-symmetric and has orbit size at most $n^{k+1}$.
\end{claim}
\begin{claimproof}
Applying $(\sigma,\tau)\in\Gamma_n$ to the left and right images, respectively, gives an extending automorphism. Every extending automorphism preserves layers, since the layer number is the directed distance from the source. Each layer contains at most $n^{k+1}$ vertices, so every vertex orbit has at most this size.
\end{claimproof}

If $f_n=0$, use a single edge labelled $0$. Otherwise, to combine the summands without enlarging the orbit bound, give their ABPs different path lengths. For each term $c_F\hom_{F,n}$, add an initial edge labelled $c_F$ and a final edge labelled $1$. Pad with edges labelled $1$ until the branches have distinct source-to-sink path lengths, then identify their sources and sinks. The resulting path sum is $f_n$. Every extending automorphism preserves path lengths, hence each branch and its layers. The orbit bound remains $n^{k+1}$, proving $(f_n)\in\symVBP$.
\end{proof}

\section{Symmetric VNP}
\label{sec:symvnp}

We introduce $\symVNP$, the analogue of $\VNP$ for symmetric algebraic circuits.

As in the previous section, we state the definitions and results for square matrices $\Xx_n$ under $\Gamma_n$, with all symmetric classes understood for one-parameter families. The results also hold for rectangular matrices.

    \begin{definition}\label{def:symVNP}
    \lean{SymmetricVNP}
    The class $\symVNP$ consists of all families of polynomials $(f_n) \in \mathbb{K}[\Xx_n]$ for which there exist a $p$-bounded function $q \colon \mathbb{N} \to \mathbb{N}$, a family of polynomials $(g_n)$ over variables $\Xx_n \cup \Yy_n$ where $|\Yy_n| = q(n)$, and a group homomorphism $\rho_n \colon \Gamma_n \to \Sym(\Yy_n)$ such that:
    $$f_n(\Xx_n) \;=\; \sum_{\boldsymbol{e} \in \{0,1\}^{q(n)}} g_n(\Xx_n, \boldsymbol{e}),$$
    the family $(g_n)$ is computed by a circuit family $(C_n)$ of polynomially bounded orbit size, and for every $\pi \in \Gamma_n$, the permutation on $\Xx_n \cup \Yy_n$ defined by applying $\pi$ to $\Xx_n$ and $\rho_n(\pi)$ to $\Yy_n$ extends to a circuit automorphism $\sigma$ of the circuit $C_n$ computing $g_n$. Orbit size is measured with respect to this joint $\Gamma_n$-action.
    \end{definition}  

    Unlike the definitions of $\symVP$ and the other symmetric complexity classes, this definition does not require the output polynomials to be matrix-symmetric. Instead, matrix symmetry follows from the joint symmetry of the witness and summation over the entire Boolean cube, as the following observation shows. Following \cite{dps_itcs,dwivedi_lower_2026}, write $\mathfrak{G}_n\subseteq\K[\Xx_n]$ for the space of matrix-symmetric polynomials.

    \begin{observation}\label{obs:symVNP-matrix-symmetric}
        \lean{BooleanSumSymmetry}
        If $(f_n)\in\symVNP$, then $f_n\in\mathfrak{G}_n$ for every $n$.
    \end{observation}
    \begin{proof}
        Fix $n$ and write $X=\Xx_n$, $Y=\Yy_n$, $f=f_n$, $g=g_n$, and $\rho=\rho_n$ for a witness in the definition. Let $\gamma\in\Gamma_n$. The joint permutation of $X\cup Y$ extends to a circuit automorphism fixing the unique output gate. Hence
        \[
            g(\gamma X,\rho(\gamma)Y)=g(X,Y).
        \]
        Since $\rho(\gamma)$ permutes $Y$, it induces a bijection on the set of Boolean assignments $\{0,1\}^{Y}$. Reindexing the sum gives
        \[
        \begin{aligned}
            f(\gamma X)
            =\sum_{\boldsymbol{e}\in\{0,1\}^{Y}}
                g(\gamma X,\boldsymbol{e})
            =\sum_{\boldsymbol{e}'\in\{0,1\}^{Y}}
                g(\gamma X,\rho(\gamma)\boldsymbol{e}')
            =\sum_{\boldsymbol{e}'\in\{0,1\}^{Y}}g(X,\boldsymbol{e}')
             =f(X).
        \end{aligned}
        \]
        Thus $f$ is matrix-symmetric, so $f\in\mathfrak{G}_n$.
    \end{proof}

Later, we will show that the permanent is complete for $\symVNP$ under symmetric projections.

\subsection{\texorpdfstring{$\symVNP$}{Symmetric VNP} contains all matrix-symmetric polynomials}

By \cref{obs:symVNP-matrix-symmetric}, every family $(f_n)\in\symVNP$ satisfies $f_n\in\mathfrak{G}_n$ for all $n$. We now prove the converse, showing that $\symVNP$ consists precisely of all families of matrix-symmetric polynomials. We express each $f\in\mathfrak{G}_n$ as a linear combination of homomorphism polynomials and encode the vertex maps using Boolean variables, giving witnesses computed by symmetric formulas of polynomial orbit size.



\begin{theorem}\label{thm:all-matrix-symmetric-symVNP}
    \lean{SymmetricVNPCharacterisation}
    Over a field $\K$ of characteristic zero, a polynomial family $(f_n)$ belongs to $\symVNP$ if and only if $f_n\in\mathfrak{G}_n$ for every $n$. More precisely, every $f\in\mathfrak{G}_n$ can be written as
    \[
        f(X)=\sum_{\boldsymbol{e}\in\{0,1\}^{2n^2}}g(X,\boldsymbol{e}),
    \]
    where $g$ admits a symmetric formula whose gate orbits have size at most $4n^2$, under a joint action on $X$ and the auxiliary variables.
\end{theorem}
\begin{proof}
    The forward implication is \cref{obs:symVNP-matrix-symmetric}. For the converse, fix $n$ and let $f\in\mathfrak{G}_n$. By the argument of \cite[Lemma~4]{dps_itcs}, $f$ is a finite linear combination of homomorphism polynomials of patterns with at most $n$ vertices on each side. The argument applies over $\K$ since it only requires division by nonzero integers. If a pattern $F$ has $a$ left and $b$ right vertices, padding it with isolated vertices to obtain $\widehat F$ with $n$ vertices on each side gives $\hom_{\widehat F,n}=n^{2n-a-b}\hom_{F,n}$ by \cref{obs:no-isolated-vertices} and its right-side analogue. Absorbing these nonzero factors into the coefficients and denoting the padded patterns again by $F$, we can therefore write
    \[
        f(X)=\sum_F c_F\hom_{F,n}(X),
    \]
    with all patterns $F$ on the same bipartition $A\uplus B$, where $|A|=|B|=n$.

    Introduce variables $y_{a,i}$ for $(a,i)\in A\times[n]$ and $y_{b,j}$ for $(b,j)\in B\times[n]$. Let $Y$ be this common set of $2n^2$ variables. For each pattern $F$, put
    \lean{SymmetricFormulaWitness}
    \begin{equation}\label{eq:symvnp-edge-product}
        W_F(X,Y)\;=\;\prod_{ab\in E(F)}\;
        \left(\;\sum_{i=1}^n\sum_{j=1}^n
            y_{a,i}\,y_{b,j}\,x_{i,j}\;\right),
    \end{equation}
    counting edges with multiplicity, and define
    \[
    \begin{gathered}
        V_A(Y)=\prod_{a\in A}\left[
            \underbrace{\left(\sum_{i=1}^n y_{a,i}\right)}_{\substack{\text{rules out}\\\text{all zeros}}}
            \;\underbrace{\prod_{1\leq i<k\leq n}(1-y_{a,i}y_{a,k})}_{\text{at most one }1}\right],
        \quad
        V_B(Y)=\prod_{b\in B}\left[
            \left(\sum_{j=1}^n y_{b,j}\right)
            \prod_{1\leq j<\ell\leq n}(1-y_{b,j}y_{b,\ell})\right],\\[2ex]
        g_F(X,Y)\;=\;W_F(X,Y)\,V_A(Y)\,V_B(Y).
    \end{gathered}
    \]
    On Boolean assignments, $V_AV_B$ is $1$ precisely when each vertex chooses exactly one image, and is $0$ otherwise: the sum rules out choosing no image, and the pairwise factors rule out choosing two. Such assignments correspond bijectively to maps $h\colon A\to[n]$, $h\colon B\to[n]$. On the assignment encoding $h$, each edge factor in $W_F$ equals $x_{h(a),h(b)}$. Hence
    \[
        \sum_{\boldsymbol{e}\in\{0,1\}^{Y}}g_F(X,\boldsymbol{e})
        =\sum_{\substack{h\colon A\to[n]\\h\colon B\to[n]}}
          \prod_{ab\in E(F)}x_{h(a),h(b)}
        =\hom_{F,n}(X).
    \]

    \begin{claim}
        Each $\hom_{F,n}$ above is a Boolean-cube sum of a polynomial $g_F$ admitting a symmetric formula with orbit size at most $4n^2$.
    \end{claim}
    \begin{claimproof}
        The Boolean-sum identity is established above, and $|Y|=2n^2$. Compute the displayed expression as a formula, using fresh copies for every occurrence of an internal gate and sharing only input gates. In \cref{eq:symvnp-edge-product}, every edge occurrence contributes one factor $\sum_{i,j}y_{a,i}y_{b,j}x_{i,j}$. If two edges are parallel, their factors are identical polynomials, but we compute them with separate copies.

        For each copy of the edge factor $P=\sum_{i,j}y_{a,i}y_{b,j}x_{i,j}$ in \cref{eq:symvnp-edge-product}, compute $t^{-1}(tP)$ instead. This adds two multiplication gates without changing $P$. Choose all constants $t,t^{-1}$ distinct and absent from the original formula. This is possible since $\K$ is infinite. Every extending automorphism fixes these constants. Each has only one parent, so both new gates are fixed. The root of this copy of $P$ is therefore fixed too. Thus identical copies cannot be exchanged.

        Extend $(\sigma,\tau)\in\Gamma_n$ to $Y$ by
        \[
            y_{a,i}\mapsto y_{a,\sigma(i)},\qquad
            y_{b,j}\mapsto y_{b,\tau(j)},
        \]
        keeping the pattern vertices fixed. Together with its action on $X$, this permutes the gates computing $y_{a,i}y_{b,j}x_{i,j}$ for each fixed $ab$, and the computations of $1-y_{a,i}y_{a,k}$ and $1-y_{b,j}y_{b,\ell}$ for each fixed $a$ and $b$. Conversely, every extending automorphism preserves the marked edge-factor copies. Within each copy, the variable inputs determine the term indices; in the vertex factors they determine the vertex and the unordered pair of indices. Thus the bounds apply to all extending automorphisms: the term gates have orbits of size at most $n^2$, and the pair-indexed gates have orbits of size at most $\binom{n}{2}$. Every addition or multiplication combining all these terms stays fixed, since its inputs are merely permuted. Input gates have orbits of size at most $n^2$ or $n$, and constants are fixed, so every gate orbit has size at most $4n^2$.
    \end{claimproof}

    Finally, set $g=\sum_F c_Fg_F$, using the same variables $Y$ in every summand. By linearity,
    \[
        \sum_{\boldsymbol{e}\in\{0,1\}^{Y}}g(X,\boldsymbol{e})
        =\sum_F c_F\hom_{F,n}(X)=f(X).
    \]
    Take a separate formula for each $g_F$, sharing only input gates. Compute its contribution as $t_F^{-1}(t_Fc_Fg_F)$. Choose all constants $t_F,t_F^{-1}$ distinct and absent from these formulas and the coefficients $c_F$. Every extending automorphism fixes these constants and their unique parent gates. Hence it preserves each summand. Adding one sum gate gives a formula for $g$. The added gates are fixed, so every gate orbit still has size at most $4n^2$. Since $|Y|=2n^2$, this proves the theorem.
\end{proof}

\subsection{Hypercube sums of symmetric formulas}
\label{subsec:symvnp-formulas}

In classical algebraic complexity, $\VNP$ is the class of hypercube sums of $\VP$ witnesses. Valiant~\cite{valiant_reducibility_1982} proved that using $\VF$ witnesses gives the same class; see also \cite[Theorem~2.13]{burgisser_completeness_2000} for a proof. Malod and Portier~\cite{MP2008} gave an alternative proof based on parse trees. The following corollary shows that the same phenomenon holds for $\symVNP$.

Here $\Sigma_{\{0,1\}}\symVP$ denotes Boolean sums of circuit witnesses satisfying the joint symmetry and orbit-size conditions in \cref{def:symVNP}, with polynomially many auxiliary variables. We use $\Sigma_{\{0,1\}}\symVF$ for the same definition with formula witnesses.

\begin{corollary}\label{cor:symvnp-formula-witnesses}
    \lean{FormulaWitnessEquivalence}
    Over a field $\K$ of characteristic zero, let $(f_n)\in\symVNP$. There exist a $p$-bounded function $q\colon\mathbb{N}\to\mathbb{N}$ and polynomials $h_n\in\K[\Xx_n\cup Y_n]$, with $|Y_n|=q(n)$, such that
    \[
        f_n(X)=\sum_{\boldsymbol{e}\in\{0,1\}^{Y_n}}
            h_n(X,\boldsymbol{e}).
    \]
    Each $h_n$ admits a formula symmetric under a joint $\Gamma_n$-action on $\Xx_n\cup Y_n$ extending the usual row and column action on $\Xx_n$. These formulas have polynomially bounded gate orbits. In short, we write
    \[
        \symVNP
        =\Sigma_{\{0,1\}}\symVP
        =\Sigma_{\{0,1\}}\symVF.
    \]
\end{corollary}
\begin{proof}
    The first equality is the definition of $\symVNP$. Since every symmetric formula is a symmetric circuit, $\Sigma_{\{0,1\}}\symVF\subseteq\symVNP$. Conversely, let $(f_n)\in\symVNP$. By \cref{obs:symVNP-matrix-symmetric}, each $f_n$ belongs to $\mathfrak{G}_n$. By \cref{thm:all-matrix-symmetric-symVNP}, it has a symmetric formula witness using $2n^2$ Boolean variables and with gate orbits of size at most $4n^2$. Thus $(f_n)\in\Sigma_{\{0,1\}}\symVF$, proving the reverse inclusion.
\end{proof}

\subsection{Completeness of the permanent under symmetric projections}
\label{sec:permanent-completeness}

For an $n\times n$ variable matrix $X$, the \emph{permanent polynomial} is
\[
    \perm_n(X)=\sum_{\pi\in\Sym_n}\prod_{i=1}^n x_{i,\pi(i)}.
\]

    \begin{proposition}\label{prop:permanent-symVNP}
        \lean{PermanentCompleteness}
        The permanent family belongs to $\symVNP$ over any field $\K$.
    \end{proposition}
    \begin{proof}
        Ryser's formula gives
        \[
            \perm_n(X)=\sum_{S\subseteq[n]}(-1)^{n-|S|}
                \prod_{i=1}^n\left(\sum_{j\in S}x_{i,j}\right).
        \]
        Indeed, expanding the products, the monomial $\prod_i x_{i,h(i)}$ for a map $h\colon[n]\to[n]$ has coefficient $\sum_{S\supseteq h([n])}(-1)^{n-|S|}$, which is $1$ if $h$ is surjective and $0$ otherwise. The surviving maps are exactly the permutations.

        Introduce $Y=\{y_1,\dots,y_n\}$ and put
        \[
            g_n(X,Y)=\left(\prod_{j=1}^n(2y_j-1)\right)
                \prod_{i=1}^n\left(\sum_{j=1}^n x_{i,j}y_j\right).
        \]
        A Boolean assignment $\boldsymbol{e}$ encodes the set $S=\{j:e_j=1\}$. On this assignment, $\prod_j(2e_j-1)=(-1)^{n-|S|}$ and $\sum_j x_{i,j}e_j=\sum_{j\in S}x_{i,j}$. Hence
        \[
            \perm_n(X)=\sum_{\boldsymbol{e}\in\{0,1\}^n}g_n(X,\boldsymbol{e}).
        \]

        The circuit given by this expression is symmetric under the action $\rho_n(\sigma,\tau)(y_j)=y_{\tau(j)}$ on $Y$. Together with $x_{i,j}\mapsto x_{\sigma(i),\tau(j)}$, this sends the gate computing $x_{i,j}y_j$ to the one computing $x_{\sigma(i),\tau(j)}y_{\tau(j)}$, the sum for row $i$ to the sum for row $\sigma(i)$, and the computation of $2y_j-1$ to that of $2y_{\tau(j)}-1$. The two products and the output gate are fixed. These maps preserve operations and wires and extend every $(\sigma,\tau)$ to a circuit automorphism.

        The gates indexed by $(i,j)$ have orbits of size at most $n^2$, those indexed by $i$ or $j$ have orbits of size at most $n$, and all other gates are fixed. Thus $g_n$ has polynomial orbit size (indeed, total size $O(n^2)$) and uses only $n$ auxiliary variables.
    \end{proof}

We now show that the permanent is also complete for $\symVNP$ under symmetric projections. The construction uses one directed cycle for each monomial and all its relabellings.

\begin{definition}[Symmetric projection to the permanent]\label{def:symmetric-projection}
    \lean{Permanent}
    Let $\Gamma=\Gamma_n$ act on $X=\Xx_n$, and fix an action $\rho\colon\Gamma\to\Sym(Y)$ on a finite auxiliary variable set $Y$. A polynomial $h\in\K[X\cup Y]$ is a \emph{symmetric projection of the permanent} if $h=\perm_A(M)$ for some $A\geq1$ and an $A\times A$ matrix $M$ of affine linear forms over $\K$ in $X\cup Y$, and every $\gamma\in\Gamma$ admits permutations $\alpha_\gamma,\beta_\gamma\in\Sym_A$ such that
    \[
        M_{\alpha_\gamma(i),\beta_\gamma(j)}
        =\gamma\cdot M_{i,j}
        \qquad(i,j\in[A]).
    \]
    Here $\gamma$ acts on $Y$ by $\rho(\gamma)$ and fixes field constants. For a polynomial in $\K[X]$, take $Y=\emptyset$.
\end{definition}

For polynomial families, we apply this definition to each $f_n$, without any bound on the dimension $A = A(n)$ of the matrix $M$. The permanent family is \emph{complete} for $\symVNP$ under symmetric projections if it belongs to $\symVNP$ and every family $(f_n)\in\symVNP$ admits such projections. The construction below uses only variables and field constants as matrix entries.

\begin{theorem}[Symmetric completeness of the permanent]\label{thm:permanent-completeness}
    \lean{PermanentCompleteness}
    Over a field $\K$ of characteristic zero, the permanent family is complete for $\symVNP$ under symmetric projections.
\end{theorem}

Recall that a \emph{directed cycle cover} is a collection of vertex-disjoint directed cycles covering all vertices, with loops allowed. The permanent of a weighted adjacency matrix sums the weights of these covers, where each weight is the product of the selected edge labels.

\begin{proof}
    By \cref{obs:symVNP-matrix-symmetric}, it suffices to construct a symmetric projection for each matrix-symmetric polynomial $f\in\K[X]$, where $X=\Xx_n$. Put $\Gamma=\Gamma_n$ and write
    \[
        f=\sum_{r\in R}c_r\prod_{j=1}^{d_r}z_{r,j},
        \qquad z_{r,j}\in X.
    \]
    We encode the average of the polynomials $\gamma(f)$ using the standard correspondence between paths and cycle covers~\cite[Section~6.2]{MahajanVinay1997}.

    Start with a vertex $v$. For each $r\in R$ and $\gamma\in\Gamma$, add a directed cycle $C_{r,\gamma}$ through $v$, with edge labels
    \[
        \frac{c_r}{|\Gamma|},\quad
        \gamma(z_{r,1}),\ldots,\gamma(z_{r,d_r}),\quad 1.
    \]
    These cycles share only $v$. Keep separate copies for all $\gamma\in\Gamma$, even when their monomials agree, so that group multiplication can permute the copies. Add a loop of weight $1$ at every vertex other than $v$, and let $M$ be the weighted adjacency matrix, with absent edges represented by $0$. A constant monomial gives a two-cycle; for $f=0$, take the $1\times1$ zero matrix.

    Every cycle cover selects exactly one cycle $C_{r,\gamma}$ through $v$ and uses loops at all remaining vertices. Summing their weights gives
    \[
        \perm(M)
        =\sum_{\gamma\in\Gamma}\sum_{r\in R}
        \frac{c_r}{|\Gamma|}\prod_{j=1}^{d_r}\gamma(z_{r,j})
        =\frac1{|\Gamma|}\sum_{\gamma\in\Gamma}\gamma(f)
        =f,
    \]
    where the last equality holds because $f$ is matrix-symmetric.

    For $\delta\in\Gamma$, let $\eta_\delta$ fix $v$ and send each other vertex of $C_{r,\gamma}$ to the vertex in the same position on $C_{r,\delta\gamma}$, counting from $v$ along the cycle. An edge labelled $\gamma(z_{r,j})$ is then sent to one labelled $\delta\gamma(z_{r,j})$, its image under $\delta$. Constant labels and the added loops are preserved. Thus
    \[
        M_{\eta_\delta(u),\eta_\delta(w)}
        =\delta\cdot M_{u,w}.
    \]
    Taking $\alpha_\delta=\beta_\delta=\eta_\delta$ in \cref{def:symmetric-projection} proves that $f$ is a symmetric projection of the permanent.

    Applying the construction for every $n$ and using the permanent's membership in $\symVNP$ from \cref{prop:permanent-symVNP} proves completeness.
\end{proof}

    \addsec{Acknowledgements}

    \emph{Prateek Dwivedi} is supported by the PRIME programme of the German Academic Exchange Service (DAAD) with funds from the German Federal Ministry of Education and Research (BMBF).
    He also thanks Université Savoie Mont Blanc for its hospitality and the University of Cambridge for hosting his visit to Anuj Dawar and Benedikt Pago, during which the ideas for this project first took shape.

    \emph{Tim Seppelt} is funded by the European Union (CountHom, No.
101077083). Views and opinions expressed are however those of the author(s) only and do not
necessarily reflect those of the European Union or the European
Research Council Executive Agency. Neither the European Union
nor the granting authority can be held responsible for them.

\emph{Anuj Dawar} and \emph{Benedikt Pago} are funded by UK Research and Innovation (UKRI) under the UK government’s Horizon Europe funding guarantee: grant number EP/X028259/1.

    \section*{AI statement}
    We used Claude Code with Sonnet~5.5, Opus~5.5 and Fable~5.1 to assist with carrying out technical arguments following human instructions, literature review, and formalising the results in Lean \cite{lax-461425}.
    The tool materially affected \cref{sec:depth-hierarchy,sec:depth-reduction2,sec:depth-reduction3} and the corresponding appendices. The authors verified the correctness and originality of all content, including references.

    \newpage
	\printbibliography[heading=bibintoc]

    \newpage
    \appendix
    \section{Reduced circuits and supports}\label{app:supports}

    We prove \cref{lem:reduced-rigid,lem:reduction,thm:support}.
    For a gate $g$, write $M(g)$ for the multiset of its children and $m(g,h)$ for the multiplicity of the wire from $h$ to $g$.
    The \emph{height} of a gate is the maximal number of wires on a path from it to an input gate.

    \restatethm{lem:reduced-rigid}{\lemReducedRigid}
    \begin{proof}
        Let $C$ be a $\Gamma_n$-symmetric reduced circuit.
        Two automorphisms extending the same element of $\Gamma_n$ differ by an automorphism $\rho$ of $C$ that fixes all input gates, since input labels are distinct.
        By induction on the height, $\rho$ fixes every gate $g$:
        the gate $\rho(g)$ has the same label as $g$ and the multiset of children $\rho(M(g)) = M(g)$, so $\rho(g) = g$ since $C$ is reduced. Thus, $\rho$ is the identity automorphism, and hence, there do not exist two distinct automorphisms of $C$ that extend the same element of $\Gamma_n$. 
    \end{proof}

    \restatethm{lem:reduction}{\lemReduction}
    \begin{proof}
        We define an equivalence relation $\equiv$ on the gates of $C$ by induction on the height.
        Input gates are equivalent if they have the same label.
        Internal gates $g$ and $g'$ are equivalent if they have the same height and label and $m(g, c) \coloneqq \sum_{h \in c} m(g, h) = m(g', c)$ for every class $c$ of gates of smaller height.
        Let $C'$ have as gates the classes from which the class $[\mathrm{out}]$ of the output gate can be reached, 
        where $[g]$ carries the label of $g$ and has a wire of multiplicity $m(g,c)$ from every class $c$.
        By induction on the height, all gates in a class compute the same polynomial that is computed by the class in $C'$.
        The circuit $C'$ is reduced since classes with the same label and the same children have the same height in $C$ and are collapsed into a single gate in $C'$.
        Moreover, $C'$ has at most as many gates and wires as $C$, and the same depth.

        Every automorphism $\sigma$ of $C$ preserves heights, labels, and multiplicities, and thus maps classes to classes.
        It induces an automorphism $\sigma'$ of $C'$, which acts on the input labels as $\sigma$ does.
        Hence, $C'$ is matrix-symmetric.
        Since $C'$ is rigid by \cref{lem:reduced-rigid}, $\sigma'$ is the unique extension of the element of $\Gamma_n$ extended by $\sigma$.
        Hence, the map $g \mapsto [g]$ maps every $\Aut_\Gamma(C)$-orbit onto an orbit of $C'$, and every orbit of $C'$ arises this way.
        So the orbit size does not increase.
    \end{proof}

    The only group-theoretic machinery that we are going to need is a classical result on subgroups of small index.
    Write $\mathbf{Alt}_{(T)} \leq \Sym_{(T)} \leq \Sym_n$ for the pointwise stabilisers of $T \subseteq [n]$ in the alternating and the symmetric group.

    \begin{theorem}[\protect{cf.\ \cite[Theorem~5.2B]{dixon_permutation_1996}}]\label{thm:dixon-mortimer-appendix}
        \lean{DixonMortimer}
        Let $n \geq 9$ and $1 \leq r \leq n/2$.
        If $H \leq \Sym_n$ has index less than $\min\{\frac12 \binom{n}{\lfloor n/2 \rfloor}, \binom{n}{r}\}$,
        then $\mathbf{Alt}_{(T)} \leq H$ for some $T \subseteq [n]$ with $|T| < r$.
    \end{theorem}
    \begin{proof}
        By \cite[Theorem~5.2B]{dixon_permutation_1996}, 
        since $n \geq 5$ and the index of $H$ is less than $\binom{n}{r}$, 
        one of the following holds:
        (i) $\mathbf{Alt}_{(T)} \leq H$ for some $T \subseteq [n]$ with $|T| < r$;
        (ii) $n = 2m$ is even and $H$ has index $\frac12 \binom{n}{m}$; or
        (iii) one of six exceptional cases with $n \leq 8$ holds.
        Case~(ii) is excluded by the bound on the index, and case~(iii) by $n \geq 9$.
    \end{proof}

Recall the definition of $\kappa_f(n)$ from Section \ref{sec:supports}:
\begin{equation}
    \kappa_f(n) \coloneqq \min\left\{ r\in \mathbb{N} : \binom{n}{r} > f(n) \right\}.
\end{equation}

    \restatethm{thm:support}{\thmSupport}
    \begin{proof}
        Write $\kappa \coloneqq \kappa_f(n)$, and let $C$ be a reduced matrix-symmetric circuit of orbit size at most $f(n)$.
        Note that $f(n) < \frac12 \binom{n}{\lfloor n/2 \rfloor}$ and $\kappa < n/2$ as $n \geq 20$ and $3\kappa \leq n+2$.
        If $C$ contains no variable, then all gates are fixed by $\Gamma_n$ and have support $\emptyset$.
        Otherwise, an input gate $x_{i,j}$ has orbit size $n^2 \leq f(n)$, so $\kappa \geq 2$.

        By induction on the height, we construct supports $S_g$ with $|(S_g)_L|, |(S_g)_R| \leq \kappa - 1$.
        For input gates $x_{i,j}$, let $S_g \coloneqq \{i\} \uplus \{j\}$, and for constants, let $S_g \coloneqq \emptyset$.
        Let $g$ be an internal gate and $H \coloneqq \{\pi \in \Sym_n \mid (\pi, \id ) \in \Stab(g)\}$.
        Then $[\Sym_n : H] \leq [\Gamma_n : \Stab(g)] \leq f(n) < \min\{\frac12 \binom{n}{\lfloor n/2 \rfloor}, \binom{n}{\kappa}\}$.
        By Theorem \ref{thm:dixon-mortimer-appendix} for $r \coloneqq \kappa$, there is $T_L \subseteq [n]$ with $|T_L| \leq \kappa - 1$ and $A \coloneqq \mathbf{Alt}_{(T_L)} \times \{\id \} \leq \Stab(g)$.

        We show that $\Sym_{(T_L)} \times \{\id \} \leq \Stab(g)$ using that $C$ is reduced.
        Let $h$ be a child of $g$.
        Since $|[n] \setminus (T_L \cup (S_h)_L)| \geq n - 2\kappa + 2 \geq 2$, 
        there is a transposition $\tau$ of two points outside $T_L \cup (S_h)_L$.
        Then $(\tau, \id) \in \StabP(S_h) \leq \Stab(h)$.
        As $\Sym_{(T_L)} \times \{ \id \} = A \cup A(\tau, \id)$, 
        for every $\sigma \in \Sym_{(T_L)} \times \{\id\}$ there is $\alpha \in A$ such that $\sigma h = \alpha h$.
        Since $\alpha$ fixes $g$, it follows that $m(g, \sigma h) = m(g, \alpha h) = m(g, h)$.
        Hence, $\sigma(M(g)) = M(g)$, so $\sigma g$ and $g$ have the same label and the same children.
        Since $C$ is reduced, $\sigma g = g$.

        Analogously, there is $T_R \subseteq [n]$ with $|T_R| \leq \kappa-1$ and $\{ \id \} \times \Sym_{(T_R)} \leq \Stab(g)$.
        Hence, $S_g \coloneqq T_L \uplus T_R$ supports $g$.
        Since $\kappa - 1 < n/2$, the minimal support $\sup(g) \subseteq S_g$ exists.

        Write $k_L \coloneqq |\sup_L(g)|$ and $k_R \coloneqq |\sup_R(g)|$, so that $k_L, k_R \leq \kappa - 1$.
        By equivariance of minimal supports, the orbit of $g$ maps onto the $\Gamma_n$-orbit of $(\sup_L(g), \sup_R(g))$.
        Hence,
        \[
            \binom{n}{k_L + k_R} 
            \leq \binom{n}{k_L} \binom{n - k_L}{k_R}
            \leq \binom{n}{k_L} \binom{n}{k_R}
            \leq f(n) 
            < \binom{n}{\kappa}.
        \]
        Suppose that $k_L + k_R \geq \kappa$.
        Since $k_L + k_R \leq 2\kappa - 2 \leq n - \kappa$ and $\binom{n}{r} \geq \binom{n}{\kappa}$ for all $\kappa \leq r \leq n - \kappa$,
        this is a contradiction.
        Hence, $k_L + k_R \leq \kappa - 1$.
    \end{proof}

    \section{Material omitted in \Cref{sec:depth-hierarchy}}

    \subsection{Circuits for patterns of bounded forestwidth}\label{app:converse}

    We construct circuits for single homomorphism polynomials, following \cite[Lemma~19]{bhargav_monotone_2025} and \cite{dps_itcs}.
    For a linear combination $\sum_i \alpha_i \hom_{F_i,n}$, one applies the construction below to every $F_i$, multiplies the output gates by the constants $\alpha_i$, adds an output sum gate, and reduces via \cref{lem:reduction}.
    The proof of \cref{thm:converse} applies verbatim, except that the depth grows by one.
    This yields the implication from 2 to 1 in \cref{thm:bounded-depth-characterisation}.

    \begin{lemma}[Normalised decompositions]\label{lem:normalised-decomposition}
        Let $\Delta, k \geq 1$ and let $F$ be a multigraph without isolated vertices with $\fw_\Delta(F) \leq k-1$.
        Then there are a tree decomposition $(T, \beta)$ of $F$ with root $r_0$ and a map $\mathrm{rep} \colon E(F) \to V(T)$ with $e \subseteq \beta(\mathrm{rep}(e))$ for every edge $e$ such that
        \begin{enumerate}
            \item $\beta(r_0) = \emptyset$, $T$ has depth at most $\Delta$, and $|\beta(t)| \leq k$ for every node $t$,
            \item for every node $t$ and every $v \in \beta(t)$, there is an edge $e$ incident with $v$ such that $\mathrm{rep}(e)$ is $t$ or a descendant of $t$,
            \item $\beta(c) \not\subseteq \beta(t)$ for every node $c \neq r_0$ with parent $t$, and
            \item $|V(T)| \leq |V(F)| + 1$.
        \end{enumerate}
    \end{lemma}
    We say that an edge $e$ is \emph{represented} at the node $\mathrm{rep}(e)$.
    So the second item asserts that every vertex in a bag is incident with an edge that is represented at this bag or below it.
    \begin{proof}
        Take a tree decomposition of $F$ of depth $\leq \Delta$ with bags of size $\leq k$ whose root $r_0$ has $\beta(r_0) = \emptyset$.
        Choose $\mathrm{rep}(e)$ as any node whose bag contains $e$, for every edge $e$, counted with multiplicity.
        The first item holds.
        For the second item, delete $v$ from $\beta(t)$ whenever it fails.
        For every vertex $v$, the nodes that keep $v$ form an upwards closed subset of the subtree of nodes whose bag contains $v$, so they still induce a subtree.
        This subtree contains $\mathrm{rep}(e)$ for every edge $e$ incident with $v$, and it is non-empty as $v$ is not isolated.
        For the third item, as long as $\beta(c) \subseteq \beta(t)$ for some node $c \neq r_0$ with parent $t$,
        attach the children of $c$ to $t$, represent the edges represented at $c$ at $t$ instead, and delete $c$.
        This preserves the first two items.
        For the fourth item, map every node $c \neq r_0$ to some vertex in $\beta(c) \setminus \beta(t)$, where $t$ is the parent of $c$.
        This map is injective since $c$ is the topmost node whose bag contains this vertex.
    \end{proof}

    \begin{theorem}\label{thm:converse}
        \lean{CircuitsForPatterns}
        Let $\Delta \geq 1$, $k \geq 2$, and $n \geq 2k+1$.
        Let $F$ be a bipartite multigraph with $\fw_\Delta(F) \leq k-1$.
        Then $\hom_{F,n}$ is computed by a reduced matrix-symmetric circuit of support size $\leq k$, support depth $\leq \Delta$, orbit size $\leq n^k$, depth $\leq 2\Delta+1$, and size $\leq 5n^k (|V(F)| + |E(F)|) + 1 $.
    \end{theorem}
    \begin{proof}
        \emph{Decompositions.}
        If $F$ has $j$ isolated vertices and $F'$ is obtained by deleting them, then $\hom_{F,n} = n^j \hom_{F',n}$ by \cref{obs:no-isolated-vertices} and $\fw_\Delta(F') \leq \fw_\Delta(F)$.
        Hence, we may suppose that $F$ has no isolated vertices: at the end, we add the constant $n^j$ as a further child of the output gate, which is a product gate, so that the depth does not grow.
        Let $A \uplus B$ be the bipartition of $F$.
        Let $(T, \beta)$ with root $r_0$ and $\mathrm{rep}$ be as in \cref{lem:normalised-decomposition}.
        Write $\mathrm{level}(t) \leq \Delta$ for the distance of a node $t$ from $r_0$,
        $t' \succeq t$ if $t'$ is a descendant of $t$ or $t' = t$, and $V_t \coloneqq \bigcup_{t' \succeq t} \beta(t')$.

        \emph{The circuit.}
        A map $\gamma \colon X \to [n] \uplus [n]$ on $X \subseteq V(F)$ is \emph{admissible} if it maps $A \cap X$ to the left and $B \cap X$ to the right copy of $[n]$.
        For every node $t$ and admissible $\gamma$ on $\beta(t)$, 
        and for every node $c \neq r_0$ with parent $t$ and admissible $\delta$ on $\beta(t)\cap\beta(c)$, define gates
        \begin{align*}
            R_{t,\gamma} &\coloneqq \prod_{e = ab \in \mathrm{rep}^{-1}(t)} x_{\gamma(a), \gamma(b)} \cdot \prod_{c \text{ child of } t} S_{c,\, \gamma|_{\beta(t)\cap\beta(c)}},
            &
            S_{c,\delta} &\coloneqq \sum_{\gamma' \supseteq \delta \text{ admissible on } \beta(c)} R_{c,\gamma'},
        \end{align*}
        where empty products are the constant $1$.
        The output gate is $R_{r_0, \emptyset}$.
        Write $P_{t,\gamma}$ and $Q_{c,\delta}$ for the polynomials computed by $R_{t,\gamma}$ and $S_{c,\delta}$.
        By induction on $T$,
        \[
            P_{t,\gamma} = \sum_{h \supseteq \gamma \text{ admissible on } V_t} \ \prod_{e \in E(F),\, \mathrm{rep}(e) \succeq t} x_{h(e)}
            \quad\text{and}\quad
            Q_{c,\delta} = \sum_{\gamma' \supseteq \delta} P_{c,\gamma'},
        \]
        where $x_{h(e)} \coloneqq x_{h(a),h(b)}$ for $e = ab$.
        For the inductive step, note that $V_c \cap V_{c'} \subseteq \beta(t)$ and $V_c \cap \beta(t) = \beta(c) \cap \beta(t)$ for distinct children $c, c'$ of $t$, so the extensions of $\gamma$ to the sets $V_c$ can be chosen independently.
        Since $V_{r_0} = V(F)$, the output gate computes $P_{r_0, \emptyset} = \hom_{F,n}$.

        \emph{Depth and size.}
        Write $d(t)$ for the depth of the subtree of $T$ rooted at $t$.
        By induction, $R_{t,\gamma}$ has height $\leq 2d(t) + 1$ and $S_{c,\delta}$ has height $\leq 2d(c) + 2$.
        Since $d(r_0) \leq \Delta$, the circuit has depth $\leq 2\Delta+1$.
        For every node $t$, there are at most $n^{|\beta(t)|} \leq n^k$ gates $R_{t,\gamma}$, each with $|\mathrm{rep}^{-1}(t)|$ wires from input gates and one wire from each child of $t$.
        For every node $c \neq r_0$, there are at most $n^k$ gates $S_{c,\delta}$, which receive at most $n^{|\beta(c)|} \leq n^k$ wires in total.
        Together with at most $n^2 \leq n^k$ input gates and the constants $1$ and $n^j$ with one wire, the size is at most
        \begin{align*}
            &n^k \bigl( |V(T)| + (|E(F)| + |V(T)| - 1) + 2(|V(T)| - 1) + 1 \bigr) + 3 \\
            &\leq n^k \bigl(4|V(F)| + |E(F)| + 3\bigr)
            \leq 5 n^k \bigl(|V(F)| + |E(F)|\bigr),
        \end{align*}
        using $|V(T)| \leq |V(F)| + 1$ by \cref{lem:normalised-decomposition}, $n^k \geq 3$, and $|V(F)| \geq 2$ and $|E(F)| \geq 1$.

        \emph{Symmetry.}
        The group $\Gamma_n$ acts on the gates via $R_{t,\gamma} \mapsto R_{t,\pi\circ\gamma}$, $S_{c,\delta} \mapsto S_{c, \pi\circ\delta}$, and naturally on input gates, and it fixes the remaining gates.
        This preserves labels and wires, so the circuit is matrix-symmetric.
        The orbit of $R_{t,\gamma}$ has at most $n^{|\beta(t)|} \leq n^k$ elements, as does the orbit of $S_{c,\delta}$, and input gates have orbits of size $\leq n^2 \leq n^k$.
        Moreover, $\pi P_{t,\gamma} = P_{t, \pi \circ \gamma}$ and $\pi Q_{c,\delta} = Q_{c,\pi\circ\delta}$ for all $\pi \in \Gamma_n$.
        Let $C$ be the reduced circuit constructed in \cref{lem:reduction}, and write $[g]$ for the gate of $C$ corresponding to a gate $g$.
        The automorphism of $C$ extending $\pi \in \Gamma_n$ maps $[g]$ to $[\pi g]$ by the proof of \cref{lem:reduction} and since $C$ is rigid.
        Hence, $C$ has orbit size $\leq n^k$.
        By \cref{lem:reduction}, the depth and size bounds carry over to $C$.

        \emph{Supports.}
        We show that $\sup([R_{t,\gamma}]) = \gamma(\beta(t))$ and $\sup([S_{c,\delta}]) = \delta(\beta(t) \cap \beta(c))$.
        Let $g \coloneqq [R_{t,\gamma}]$ and $U \coloneqq \gamma(\beta(t))$.
        Every $\pi \in \StabP(U)$ satisfies $\pi \circ \gamma = \gamma$ and thus fixes $g$.
        Since $|U| \leq k < n/2$, the minimal support $\sup(g) \subseteq U$ exists.
        Suppose that there is $p_0 \in U \setminus \sup(g)$, say on the left side, and let $v \in \beta(t)$ with $\gamma(v) = p_0$.
        Since $|U| < n$, there is a left point $p_1 \not\in U$.
        The transposition $\tau$ of $p_0$ and $p_1$ fixes $\sup(g)$ pointwise and thus fixes $g$, so $[R_{t,\tau\circ\gamma}] = g$ and $\tau P_{t,\gamma} = P_{t,\tau\circ\gamma} = P_{t,\gamma}$.
        By \cref{lem:normalised-decomposition}, every monomial of $P_{t,\gamma}$ contains a variable in row $p_0$, so every monomial of $\tau P_{t,\gamma}$ contains a variable in row $p_1$.
        However, mapping all left vertices in $V_t \setminus \beta(t)$ to $p_0$ yields a monomial of $P_{t,\gamma}$ with positive coefficient that contains no variable in row $p_1$, a contradiction.
        The same argument applies to $[S_{c,\delta}]$, using the second item of \cref{lem:normalised-decomposition} at $c$.
        The constants and the input gates $x_{i,j}$ have minimal supports $\emptyset$ and $\{i\} \uplus \{j\}$, respectively.
        In particular, $C$ has support size $\leq k$.

        \emph{Support depth.}
        The children of $[R_{t,\gamma}]$ are input gates $x_{\gamma(a),\gamma(b)}$ for $a,b \in \beta(t)$, constants, and gates $[S_{c, \gamma|_{\beta(t)\cap\beta(c)}}]$.
        Their supports are contained in $\gamma(\beta(t))$, so no wire into $[R_{t,\gamma}]$ is support-changing.
        Hence, only wires into gates $[S_{c,\delta}]$ are support-changing.
        By induction on the height of gates in $C$, 
        the support depth of $[R_{t,\gamma}]$ is at most $\Delta - \mathrm{level}(t)$ 
        and that of $[S_{c,\delta}]$ is at most $\Delta - \mathrm{level}(c) + 1$ for every representative.
        Here, we use that all representatives of a gate of $C$ have the same children in $C$.
        Since $\mathrm{level}(r_0) = 0$, the circuit $C$ has support depth $\leq \Delta$.
    \end{proof}

    \subsection{Homomorphism polynomials from symmetric circuits}\label{app:patterns}

    \label{sec:extended-differences}

    We prove the implication from 1 to 2 in \cref{thm:bounded-depth-characterisation}.
    We follow the approach of \cite[Section~4]{dps_itcs} via homomorphism polynomials of labelled graphs.
    Since the support size $k$ may grow with $n$ and we keep track of the depth of decompositions, we deviate from \cite{dps_itcs} as follows.
    \begin{itemize}
        \item We use rooted tree decompositions whose root bag consists \emph{exactly} of the labelled non-isolated vertices, and isolated vertices occur in no bag.
        In \cite{dps_itcs}, the labelled vertices are only required to lie in a common bag.
        Our convention makes gluing free in depth and adding fresh labels free in width.
        As a consequence, bags have size at most $k$ rather than $2k$ as in \cite{dps_itcs}.
        \item In \cite{dps_itcs}, labels are summed or multiplied out one at a time.
        This would cost one level of depth per label, i.e.\ up to $k$ levels per gate.
        We instead eliminate all labels of an orbit of children at once by a single restricted sum over injective assignments, at the cost of one level (\cref{lem:restricted}).
        \item To that end, products over orbits are expressed via power sums (\cref{lem:newton}), 
        replacing the orbit-counting argument for products and the induction for restricted products of \cite{dps_itcs}.
        \item We work with reduced circuits and \cref{thm:support} instead of rigid circuits of polynomial orbit size, and the depth of the decomposition grows only at support-changing wires (\cref{lem:induction}).
    \end{itemize}

    \subsubsection{Labelled graphs}\label{sec:labelled-graphs}

    Let $\ell, r \in \mathbb{N}$.
    An \emph{$(\ell,r)$-labelled graph} $\boldsymbol{F} = (F, \boldsymbol{a}, \boldsymbol{b})$ consists of a bipartite multigraph $F$ with bipartition $A \uplus B$ and tuples $\boldsymbol{a} \in A^\ell$ and $\boldsymbol{b} \in B^r$ \cite[Definition~4.1]{dps_itcs}.
    For $\boldsymbol{v} \in [n]^\ell$ and $\boldsymbol{w} \in [n]^r$, let
    \[
        \boldsymbol{F}_n(\boldsymbol{v}, \boldsymbol{w}) \coloneqq \sum_{\substack{h \colon A \uplus B \to [n] \uplus [n] \\ h(\boldsymbol{a}) = \boldsymbol{v},\ h(\boldsymbol{b}) = \boldsymbol{w}}} \prod_{ab \in E(F)} x_{h(a), h(b)},
    \]
    where $h$ maps $A$ to the left and $B$ to the right copy of $[n]$.
    For $\ell = r = 0$, this is $\hom_{F,n}$.
    Write $F^+$ for the graph obtained from $F$ by deleting all isolated vertices and $L^+(\boldsymbol{F})$ for the set of labelled vertices of $F^+$.

    For $k \geq 1$ and $\Delta \geq 0$, let $\mathcal{T}^{\Delta,k}(\ell,r)$ denote the class of all $(\ell,r)$-labelled graphs $\boldsymbol{F}$ such that $F^+$ admits a rooted tree decomposition of depth $\leq \Delta$ with bags of size $\leq k$ whose root bag is $L^+(\boldsymbol{F})$.
    We call such a decomposition \emph{root-exact}.
    Let $\mathfrak{T}^{\Delta,k}_{n}(\ell,r)$ denote the $\mathbb{K}$-linear span of the maps $\boldsymbol{F}_n \colon [n]^\ell \times [n]^r \to \mathbb{K}[\Xx_n]$ for $\boldsymbol{F} \in \mathcal{T}^{\Delta,k}(\ell,r)$.
    Clearly, $\mathfrak{T}^{\Delta,k}_{n}(\ell,r) \subseteq \mathfrak{T}^{\Delta+1,k}_{n}(\ell,r)$.
    For a set $I$ of label positions, let $\mathcal{T}^{\Delta,k}_I(\ell,r)$ denote the class of all $\boldsymbol{F} \in \mathcal{T}^{\Delta,k}(\ell,r)$ such that every vertex in $L^+(\boldsymbol{F})$ carries a label at a position in $I$,
    and let $\mathfrak{T}^{\Delta,k}_{n,I}(\ell,r)$ denote the span of the corresponding maps.
    If $\ell + r \leq k$, then taking $I$ to be the set of all positions shows that these restrictions are vacuous.

    The \emph{gluing} $\boldsymbol{F} \odot \boldsymbol{F}'$ of $\boldsymbol{F}, \boldsymbol{F}' \in \mathcal{T}^{\Delta,k}(\ell,r)$ is obtained from the disjoint union of $F$ and $F'$ by identifying $a_i$ with $a'_i$ and $b_j$ with $b'_j$ for all $i$ and $j$.
    Then $(\boldsymbol{F} \odot \boldsymbol{F}')_n$ is the pointwise product of $\boldsymbol{F}_n$ and $\boldsymbol{F}'_n$ \cite{dps_itcs}.
    Write $\boldsymbol{J}$ for the edgeless graph whose labels lie on distinct vertices, and $\boldsymbol{D}^{i,j}$ for the edgeless graph whose labels lie on distinct vertices except that the labels at the positions $i \neq j$ on the same side share a vertex.
    Then $\boldsymbol{J}_n \equiv 1$, and $\boldsymbol{D}^{i,j}_n(\boldsymbol{v},\boldsymbol{w})$ is $1$ if the entries at positions $i$ and $j$ agree and $0$ otherwise.

    \begin{lemma}\label{lem:labelled-ops}
        Let $k \geq 1$, $\Delta \geq 0$, $\ell, r \in \mathbb{N}$, and $\boldsymbol{F}, \boldsymbol{F}' \in \mathcal{T}^{\Delta,k}(\ell,r)$.
        \begin{enumerate}
            \item Every edgeless $(\ell,r)$-labelled graph is in $\mathcal{T}^{0,k}_I(\ell,r)$ for every set $I$ of positions.
            \item If $\boldsymbol{F}, \boldsymbol{F}' \in \mathcal{T}^{\Delta,k}_I(\ell,r)$ for a set $I$ of at most $k$ positions, then $\boldsymbol{F} \odot \boldsymbol{F}' \in \mathcal{T}^{\Delta,k}_I(\ell,r)$.
            \item Permuting the labels of $\boldsymbol{F}$ on each side, and adding labels on new isolated vertices, yields graphs in $\mathcal{T}^{\Delta,k}$.
            If $\boldsymbol{F} \in \mathcal{T}^{\Delta,k}_I$, then the resulting graphs are in $\mathcal{T}^{\Delta,k}_{I'}$, where $I'$ denotes the new positions of the labels at the positions in $I$.
            \item Deleting labels from $\boldsymbol{F}$ yields a graph in $\mathcal{T}^{\Delta+1,k}$.
        \end{enumerate}
        In particular, for every set $I$ of at most $k$ positions, $\mathfrak{T}^{\Delta,k}_{n,I}(\ell,r)$ is a $\mathbb{K}$-algebra under pointwise multiplication with unit $\boldsymbol{J}_n$.
        This applies to $\mathfrak{T}^{\Delta,k}_{n}(\ell,r)$ if $\ell + r \leq k$.
    \end{lemma}
    \begin{proof}
        1.~Use a single node with empty bag.
        2.~Identify the roots of root-exact decompositions of $\boldsymbol{F}$ and $\boldsymbol{F}'$ and apply the quotient map to all bags.
        Since gluing adds no edges, a vertex of $F \odot F'$ is non-isolated if and only if one of its preimages is, so the new root bag is $L^+(\boldsymbol{F} \odot \boldsymbol{F}')$.
        Every vertex in this bag carries a label at a position in $I$, and distinct vertices carry distinct labels, so the root bag has size at most $|I| \leq k$.
        The other bags do not grow.
        Unlabelled vertices occur in one of the two subtrees only, and the occurrences of a labelled non-isolated vertex are connected via the common root.
        3.~The decomposition remains valid.
        4.~Add a new root above the old one whose bag consists of the remaining labelled non-isolated vertices, a subset of the old root bag.
    \end{proof}

    \begin{lemma}\label{lem:labelled-unlabelled}
        Let $k, \Delta \geq 1$.
        An unlabelled graph $F$ is in $\mathcal{T}^{\Delta,k}(0,0)$ if, and only if, $\fw_{\Delta}(F) \leq k-1$.
    \end{lemma}
    \begin{proof}
        A root-exact decomposition of $F^+$ has an empty root bag.
        Attaching a child with bag $\{v\}$ to the root for every isolated vertex $v$ yields a decomposition witnessing $\fw_\Delta(F) \leq k-1$, since $\Delta \geq 1$.
        Conversely, deleting the isolated vertices from all bags of such a decomposition yields a root-exact decomposition of $F^+$.
    \end{proof}

    \subsubsection{Eliminating many labels at once}
    \label{sec:many-labels-at-once}

    Let $\ell, \ell', r, r' \in \mathbb{N}$.
    For $\boldsymbol{v} \in [n]^\ell$ and $\boldsymbol{w} \in [n]^r$, let $Y(\boldsymbol{v},\boldsymbol{w})$ denote the set of pairs $(\boldsymbol{y}, \boldsymbol{z}) \in [n]^{\ell'} \times [n]^{r'}$ of injective tuples such that $\boldsymbol{y}$ avoids the entries of $\boldsymbol{v}$ and $\boldsymbol{z}$ avoids the entries of $\boldsymbol{w}$.
    For a map $\phi$ on $[n]^{\ell+\ell'} \times [n]^{r+r'}$, define the \emph{restricted sum}
    \[
        (\Sigma^{\neq} \phi)(\boldsymbol{v}, \boldsymbol{w}) \coloneqq \sum_{(\boldsymbol{y},\boldsymbol{z}) \in Y(\boldsymbol{v},\boldsymbol{w})} \phi(\boldsymbol{v}\boldsymbol{y}, \boldsymbol{w}\boldsymbol{z}).
    \]

    The following expression of a product in terms of power sums is a consequence of Newton's identities.

    \begin{lemma}[Products from power sums]\label{lem:newton}
        Let $R$ be a commutative $\mathbb{Q}$-algebra, $N \geq 1$, $y_1, \dots, y_N \in R$, and $p_m \coloneqq \sum_{i=1}^N y_i^m$.
        Then
        \[
            \prod_{i=1}^N y_i = \frac{1}{N!} \sum_{\pi} \mu(\pi) \prod_{P \in \pi} p_{|P|}
            \quad\text{where}\quad
            \mu(\pi) \coloneqq \prod_{P \in \pi} (-1)^{|P|-1} (|P|-1)!
        \]
        and $\pi$ ranges over all partitions of $[N]$.
    \end{lemma}
    \begin{proof}
        For a partition $\pi$, let $g(\pi) \coloneqq \sum_{h \colon \pi \to [N]} \prod_{i} y_{h(\pi(i))} = \prod_{P \in \pi} p_{|P|}$, where $\pi(i)$ is the part containing $i$,
        and let $f(\pi)$ be the same sum restricted to injective $h$.
        Then $g(\pi) = \sum_{\sigma \geq \pi} f(\sigma)$, where $\sigma$ ranges over coarsenings of $\pi$.
        By Möbius inversion on the partition lattice, $N! \prod_i y_i = f(\hat 0) = \sum_\sigma \mu(\sigma) g(\sigma)$.
    \end{proof}

    The following lemma eliminates the last $\ell'$ left and $r'$ right labels at once.

    \begin{lemma}\label{lem:restricted}
        Let $k, c \geq 1$ and $\Delta \geq 0$ with $\ell + r \leq k$, let $I$ be a set of at most $k$ positions, and let $\phi \in \mathfrak{T}^{\Delta,k}_{n,I}(\ell+\ell', r+r')$.
        There are $\psi_\Sigma, \psi_\Pi \in \mathfrak{T}^{\Delta+1,k}_{n}(\ell, r)$ with the following property.
        Let $\boldsymbol{v} \in [n]^\ell$ and $\boldsymbol{w} \in [n]^r$ be injective, and let $\mathcal{P}$ be a partition of $Y(\boldsymbol{v},\boldsymbol{w})$ into classes each of size exactly $c$
        such that $(\boldsymbol{y},\boldsymbol{z}) \mapsto \phi(\boldsymbol{v}\boldsymbol{y}, \boldsymbol{w}\boldsymbol{z})$ is constant on every class $P \in \mathcal{P}$.
        For $P \in \mathcal{P}$, write $\phi_P \coloneqq \phi(\boldsymbol{v}\boldsymbol{y}, \boldsymbol{w}\boldsymbol{z})$ for any $(\boldsymbol{y},\boldsymbol{z}) \in P$.
        Then
        \[
            \psi_\Sigma(\boldsymbol{v},\boldsymbol{w}) = \sum_{P \in \mathcal{P}} \phi_P
            \quad\text{and}\quad
            \psi_\Pi(\boldsymbol{v},\boldsymbol{w}) = \prod_{P \in \mathcal{P}} \phi_P.
        \]
    \end{lemma}
    \begin{proof}
        Let $\delta$ be the pointwise product of the maps $\boldsymbol{J}_n - \boldsymbol{D}^{i,j}_n$ over all pairs of distinct positions $i,j$ on the same side of which at least one is among the last $\ell'$ left or $r'$ right positions.
        By \cref{lem:labelled-ops}, $\delta \in \mathfrak{T}^{0,k}_{n,I}(\ell+\ell',r+r')$, and $\delta(\boldsymbol{v}\boldsymbol{y}, \boldsymbol{w}\boldsymbol{z})$ is the indicator of $(\boldsymbol{y},\boldsymbol{z}) \in Y(\boldsymbol{v},\boldsymbol{w})$.
        For every $\phi' \in \mathfrak{T}^{\Delta,k}_{n,I}(\ell+\ell', r+r')$, the map $\Sigma^{\neq}\phi'$ is obtained from $\delta \odot \phi'$ by deleting the last $\ell'$ left and $r'$ right labels,
        so $\Sigma^{\neq}\phi' \in \mathfrak{T}^{\Delta+1,k}_{n}(\ell, r)$ by \cref{lem:labelled-ops}.

        Let $\psi_\Sigma \coloneqq \frac1c \Sigma^{\neq}\phi$.
        Under the assumptions of the lemma, $\psi_\Sigma(\boldsymbol{v},\boldsymbol{w}) = \frac1c \sum_{(\boldsymbol{y},\boldsymbol{z}) \in Y(\boldsymbol{v},\boldsymbol{w})} \phi(\boldsymbol{v}\boldsymbol{y}, \boldsymbol{w}\boldsymbol{z}) = \sum_{P \in \mathcal{P}} \phi_P$.
        For injective $\boldsymbol{v}$ and $\boldsymbol{w}$, the size of $Y(\boldsymbol{v},\boldsymbol{w})$ depends only on $n$, $\ell$, $\ell'$, $r$, and $r'$.
        If $c$ does not divide it, then no partition $\mathcal{P}$ as above exists and we let $\psi_\Pi \coloneqq 0$.
        Otherwise, let $N \coloneqq |Y(\boldsymbol{v},\boldsymbol{w})|/c$, let $Q_m \coloneqq \frac1c \Sigma^{\neq}(\phi^{\odot m})$ for $m \geq 1$, 
        and let $\psi_\Pi \coloneqq \frac1{N!} \sum_\pi \mu(\pi) \bigodot_{B \in \pi} Q_{|B|}$, where $\pi$ ranges over all partitions of $[N]$.
        Since $\phi^{\odot m} \in \mathfrak{T}^{\Delta,k}_{n,I}(\ell+\ell', r+r')$ and $\ell + r \leq k$, we have $\psi_\Pi \in \mathfrak{T}^{\Delta+1,k}_{n}(\ell, r)$ by \cref{lem:labelled-ops}.
        Under the assumptions of the lemma, $|\mathcal{P}| = N$ and $Q_m(\boldsymbol{v},\boldsymbol{w}) = \frac1c \sum_{(\boldsymbol{y},\boldsymbol{z}) \in Y(\boldsymbol{v},\boldsymbol{w})} \phi(\boldsymbol{v}\boldsymbol{y}, \boldsymbol{w}\boldsymbol{z})^m = \sum_{P \in \mathcal{P}} \phi_P^m$ for all $m \geq 1$.
        That is, $Q_m(\boldsymbol{v},\boldsymbol{w})$ is the $m$-th power sum $p_m$ of the $N$ elements $\phi_P$ for $P \in \mathcal{P}$ of $\mathbb{K}[\Xx_n]$.
        Since pointwise products are evaluated pointwise, \cref{lem:newton} applied to these elements yields
        \[
            \psi_\Pi(\boldsymbol{v},\boldsymbol{w}) 
            = \frac1{N!} \sum_\pi \mu(\pi) \prod_{B \in \pi} Q_{|B|}(\boldsymbol{v},\boldsymbol{w})
            = \frac1{N!} \sum_\pi \mu(\pi) \prod_{B \in \pi} p_{|B|}
            = \prod_{P \in \mathcal{P}} \phi_P.\qedhere
        \]
    \end{proof}

    We stress that the costs in \cref{lem:restricted} depend neither on $\ell'$ and $r'$ nor on $k$.

    \subsubsection{Induction on the circuit}

    Let $C$ be a reduced matrix-symmetric circuit with support size $k < n/2$, so that every gate $g$ has a minimal support.
    Fix an ordering $\vec{\sup}(g) = (\vec{\sup}_L(g), \vec{\sup}_R(g))$ of $\sup(g)$ for every gate $g$,
    and write $\mathrm{sc}(g)$ for the maximal number of support-changing wires on a path from $g$ to an input gate.
    For $\gamma \in \Gamma_n$, we have $p_{\gamma g} = \gamma p_g$ and $\sup(\gamma g) = \gamma \sup(g)$.

    \begin{lemma}\label{lem:induction}
        For every gate $g$ of $C$, there is $\phi_g \in \mathfrak{T}^{\mathrm{sc}(g),k}_{n}(|\sup_L(g)|, |\sup_R(g)|)$ such that $p_g = \phi_g(\vec{\sup}(g))$.
    \end{lemma}
    \begin{proof}
        \emph{Input gates.}
        The gate $x_{i,j}$ has minimal support $\{i\} \uplus \{j\}$, and $k \geq 2$.
        Let $\phi$ be the map of the $(1,1)$-labelled single edge, whose root bag is the edge.
        A constant $c$ has empty support, and we let $\phi \coloneqq c \boldsymbol{J}_n$.

        \emph{Internal gates.}
        Let $g$ be internal, $\boldsymbol{u} \coloneqq \vec{\sup}(g)$, and $G \coloneqq \StabP(\sup(g)) \leq \Stab(g)$.
        Then $G$ permutes the children of $g$ and preserves wire multiplicities.
        Let $O$ be a $G$-orbit of children of $g$, let $m_O$ be the common multiplicity of the wires from $O$ to $g$, and let $h \in O$.
        Let $N \coloneqq \sup(h) \setminus \sup(g)$, ordered as in $\vec{\sup}(h)$.
        By \cref{lem:labelled-ops}, there is $\psi_h \in \mathfrak{T}^{\mathrm{sc}(h),k}_{n,I}$, where $I$ is the set of the $|\sup(h)| \leq k$ positions of the elements of $\sup(h)$, with label positions for $\boldsymbol{u}$ followed by positions for $N$ such that
        $\psi_h(\boldsymbol{u}, \boldsymbol{y}) = \phi_h(\vec{\sup}(h)[N \mapsto \boldsymbol{y}])$ for all tuples $\boldsymbol{y}$,
        obtained by permuting the labels of $\phi_h$ and adding labels for $\sup(g) \setminus \sup(h)$ on isolated vertices.
        Here and below, we suppress the split of the label positions into left and right ones.

        Let $Y_{\boldsymbol{u}} \coloneqq Y(\boldsymbol{u})$ be the set of injective tuples $\boldsymbol{y}$ for the positions of $N$ that avoid $\sup(g)$ on each side.
        Since $G$ acts as the full symmetric group on the complement of $\sup(g)$ on each side, it acts transitively on $Y_{\boldsymbol{u}}$.
        For $\boldsymbol{y} \in Y_{\boldsymbol{u}}$, choose $\gamma \in G$ mapping $N$ to $\boldsymbol{y}$ and let $h_{\boldsymbol{y}} \coloneqq \gamma h$.
        This does not depend on the choice of $\gamma$, since two choices differ by an element of $\StabP(\sup(g) \cup N) \leq \StabP(\sup(h)) \leq \Stab(h)$.
        The map $\boldsymbol{y} \mapsto h_{\boldsymbol{y}}$ is $G$-equivariant and onto $O$, so all its fibres have the same size $c$.
        Moreover, $p_{h_{\boldsymbol{y}}} = \gamma p_h = \phi_h(\gamma \vec{\sup}(h)) = \psi_h(\boldsymbol{u},\boldsymbol{y})$.

        If $N = \emptyset$, then $O = \{h\}$ since $G \leq \StabP(\sup(h))$, and we let $\psi^O_\Sigma \coloneqq \psi^O_\Pi \coloneqq \psi_h \in \mathfrak{T}^{\mathrm{sc}(g),k}_{n}$.
        Otherwise, the wires from $O$ to $g$ are support-changing, so $\mathrm{sc}(g) \geq \mathrm{sc}(h) + 1$.
        By \cref{lem:restricted}, applied to $\psi_h$ with the fibres of $\boldsymbol{y} \mapsto h_{\boldsymbol{y}}$ as classes,
        there are $\psi^O_\Sigma$ and $\psi^O_\Pi$ in $\mathfrak{T}^{1+\mathrm{sc}(h),k}_{n} \subseteq \mathfrak{T}^{\mathrm{sc}(g),k}_{n}$ such that
        \[
            \psi^O_\Sigma(\boldsymbol{u}) = \sum_{h' \in O} p_{h'}
            \quad\text{and}\quad
            \psi^O_\Pi(\boldsymbol{u}) = \prod_{h' \in O} p_{h'}.
        \]
        If $g$ is a sum gate, let $\phi_g \coloneqq \sum_O m_O \psi^O_\Sigma$.
        If $g$ is a product gate, let $\phi_g$ be the pointwise product of the maps $(\psi^O_\Pi)^{\odot m_O}$.
        Then $\phi_g \in \mathfrak{T}^{\mathrm{sc}(g),k}_{n}$ by \cref{lem:labelled-ops}, since $|\sup(g)| \leq k$, and $\phi_g(\boldsymbol{u}) = p_g$.
    \end{proof}

    \begin{theorem}\label{thm:patterns}
        \lean{PatternsFromCircuits}
        Let $\Delta \geq 1$, $k \geq 2$, and $n \geq 2k+1$.
        If $p$ is computed by a reduced matrix-symmetric circuit of support size $\leq k$ and support depth $\leq \Delta$,
        then $p$ is a linear combination of homomorphism polynomials $\hom_{F,n}$ for patterns $F$ with $\fw_\Delta(F) \leq k-1$.
    \end{theorem}
    \begin{proof}
        Since $k < n/2$, \cref{lem:induction} applies.
        The output gate is fixed by $\Gamma_n$, so its minimal support is empty.
        Hence, $p \in \mathfrak{T}^{\Delta,k}_{n}(0,0)$, and the claim follows from \cref{lem:labelled-unlabelled}.
    \end{proof}

    \subsection{Combinatorics of bounded-depth tree decompositions}
    \label{sec:combinatorics-delta-tw}

    In this section, we introduce a dual notion to $\Delta$-forestwidth certifying that a graph has high $\Delta$-forestwidth.
    We then show that this notion is reflected by weak oddomorphisms.
    These are graph homomorphisms with certain parity constraints introduced by \textcite{roberson_oddomorphisms_2022} arising in the analysis of CFI graphs.
    We use such graphs in \cref{thm:non-uniform-non-uniform}, the main tool for proving symmetric lower bounds on homomorphism polynomials.

    \subsubsection{Richness}
    We introduce a dual definition to $\Delta$-forestwidth following \cite[Lemma~18]{bhargav_monotone_2025}.
    \begin{definition}
        Let $k \geq 1$.
        A graph $F$ is \emph{$(1,k)$-rich} if some connected component of $F$ has at least $k+1$ vertices.
        For $\Delta \geq 2$, a connected graph $F$ is \emph{$(\Delta, k)$-rich} if it contains $k+1$ pairwise vertex-disjoint connected subgraphs each of which is $(\Delta-1, k)$-rich.
        A disconnected graph is \emph{$(\Delta, k)$-rich} if it has a $(\Delta,k)$-rich connected component.
    \end{definition}

    \begin{lemma}\label{lem:rich-fw}
        Let $\Delta, k \geq 1$.
        If a graph $F$ is $(\Delta,k)$-rich, then $\fw_\Delta(F) \geq k$.
    \end{lemma}
    \begin{proof}
        The proof essentially follows \cite[Lemma~18]{bhargav_monotone_2025}.
        It suffices to treat a rich connected component, so let $F$ be connected. Suppose that $F$ has a tree decomposition of depth at most $\Delta$ with empty root bag and all bags of size at most $k$. As the root bag is empty and $F$ is connected, $F$ lies in the bags of the subtree $(T,r,\beta)$ rooted at a single child~$r$ of the root, which has depth at most $\Delta-1$. Induction on $\Delta$. For $\Delta = 1$, $T$ is a single node, so $|V(F)| \geq k+1$ forces a bag of size at least $k+1$.

    Let $\Delta \geq 2$ and $R \coloneqq \beta(r)$, so $|R| \leq k$. Among the $k+1$ disjoint $(\Delta-1,k)$-rich connected subgraphs $F_1, \dots, F_{k+1}$ of $F$, some $F_i$ is disjoint from $R$. A vertex outside $R$ occurs in the bags of exactly one child subtree of $r$, and two adjacent vertices outside $R$ occur in the same one, because a bag containing both cannot be $\beta(r)$. Hence $V(F_i)$ lies in the bags of one child subtree $T_c$, and adding a root with an empty bag above $(T_c, c, \beta|_{T_c} \cap V(F_i))$ yields a tree decomposition of $F_i$ of depth at most $\Delta-1$ with empty root bag and bags of size at most $k$. So $\fw_{\Delta-1}(F_i) \leq k-1$, contradicting the induction hypothesis.
    \end{proof}

    Let $P_m$ denote the $m$-vertex path.
    \begin{example}\label{ex:path-rich}
        Let $\Delta, k\geq 1$.
        The path $P_{(k+1)^\Delta}$ is $(\Delta, k)$-rich.
    \end{example}
    \begin{proof}
        We show by induction on $\Delta$ that every path with at least $(k+1)^\Delta$ vertices is $(\Delta,k)$-rich.
        For $\Delta = 1$, the path is connected and has at least $k+1$ vertices.
        For $\Delta \geq 2$, the path is connected and splits into $k+1$ pairwise vertex-disjoint subpaths with at least $(k+1)^{\Delta-1}$ vertices each, which are $(\Delta-1,k)$-rich by induction.
    \end{proof}

    \begin{example}[\cite{bhargav_monotone_2025}]\label{ex:rich-tree}
        Let $\Delta, k \geq 1$.
        The full $(k+1)$-ary tree $T_{\Delta, k+1}$ of depth~$\Delta$ is $(\Delta, k)$-rich.
    \end{example}
    \begin{proof}
        This follows by induction on $\Delta$:
        The tree $T_{1,k+1}$ is connected and has $k+2$ vertices.
        The tree $T_{\Delta+1, k+1}$ is connected and the subtrees rooted at the $k+1$ children of its root are disjoint copies of $T_{\Delta,k+1}$.
        In particular, $\fw_{\Delta}(T_{\Delta, k+1}) \geq k$ by \cref{lem:rich-fw}, cf.\ \cref{example:bccd17}.
    \end{proof}

     \begin{lemma}\label{lem:rich-minor-mono}
    Let $\Delta, k \geq 1$.
    If a connected graph $F$ has a $(\Delta,k)$-rich minor, then $F$ is $(\Delta,k)$-rich.
    \end{lemma}
    \begin{proof}
    Let $P$ be a rich minor of $F$ with a model $(X_v)_{v \in V(P)}$ of disjoint connected branch sets in $F$. We may assume $P$ connected. For $\Delta = 1$, $|V(F)| \geq |V(P)| \geq k+1$. For $\Delta \geq 2$, let $P_1, \dots, P_{k+1}$ be disjoint $(\Delta-1,k)$-rich connected subgraphs of $P$; then $F_i \coloneqq F[\bigcup_{v \in V(P_i)} X_v]$ are disjoint connected subgraphs of $F$ having $P_i$ as a minor, hence $(\Delta-1,k)$-rich by induction, and $F$ is connected.
    \end{proof}

    \begin{lemma}\label{lem:rich-tree}
    Let $\Delta, k \geq 1$. Every $(\Delta,k)$-rich graph $F$ has a $(\Delta,k)$-rich tree $P$ as a minor with $|V(P)| = (k+1)^\Delta$ and $\operatorname{maxdeg}(P) \leq \Delta k$.
    \end{lemma}
    \begin{proof}
    Induction on $\Delta$, within a rich component $F$. For $\Delta = 1$, any subtree of $F$ with $k+1$ vertices is $(1,k)$-rich and has maximum degree at most $k$.

    Let $\Delta \geq 2$ and let $H_1, \dots, H_{k+1}$ be disjoint $(\Delta-1,k)$-rich connected subgraphs of $F$. By induction, each $H_i$ has a $(\Delta-1,k)$-rich tree minor $P_i$ with $(k+1)^{\Delta-1}$ vertices and maximum degree at most $(\Delta-1)k$. The branch sets of models of the $P_i$ in the $H_i$ are pairwise disjoint connected subsets of $V(F)$. Since $F$ is connected, we may enlarge them to a partition of $V(F)$ into connected sets by repeatedly adding an uncovered vertex to a branch set containing one of its neighbours. The enlarged sets still form models of the $P_i$ in $F$; let $Y_i$ be the union of the branch sets of the model of $P_i$, a connected set. The graph obtained from $F$ by contracting each of $Y_1, \dots, Y_{k+1}$ to a single vertex is connected, so it has a spanning tree $T$ with $k$ edges. For each edge $ij$ of $T$, pick an edge of $F$ between $Y_i$ and $Y_j$; it joins the branch set of some vertex of $P_i$ to the branch set of some vertex of $P_j$. Let $P$ be the disjoint union of the $P_i$ together with the $k$ edges between the corresponding vertices. Then $P$ is a minor of $F$ and a tree, since it is obtained from the tree $T$ by replacing every vertex by a tree. It contains the disjoint $(\Delta-1,k)$-rich subtrees $P_i$, so it is $(\Delta,k)$-rich, and it has $(k+1) \cdot (k+1)^{\Delta-1} = (k+1)^\Delta$ vertices. As $T$ has $k+1$ vertices, at most $k$ of the added edges are incident to vertices of any fixed $P_i$, so the maximum degree of $P$ is at most $(\Delta-1)k + k = \Delta k$.
    \end{proof}

    \subsubsection{Oddomorphisms reflect richness}

    \begin{definition}[\cite{roberson_oddomorphisms_2022}]
        \lean{Oddomorphisms}
        Let $F$ and $G$ be graphs.
        Let $\phi \colon F\to G$ be a homomorphism.
        A vertex $a \in V(F)$ is \emph{$\phi$-even} /  \emph{$\phi$-odd} if $|N_F(a) \cap \phi^{-1}(x)|$ is even/odd for every $x \in N_G(\phi(a))$.
        The homomorphism $\phi$ is an \emph{oddomorphism}
        if 
        \begin{enumerate}
            \item every vertex $a\in V(F)$ is $\phi$-even or $\phi$-odd,
            \item for every $x \in V(G)$, the set $\phi^{-1}(x)$ contains an odd number of $\phi$-odd vertices.
        \end{enumerate}
        The homomorphism $\phi$ is a \emph{weak oddomorphism} if there is a subgraph $F' \subseteq F$ such that $\phi|_{F'} \colon F' \to G$ is an oddomorphism.
    \end{definition}

    \begin{lemma}[\protect{\cite[Theorem~3.13, Lemma~5.2]{roberson_oddomorphisms_2022}}]\label{lem:parity-simplified}
    Let $\phi \colon F \to G$ be an oddomorphism.
    \begin{enumerate}
        \item If $G$ is connected,
        then there exists a connected subgraph $F'$ of $F$ such that $\phi|_{F'} \colon F' \to G$ is an oddomorphism.
        \item For every $U \subseteq V(G)$, the restriction of $\phi$ to $F[\phi^{-1}(U)]$ is an oddomorphism $F[\phi^{-1}(U)] \to G[U]$.
    \end{enumerate}
\end{lemma}
\begin{proof}
    For the second item, let $F_U \coloneqq F[\phi^{-1}(U)]$.
    For $a \in \phi^{-1}(U)$ and $x \in U$, it holds that $N_{F_U}(a) \cap \phi^{-1}(x) = N_F(a) \cap \phi^{-1}(x)$.
    Hence, every $\phi$-even or $\phi$-odd vertex of $F$ in $\phi^{-1}(U)$ is even or odd, respectively, for the restriction of $\phi$ to $F_U$.
    Moreover, the fibres of $\phi$ over the vertices in $U$ are unchanged.
\end{proof}

    \begin{theorem}\label{thm:rich-oddo}
    Let $\Delta, k \geq 1$. 
    Let $F \to G$ be a weak oddomorphism.
    If $G$ is $(\Delta,k)$-rich, then $F$ is $(\Delta,k)$-rich.
    \end{theorem}

    The proof is similar to that of \cite[Theorem~3.6]{neuen_distinguishing_2026}.

    \begin{proof}
    A weak oddomorphism restricts to an oddomorphism from a subgraph of $F$, and a graph with a rich subgraph is rich, so it suffices to show: if $\varphi\colon F \to G$ is an oddomorphism and $G$ is $(\Delta,k)$-rich, then $F$ is $(\Delta,k)$-rich. We induct on $\Delta$. Let $G_0$ be a rich component of $G$. By \cref{lem:parity-simplified}(2) with $U = V(G_0)$ and then \cref{lem:parity-simplified}(1), some connected subgraph $F_0 \subseteq F$ admits an oddomorphism onto $G_0$; it suffices to show that $F_0$ is rich. So we may assume that $F$ and $G$ are connected.

    For $\Delta = 1$, every fibre of an oddomorphism is non-empty, so $|V(F)| \geq |V(G)| \geq k+1$. For $\Delta \geq 2$, let $G_1, \dots, G_{k+1}$ be pairwise disjoint $(\Delta-1,k)$-rich connected subgraphs of $G$. For each $i$, by \cref{lem:parity-simplified}(2) with $U = V(G_i)$, the restriction of $\varphi$ to $F_i \coloneqq F[\varphi^{-1}(V(G_i))]$ is an oddomorphism onto $G[V(G_i)]$, which is connected and $(\Delta-1,k)$-rich. By induction $F_i$ is $(\Delta-1,k)$-rich, so it contains a $(\Delta-1,k)$-rich connected subgraph $D_i$. The $D_i$ are pairwise disjoint, since they lie over disjoint vertex sets, and $F$ is connected. Hence $F$ is $(\Delta,k)$-rich.
    \end{proof}

    We collect the facts about $(\Delta, k)$-rich graphs proven here and thereby verify the hypotheses of \cref{thm:non-uniform-non-uniform}.

    \begin{corollary}\label{cor:non-rich-class}
        Let $\Delta, k \geq 1$.
        The class $\mathcal{F}$ of all graphs that are not $(\Delta, k)$-rich contains $K_1$ and is closed under taking minors and weak oddomorphisms.
    \end{corollary}
    \begin{proof}
        Every $(\Delta, k)$-rich graph has at least $k+1 \geq 2$ vertices. Hence, $K_1 \in \mathcal{F}$.
        If a graph $F$ has a $(\Delta, k)$-rich minor, then a rich connected component of this minor is a minor of a connected component of $F$, which is $(\Delta,k)$-rich by \cref{lem:rich-minor-mono}. Hence, $\mathcal{F}$ is minor-closed.
        By \cref{thm:rich-oddo}, $\mathcal{F}$ is closed under weak oddomorphisms.
    \end{proof}

    \subsection{Non-uniform descriptive complexity monotonicity}

    We prove \cref{thm:non-uniform-non-uniform}.

    \restatethm{thm:non-uniform-non-uniform}{\thmNonUnfirom}

    The proof of \cref{thm:non-uniform-non-uniform} follows \cite[Theorem~6.1]{dps_itcs}, where a graph $F \not\in \mathcal{F}$ is distinguished from $\mathcal{F}$ by means of the CFI graphs of an \emph{induced} minor of $F$ that does not belong to $\mathcal{F}$.
    The size of these CFI graphs is exponential in the maximum degree of the induced minor, which may be much larger than the maximum degree of the forbidden minor~$G$ it contains.
    We avoid this blow-up by modifying the CFI graphs such that they can be used for minors which are not necessarily induced.
    Furthermore, we work with bipartite graphs with fixed bipartitions throughout, which renders the dependence of the bound in \cite{dps_itcs} on the chromatic number of $F$ obsolete.

    \subsubsection{CFI graphs for non-induced minors}

    We recall the CFI graphs of \cite{roberson_oddomorphisms_2022}, cf.\ \cite{cai_optimal_1992}.
    For a connected graph $G$ and $v \in V(G)$, write $E(v) \subseteq E(G)$ for the set of edges incident to $v$.
    For $W \in \mathbb{Z}_2^{V(G)}$, the graph $G_W$ has vertices $(v, T)$ for $v \in V(G)$ and $T \in \mathbb{Z}_2^{E(v)}$ such that $\sum_{e \in E(v)} T(e) = W(v)$.
    Two vertices $(u, S)$ and $(v, T)$ are adjacent if $uv \in E(G)$ and $S(uv) = T(uv)$.
    Up to isomorphism, $G_W$ depends only on $\sum_{v \in V(G)} W(v) \in \mathbb{Z}_2$ \cite[Corollary~3.7]{roberson_oddomorphisms_2022}.
    We write $G_0$ and $G_1$ for the two resulting graphs.
    Each of them has $\gamma(G)$ vertices.
    The map $\rho \colon G_i \to G$, $(v, T) \mapsto v$ is a homomorphism.
    For a homomorphism $\psi \colon F \to G$, write $\hom_\psi(F, G_i)$ for the number of homomorphisms $h \colon F \to G_i$ such that $\rho \circ h = \psi$.

    \begin{lemma}[\protect{\cite[Theorem~3.6 and proof of Theorem~3.13]{roberson_oddomorphisms_2022}}]\label{lem:cfi-psi}
        Let $G$ be a connected graph and $\psi \colon F \to G$ be a homomorphism.
        Then $\hom_\psi(F, G_0) \geq \hom_\psi(F, G_1)$ with strict inequality if, and only if, $\psi$ is a weak oddomorphism.
    \end{lemma}

    For $i \in \{0,1\}$,
    let $\widehat{G}_i$ denote the graph obtained from $G_i$ by adding all edges $(u,S)(v,T)$ for distinct non-adjacent vertices $u, v \in V(G)$.
    That is, the non-edges of $G$ are replaced by complete bipartite graphs between the respective gadgets.
    Clearly, $\widehat{G}_i$ has $\gamma(G)$ vertices.

    \begin{lemma}\label{lem:filled-cfi}
        Let $G$ be a connected graph and $F$ be a graph.
        Then $\hom(F, \widehat{G}_0) \geq \hom(F, \widehat{G}_1)$ with strict inequality if, and only if, 
        there exist a subgraph $F'$ of $F$ with $V(F') = V(F)$ and a weak oddomorphism $\varphi \colon F' \to G$ such that $\varphi(a) \neq \varphi(b)$ for all $ab \in E(F)$.
    \end{lemma}
    \begin{proof}
        Write $K$ for the complete graph on $V(G)$.
        Since $\widehat{G}_i$ has no edges between vertices $(v, S)$ and $(v, T)$, the map $\rho \colon \widehat{G}_i \to K$, $(v, T) \mapsto v$ is a homomorphism.
        For a homomorphism $\psi \colon F \to K$, 
        let $F_\psi$ denote the subgraph of $F$ with vertex set $V(F)$ and edge set $\{ab \in E(F) \mid \psi(a)\psi(b) \in E(G)\}$.
        Then $\psi$ is a homomorphism $F_\psi \to G$.
        Let $h \colon V(F) \to V(\widehat{G}_i)$ be a map such that $\rho \circ h = \psi$.
        If $ab \in E(F) \setminus E(F_\psi)$, then $\psi(a)$ and $\psi(b)$ are distinct and non-adjacent in $G$, so $h(a)h(b) \in E(\widehat{G}_i)$.
        If $ab \in E(F_\psi)$, then $h(a)h(b) \in E(\widehat{G}_i)$ if, and only if, $h(a)h(b) \in E(G_i)$.
        Hence, $h$ is a homomorphism $F \to \widehat{G}_i$ if, and only if, it is a homomorphism $F_\psi \to G_i$.
        It follows that
        \begin{equation}\label{eq:filled-cfi}
            \hom(F, \widehat{G}_i) = \sum_{\psi \colon F \to K} \hom_\psi(F_\psi, G_i).
        \end{equation}
        By \cref{lem:cfi-psi}, this implies that $\hom(F, \widehat{G}_0) \geq \hom(F, \widehat{G}_1)$ 
        and that the inequality is strict if, and only if, there is a homomorphism $\psi \colon F \to K$ such that $\psi \colon F_\psi \to G$ is a weak oddomorphism.
        If such a $\psi$ exists, then $F' \coloneqq F_\psi$ and $\varphi \coloneqq \psi$ are as desired.
        Conversely, let $F'$ and $\varphi$ be as in the lemma.
        Then $\varphi$ is a homomorphism $F \to K$ and $F'$ is a subgraph of $F_\varphi$.
        Since $\varphi$ restricts to an oddomorphism on a subgraph of $F'$, 
        the homomorphism $\varphi \colon F_\varphi \to G$ is a weak oddomorphism.
    \end{proof}

    Next, we lift \cref{lem:filled-cfi} to bipartite graphs containing $G$ as a minor.
    For bipartite graphs $F$ and $B$ with fixed bipartitions,
    write $\hom_{\mathrm{bip}}(F, B)$ for the number of homomorphisms $F \to B$ that map the left side of $F$ to the left side of $B$ and the right side of $F$ to the right side of $B$.
    For a simple graph $H$, 
    let $B(H)$ denote the simple bipartite graph with left side $V(H) \times \{0\}$ and right side $V(H) \times \{1\}$ in which $(u, 0)$ and $(v, 1)$ are adjacent if $u = v$ or $uv \in E(H)$.
    That is, $B(H)$ is obtained from the bipartite double cover of $H$ by adding the perfect matching $(v, 0)(v,1)$ for $v \in V(H)$.
    For a graph $F$, write $\Gamma(F)$ for the set of all partitions $\mathcal{R}$ of $V(F)$ such that $F[R]$ is connected for all $R \in \mathcal{R}$.
    For $\mathcal{R} \in \Gamma(F)$, 
    the simple graph $F/\mathcal{R}$ has vertex set $\mathcal{R}$ and two distinct parts are adjacent if there is an edge of $F$ between them.

    \begin{lemma}\label{lem:bip-contract}
        For every simple bipartite graph $F$ with fixed bipartition and every simple graph $H$,
        \begin{equation}\label{eq:bip-contract}
            \hom_{\mathrm{bip}}(F, B(H)) = \sum_{\mathcal{R} \in \Gamma(F)} \hom(F/\mathcal{R}, H).
        \end{equation}
    \end{lemma}
    Note that the graphs $F/\mathcal{R}$ and $H$ on the right-hand side of \cref{eq:bip-contract} are not necessarily bipartite and that the right-hand side does not depend on the bipartition of $F$.
    \begin{proof}
        Write $A \uplus A'$ for the fixed bipartition of $F$.
        Call a map $h \colon V(F) \to V(H)$ \emph{admissible} if $h(a) = h(a')$ or $h(a)h(a') \in E(H)$ for all $aa' \in E(F)$.
        By definition of $B(H)$,
        the bipartition-respecting homomorphisms $F \to B(H)$ correspond bijectively to the admissible maps~$h$
        via $a \mapsto (h(a), 0)$ for $a \in A$ and $a' \mapsto (h(a'), 1)$ for $a' \in A'$.
        It remains to give a bijection between the admissible maps and the pairs $(\mathcal{R}, g)$ of a partition $\mathcal{R} \in \Gamma(F)$ and a homomorphism $g \colon F/\mathcal{R} \to H$.

        Given an admissible map $h$, 
        let $\mathcal{R}$ be the set of vertex sets of the connected components of the graphs $F[h^{-1}(v)]$ for $v \in V(H)$.
        Then $\mathcal{R} \in \Gamma(F)$ and $h$ is constant on every part.
        Let $g \colon \mathcal{R} \to V(H)$ send a part to the image of its vertices under $h$.
        If two distinct parts $R, S \in \mathcal{R}$ are adjacent in $F/\mathcal{R}$, then there is an edge $rs \in E(F)$ with $r \in R$ and $s \in S$.
        Since $r$ and $s$ are adjacent and lie in distinct connected components of the fibres of $h$, we have $h(r) \neq h(s)$ and thus $g(R)g(S) = h(r)h(s) \in E(H)$ as $h$ is admissible.
        Hence, $g$ is a homomorphism $F/\mathcal{R} \to H$.

        Conversely, given $\mathcal{R} \in \Gamma(F)$ and a homomorphism $g \colon F/\mathcal{R} \to H$,
        let $h$ send every vertex in a part $R \in \mathcal{R}$ to $g(R)$.
        This map is admissible: the endpoints of an edge of $F$ lie in the same part or in two parts that are adjacent in $F/\mathcal{R}$.
        It remains to observe that the partition constructed from $h$ as above is $\mathcal{R}$.
        Every part $R \in \mathcal{R}$ induces a connected subgraph on which $h$ is constant.
        If there is an edge of $F$ between two distinct parts $R$ and $S$, then $g(R) \neq g(S)$ since $H$ does not have loops.
        Hence, the connected components of the fibres of $h$ are precisely the parts of $\mathcal{R}$.
        The two constructions are mutually inverse.
    \end{proof}

    \Cref{lem:bip-contract} is a variant of the identity
    \begin{equation}\label{eq:lexprod}
        \hom(F, H \cdot K) = \sum_{\mathcal{R} \in \Gamma(F)} \hom(F/\mathcal{R}, H) \prod_{R \in \mathcal{R}} \hom(F[R], K)
    \end{equation}
    of \cite[Theorem~16]{seppelt_logical_2024} for the lexicographic product $H \cdot K$ of simple graphs $H$ and $K$, 
    which is used in \cite{dps_itcs} for $K$ a clique whose size is the chromatic number of $F$.
    Here, $H \cdot K$ has vertex set $V(H) \times V(K)$ and $(u, i)$ and $(v, j)$ are adjacent if $uv \in E(H)$, or $u = v$ and $ij \in E(K)$.
    The proofs of both identities are analogous.
    For bipartite $F$ and $K = K_2$, 
    the factors $\hom(F[R], K_2)$ in \cref{eq:lexprod} are equal to $2$, 
    so the two right-hand sides differ only in the positive weights $2^{|\mathcal{R}|}$.
    The left-hand sides differ as follows:
    The graph $B(H)$ is the spanning subgraph of $H \cdot K_2$ comprising the edges between $V(H) \times \{0\}$ and $V(H) \times \{1\}$.
    While a homomorphism $F \to H \cdot K_2$ may choose for every part $R \in \mathcal{R}$ which side of the bipartition of $F[R]$ is mapped to $V(H) \times \{0\}$, 
    this choice is dictated by the fixed bipartition of $F$ for the homomorphisms counted by $\hom_{\mathrm{bip}}(F, B(H))$.
    The advantage of \cref{lem:bip-contract} lies in the size of the host graphs:
    in order to evaluate a homomorphism polynomial at $H \cdot K_2$, one has to pass to its bipartite double cover, which has $2|V(H)|$ vertices on either side,
    whereas $B(H)$ has only $|V(H)|$ vertices on either side.
    This saves a factor of~$2$ in \cref{thm:non-uniform-non-uniform}.

    \begin{lemma}\label{lem:minor-lifting}
        Let $\mathcal{C}$ be a minor-closed graph class.
        Let $G$ be a connected graph such that no graph in $\mathcal{C}$ admits a weak oddomorphism to $G$.
        Let $B_i \coloneqq B(\widehat{G}_i)$ for $i \in \{0,1\}$.
        Then
        \begin{enumerate}
            \item $B_0$ and $B_1$ are simple bipartite graphs with $\gamma(G)$ vertices on either side,
            \item $\hom_{\mathrm{bip}}(F', B_0) = \hom_{\mathrm{bip}}(F', B_1)$ for all bipartite $F' \in \mathcal{C}$ with any fixed bipartition, and
            \item $\hom_{\mathrm{bip}}(F, B_0) > \hom_{\mathrm{bip}}(F, B_1)$ for every bipartite graph $F$ with fixed bipartition that contains $G$ as a minor.
        \end{enumerate}
    \end{lemma}
    \begin{proof}
        The first assertion is immediate.
        By \cref{lem:bip-contract},
        for every bipartite graph $F'$ with fixed bipartition and $i \in \{0,1\}$,
        \begin{equation}\label{eq:minor-lifting}
            \hom_{\mathrm{bip}}(F', B_i) = \sum_{\mathcal{R} \in \Gamma(F')} \hom(F'/\mathcal{R}, \widehat{G}_i).
        \end{equation}
        If $F' \in \mathcal{C}$, then all subgraphs of all graphs $F'/\mathcal{R}$ are minors of $F'$ and hence in $\mathcal{C}$.
        By assumption, none of them admits a weak oddomorphism to $G$.
        By \cref{lem:filled-cfi}, 
        $\hom(F'/\mathcal{R}, \widehat{G}_0) = \hom(F'/\mathcal{R}, \widehat{G}_1)$ for all $\mathcal{R}$.
        This yields the second assertion.

        For the third assertion, 
        first observe that, by \cref{eq:minor-lifting,lem:filled-cfi}, 
        $\hom_{\mathrm{bip}}(F', B_0) \geq \hom_{\mathrm{bip}}(F', B_1) \geq 1$ for every bipartite graph $F'$ with fixed bipartition.
        The second inequality holds since, for every vertex $x$ of $\widehat{G}_1$, mapping the left side of $F'$ to $(x, 0)$ and the right side to $(x, 1)$ is a homomorphism.
        Let $F$ be a bipartite graph with fixed bipartition containing $G$ as a minor.
        Since $G$ is connected, it is a minor of a connected component $C$ of $F$.
        Since $\hom_{\mathrm{bip}}(F, B_i)$ is the product of the numbers $\hom_{\mathrm{bip}}(C', B_i)$ over all connected components $C'$ of $F$,
        it suffices to show that $\hom_{\mathrm{bip}}(C, B_0) > \hom_{\mathrm{bip}}(C, B_1)$.
        To that end, it suffices to argue that $\hom(C/\mathcal{R}, \widehat{G}_0) > \hom(C/\mathcal{R}, \widehat{G}_1)$ for some $\mathcal{R} \in \Gamma(C)$.

        Since $G$ is a minor of $C$, there exist pairwise disjoint sets $X_v \subseteq V(C)$ for $v \in V(G)$ such that each $C[X_v]$ is connected and there is an edge of $C$ between $X_u$ and $X_v$ for all $uv \in E(G)$.
        Since $C$ is connected, we may suppose that the $X_v$ partition $V(C)$ by repeatedly adding an uncovered vertex to a set $X_v$ containing one of its neighbours.
        Let $\mathcal{R} \coloneqq \{X_v \mid v \in V(G)\} \in \Gamma(C)$.
        Identifying $X_v$ with $v$, the graph $C/\mathcal{R}$ has vertex set $V(G)$ and its edge set contains $E(G)$.
        By \cref{lem:filled-cfi}, $\hom(C/\mathcal{R}, \widehat{G}_0) > \hom(C/\mathcal{R}, \widehat{G}_1)$.
    \end{proof}

    In contrast to \cite[Theorem~6.1]{dps_itcs}, the graph $C/\mathcal{R}$ in the last step of the proof is an induced minor of $F$ which contains $G$ as a spanning subgraph but may have many more edges.
    These edges are irrelevant for $\widehat{G}_i$ by \cref{lem:filled-cfi}.
    Furthermore, 
    the matching edges $(v,0)(v,1)$ of $B(\widehat{G}_i)$ replace the lexicographic products with cliques and the additional universal vertices of \cite{dps_itcs}, whose purpose is to account for contractions and deletions.
    This is possible since the patterns and the hosts are bipartite graphs with fixed bipartitions:
    a connected subgraph of the pattern may be mapped onto a single matching edge.

    \subsubsection{Proof of \cref{thm:non-uniform-non-uniform}}

    \begin{proof}[Proof of \cref{thm:non-uniform-non-uniform}]
        Since $K_1 \in \mathcal{F}$, the graph $G$ has at least two vertices.
        Since $\mathcal{F}$ is closed under weak oddomorphisms and $G \not\in \mathcal{F}$, no graph in $\mathcal{F}$ admits a weak oddomorphism to $G$.
        Let $F^\circ$ denote the underlying simple graph of $F$ with the bipartition of $F$.
        Write $F^\circ = F' + \ell K_1$ where $F'$ does not have any isolated vertices and $\ell \geq 0$.
        Since $G$ is connected and has at least two vertices, $G$ is a minor of $F'$.

        Let $B_0$ and $B_1$ be the bipartite graphs from \cref{lem:minor-lifting} for $\mathcal{C} \coloneqq \mathcal{F}$ and $G$.
        Let $m \coloneqq N - \gamma(G)$ and 
        let $B'_i$ be obtained from $B_i$ by adding $m$ isolated vertices on either side, for $i \in \{0,1\}$.
        We may suppose that $B'_0$ and $B'_1$ have left side $[N] \times \{0\}$ and right side $[N] \times \{1\}$.

        \begin{claim}\label{cl:nunu1}
            $\hom_{\mathrm{bip}}(F'', B'_0) = \hom_{\mathrm{bip}}(F'', B'_1)$ for all bipartite $F'' \in \mathcal{F}$ with any fixed bipartition.
        \end{claim}
        \begin{claimproof}
            Write $C_1, \dots, C_r$ for the connected components of $F''$.
            Since $\mathcal{F}$ is minor-closed, $C_1, \dots, C_r \in \mathcal{F}$.
            Let $e_j \coloneqq m$ if $C_j$ is a single vertex and $e_j \coloneqq 0$ otherwise.
            This is the number of bipartition-respecting homomorphisms from $C_j$ to the $2m$ additional vertices of $B'_i$.
            By \cref{lem:minor-lifting},
            \[
                \hom_{\mathrm{bip}}(F'', B'_0) 
                = \prod_{j=1}^r \left( \hom_{\mathrm{bip}}(C_j, B_0) + e_j \right)
                = \prod_{j=1}^r \left( \hom_{\mathrm{bip}}(C_j, B_1) + e_j \right)
                = \hom_{\mathrm{bip}}(F'', B'_1). \qedhere
            \]
        \end{claimproof}

        \begin{claim}\label{cl:nunu2}
            $\hom_{\mathrm{bip}}(F^\circ, B'_0) \neq \hom_{\mathrm{bip}}(F^\circ, B'_1)$.
        \end{claim}
        \begin{claimproof}
            Since $F'$ does not have isolated vertices, no homomorphism $F' \to B'_i$ has an isolated vertex of $B'_i$ in its image.
            Hence, $\hom_{\mathrm{bip}}(F^\circ, B'_i) = N^\ell \hom_{\mathrm{bip}}(F', B'_i) = N^\ell \hom_{\mathrm{bip}}(F', B_i)$.
            Since $G$ is a minor of $F'$, 
            the claim follows from \cref{lem:minor-lifting}.
        \end{claimproof}

        Let $X_i \in \{0,1\}^{N \times N}$ denote the bi-adjacency matrix of $B'_i$ for $i \in \{0,1\}$.
        Since the entries of $X_i$ are $0$ or $1$, evaluating the homomorphism polynomial of a bipartite multigraph at $X_i$ yields the number of bipartition-respecting homomorphisms from its underlying simple graph to $B'_i$.
        By \cref{cl:nunu1}, the linear map $p \mapsto p(X_0) - p(X_1)$ vanishes on all generators of $\mathfrak{H}_N(\mathcal{F})$ and hence on $\mathfrak{H}_N(\mathcal{F})$.
        By \cref{cl:nunu2} and since $\mathbb{K}$ has characteristic zero, it does not vanish on $\hom_{F, N}$.
        Hence, $\hom_{F,N} \not\in \mathfrak{H}_N(\mathcal{F})$.
    \end{proof}

    \subsection{Separation}

We now put the pieces together and prove \cref{thm:depth-hierarchy}.
To that end, we first deduce the following corollary of \cref{cor:non-rich-class,thm:non-uniform-non-uniform}.

    \begin{corollary}\label{thm:single-hom}
        Let $\Delta \geq 1$ and $k \geq 2$.
        Let $G$ be a connected $(\Delta, k)$-rich graph and $F$ be a bipartite multigraph containing $G$ as a minor.
        Let $N \geq \max\{\gamma(G), 4(k+1), 20\}$.
        Then every matrix-symmetric circuit of support depth $\leq \Delta$ computing $\hom_{F, N}$ has orbit size at least $\binom{N}{k+1}$.
    \end{corollary}
    \begin{proof}
        Towards a contradiction, let $C$ be such a circuit of orbit size less than $\binom{N}{k+1}$.
        Let $C'$ be the reduced circuit constructed from $C$ in \cref{lem:reduction}.
        Since $N \geq 4(k+1) \geq 12$, 
        the orbit size of $C'$ is at most $f \coloneqq \binom{N}{k+1} - 1 < \binom{N}{k+2}/2 \leq \binom{N}{\lfloor N/2 \rfloor}/2$ and $\kappa_f(N) \leq k+1 \leq N/4$.
        By \cref{thm:support}, $\maxsup(C') \leq k < N/2$.
        Hence, $C'$ has support depth $\leq \Delta$ by definition.
        By \cref{thm:bounded-depth-characterisation} for $C'$, the polynomial
        $\hom_{F,N}$ is a linear combination of homomorphism polynomials $\hom_{F', N}$ for patterns $F'$ with $\fw_\Delta(F') \leq k-1$.
        Let $\mathcal{F}$ denote the class of all simple graphs that are not $(\Delta, k)$-rich.
        By \cref{lem:rich-fw}, $\hom_{F, N} \in \mathfrak{H}_N(\mathcal{F})$.
        Since $G \not\in \mathcal{F}$ is a connected minor of $F$ and $N \geq \gamma(G)$, 
        this contradicts \cref{thm:non-uniform-non-uniform}, which is applicable by \cref{cor:non-rich-class}.
    \end{proof}

    \restatethm{thm:depth-hierarchy}{\thmDepthHierarchy}

    \label{proof:depth-hierarchy}
    \begin{proof}
        For the second assertion, 
        recall that $\fw_{\Delta+1}(T_{\Delta, n}) = 1$ by \cref{example:bccd17}.
        By \cref{thm:converse} for $\Delta+1$ and $k \coloneqq 2$,
        the polynomial $\hom_{T_{\Delta,n},n}$ admits a matrix-symmetric circuit of support depth $\leq \Delta+1$, orbit size $\leq n^2$, depth $\leq 2(\Delta+1)+1 = 2\Delta+3$, and size $\leq 5n^2(|V(T_{\Delta,n})| + |E(T_{\Delta,n})|) + 1$.
        Since $n \geq 20$, we have $|V(T_{\Delta,n})| = \sum_{j=0}^{\Delta} n^j \leq \frac{n}{n-1} n^\Delta \leq \frac{20}{19} n^\Delta$ and $|E(T_{\Delta,n})| = |V(T_{\Delta,n})| - 1$.
        Hence, the size is at most $5n^2 \cdot \frac{40}{19} n^\Delta + 1 \leq 13 n^{\Delta+2}$.

        For the first assertion,
        let $\Delta \geq 1$ and $n \geq 20$.
        Write $L \coloneqq \log_2 n$ and $\ell \coloneqq \log_2 L$, and let $k \coloneqq \lfloor L - (\Delta-1) \ell - 3 \rfloor$.
        If $k < 2$, then $L - \Delta\ell - 3 < 2 - \ell \leq 0$ since $\ell \geq 2$, and the asserted bound is trivial.
        Hence, suppose that $k \geq 2$.
        Then $n \geq 32$ and $k + 1 \leq L \leq n/4$.

        Let $G \coloneqq T_{\Delta, k+1}$, which is a subgraph of $T_{\Delta, n}$.
        By \cref{ex:rich-tree}, $G$ is $(\Delta, k)$-rich.
        The root of $G$ has degree $k+1$,
        its $\sum_{j=1}^{\Delta-1} (k+1)^j \leq 2(k+1)^{\Delta-1}$ other inner vertices have degree $k+2$,
        and its $(k+1)^{\Delta}$ leaves have degree $1$.
        Hence, using $k + 1 \leq 2^k$,
        \[
            \gamma(G) 
            \leq 2^k + 2^{k+2} (k+1)^{\Delta-1} + (k+1)^{\Delta}
            \leq 2^{k+3} (k+1)^{\Delta-1}
            \leq 2^{L - (\Delta-1)\ell} L^{\Delta-1}
            = n.
        \]

        Let $C$ be a matrix-symmetric circuit of support depth $\leq \Delta$ computing $\hom_{T_{\Delta,n},n}$.
        Since $G$ is a connected $(\Delta, k)$-rich subgraph of $T_{\Delta, n}$ and $n \geq \max\{\gamma(G), 4(k+1)\}$,
        \cref{thm:single-hom} for $\Delta$ and $k$ applies.
        Hence, the orbit size of $C$ is at least $\binom{n}{k+1} \geq \left(\frac{n}{k+1}\right)^{k+1}$.
        Since $L - (\Delta-1)\ell - 3 < k + 1 \leq L$,
        \[
            \log_2 \binom{n}{k+1} 
            \geq (k+1)(L - \ell)
            \geq \left(L - (\Delta-1)\ell - 3\right)(L - \ell)
            \geq L \left(L - \Delta\ell - 3\right).
        \]
        Thus, the orbit size of $C$ is at least $n^{L - \Delta\ell - 3}$.
    \end{proof}

\section{Material omitted in \Cref{sec:depth-reduction}}

\subsection{Omitted material on reduction to logarithmic depth}
\label{app:vsbr}

\restatethm{cor:vsbr}{\thmVSBR}
\begin{proof}
    Let $n$ be sufficiently large, and write $p \coloneqq p_n$, $d \coloneqq d(n)$, $s \coloneqq \max\{s(n), n^2\}$, $\kappa \coloneqq \kappa_s(n)$, and $r \coloneqq \lceil n^{1-\varepsilon} \rceil$.
    Since $s \geq n^2 > \binom{n}{2}$, we have $\kappa \geq 3$.
    Since $\binom{n}{r} \geq (n/r)^r \geq (n^\varepsilon/2)^{n^{1-\varepsilon}} > 2^{n^{1-\varepsilon}} \geq s$, we have $\kappa \leq r$, i.e.\ $\kappa - 1 \leq n^{1-\varepsilon}$.
    Hence $n \geq 20$ and $n \geq 6\kappa - 5$.
    By minimality of $\kappa$, $(n/(\kappa-1))^{\kappa-1} \leq \binom{n}{\kappa-1} \leq s$, and $n/(\kappa-1) \geq n^{\varepsilon}$ yields $n^{\varepsilon(\kappa-1)} \leq s$, i.e.\ $n^{3\kappa-3} \leq s^{3/\varepsilon}$.

    Let $C$ be a matrix-symmetric circuit of orbit size $\leq s$ computing $p$.
    By \cref{lem:reduction}, we may suppose that $C$ is reduced.
    Let $\Delta \geq 1$ bound its depth, so that $C$ has support depth $\leq \Delta$.
    As $3\kappa \leq n+2$, \cref{thm:support} shows that $C$ has support size $\leq \kappa - 1$.
    By \cref{thm:bounded-depth-characterisation} with $k \coloneqq \kappa - 1 \geq 2$, we have $p = \sum_F \alpha_F \hom_{F,n}$ for patterns $F$ with $\fw_\Delta(F) \leq \kappa - 2$.
    In particular, $\tw(F) \leq \kappa - 2$.
    Since $\hom_{F,n}$ is homogeneous of degree $|E(F)|$, comparing homogeneous components shows that we may keep only the patterns with $|E(F)| \leq d$.
    Patterns without edges contribute constants.
    For the others, we may delete isolated vertices by \cref{obs:no-isolated-vertices}, so that $|V(F)| \leq 2d$.

    By \cref{thm:bodlander}, every such $F$ has a rooted tree decomposition with bags of size $\leq 3(\kappa - 1) = 3\kappa - 3$ and depth $\leq q \coloneqq 2\lceil \log_{5/4}(4d) \rceil$.
    Adding a root with an empty bag shows $\fw_{q+1}(F) \leq 3\kappa - 4$.
    By \cref{thm:converse} with $k \coloneqq 3\kappa - 3$, using $n \geq 6\kappa - 5 = 2k+1$,
    $\hom_{F,n}$ is computed by a reduced matrix-symmetric circuit of depth $\leq 2q + 3$ and orbit size $\leq n^{3\kappa-3} \leq s^{3/\varepsilon}$.
    Taking the linear combination as in \cref{app:converse} increases the depth by one, so the depth is $\leq 2q + 4 \in O(\log d)$.
\end{proof}

Let $P_m$ denote the $m$-vertex path.

\begin{lemma}\label{lem:path-lower}
    Let $\Delta, k \geq 2$, $d \geq 1$, and $N \geq \max\{4(k+1), 20\}$ be such that $(k+1)^\Delta \leq \min\{d, N/2\}$.
    Then every matrix-symmetric circuit of support depth $\leq \Delta$ computing $\hom_{P_d, N}$ has orbit size at least $\binom{N}{k+1}$.
\end{lemma}
\begin{proof}
    Let $G \coloneqq P_{(k+1)^\Delta}$.
    By \cref{ex:path-rich}, $G$ is $(\Delta, k)$-rich.
    Since $G$ has maximum degree~$2$, $\gamma(G) \leq 2(k+1)^\Delta \leq N$.
    The claim follows from \cref{thm:single-hom}.
\end{proof}

\restatethm{cor:log-depth-optimal}{\thmDepthLogOptimal}
\begin{proof}
    Write $d \coloneqq d(n)$.
    For the upper bound, decompose $P_d$ recursively: a node for a subpath with endpoints $a$ and $b$ and midpoint $m$ has the bag $\{a, m, b\}$ and children for the subpaths from $a$ to $m$ and from $m$ to $b$, and subpaths with one edge are leaves.
    Together with a root with an empty bag, this shows $\fw_{\lceil \log_2 d \rceil + 1}(P_d) \leq 2$.
    By \cref{thm:converse} with $k \coloneqq 3$, the polynomial $\hom_{P_d,n}$ is computed by a matrix-symmetric circuit of orbit size $\leq n^3$ and depth $\leq 2\lceil \log_2 d \rceil + 3$ for $n \geq 7$.

    For the lower bound, we may suppose that $c \geq 2$ is an integer.
    Let $n$ be sufficiently large, so that $n^c < \binom{n}{c+1}$ and $n \geq \max\{4(c+1), 20\}$, and let $\Delta \coloneqq \lfloor \log_{c+1} \min\{d, n/2\} \rfloor$.
    If $\Delta \geq 2$, then \cref{lem:path-lower} for $k \coloneqq c$ shows that the support depth is $> \Delta \geq \log_{c+1} \min\{d, n/2\} - 1$, since $(c+1)^\Delta \leq \min\{d, n/2\}$.
    If $\Delta \leq 1$, then $\log_{c+1} \min\{d, n/2\} - 1 < 1$, and the support depth is $\geq 1$ since $\hom_{P_d,n}$ is not constant.
    Finally, if $d > n/2$, then $\log_{c+1}(n/2) \geq \log_{c+1} n - 1 \geq \frac1a \log_{c+1} d - 1$ as $d \leq n^a$.
\end{proof}

\subsection{Material omitted on bounded pathwidth and logarithmic depth}
\label{app:vbp-vac1}

Let $c \geq 0$ and let $\Sigma \coloneqq \{\ell_1, \dots, \ell_{c+1}, r_1, \dots, r_{c+1}\}$ be a set of \emph{slots}, where the $\ell_i$ are \emph{left} and the $r_j$ are \emph{right}.
A \emph{bilabelled graph} $\boldsymbol{F} = (F, \alpha, \beta)$ consists of a bipartite multigraph $F$ and maps $\alpha, \beta \colon \Sigma \to V(F)$, the \emph{in-} and \emph{out-labels}, sending left slots to left and right slots to right vertices.
Let $L_n \coloneqq [n]^\Sigma$ and $N \coloneqq |L_n| = n^{2(c+1)}$.
The \emph{homomorphism matrix} $\boldsymbol{F}_n \in \mathbb{K}[\Xx_n]^{L_n \times L_n}$ is
\[
    (\boldsymbol{F}_n)_{a,b} \coloneqq \sum_{\substack{h \colon V(F) \to [n] \uplus [n] \\ h \circ \alpha = a,\ h \circ \beta = b}} \prod_{xy \in E(F)} x_{h(x), h(y)},
\]
where $h$ maps left vertices to the left and right vertices to the right copy of $[n]$.
That is, $\boldsymbol{F}_n$ is the map of the $(2(c+1), 2(c+1))$-labelled graph with labels $\alpha$ and $\beta$, cf.\ \cref{sec:labelled-graphs}, arranged as a matrix.
Clearly, $\boldsymbol{1}^T \boldsymbol{F}_n \boldsymbol{1} = \hom_{F,n}$ and $\tr(\boldsymbol{F}_n) = \hom_{F^\circ, n}$, where $F^\circ$ is obtained from $F$ by identifying $\alpha(s)$ with $\beta(s)$ for all $s \in \Sigma$.

The \emph{series composition} $\boldsymbol{F} \cdot \boldsymbol{F}'$ of $\boldsymbol{F} = (F, \alpha, \beta)$ and $\boldsymbol{F}' = (F', \alpha', \beta')$ is obtained from the disjoint union of $F$ and $F'$ by identifying $\beta(s)$ with $\alpha'(s)$ for all $s \in \Sigma$, with in-labels $\alpha$ and out-labels $\beta'$.
Then $(\boldsymbol{F} \cdot \boldsymbol{F}')_n = \boldsymbol{F}_n \boldsymbol{F}'_n$, since homomorphisms from $F \cdot F'$ correspond to pairs of homomorphisms from $F$ and $F'$ that agree on the identified vertices, and summing over their common value $b = h \circ \beta = h' \circ \alpha'$ is matrix multiplication.

Let $\boldsymbol{I}$ be the edgeless bilabelled graph with vertex set $\Sigma$ and $\alpha = \beta = \mathrm{id}$.
The \emph{atomic} graphs are the following $m \coloneqq (c+1)^2 + 2(c+1)$ bilabelled graphs.
For a left slot $s$ and a right slot $t$, let $\boldsymbol{E}_{s,t}$ be $\boldsymbol{I}$ with an edge $st$ added.
For a slot $s$, let $\boldsymbol{R}_s$ be $\boldsymbol{I}$ with a new vertex $s'$ on the side of $s$ added and $\beta(s) \coloneqq s'$.
Then $(\boldsymbol{E}_{s,t})_n$ is the diagonal matrix with entries $x_{a(s), a(t)}$, and $(\boldsymbol{R}_s)_{n}$ is the $0/1$-matrix whose $(a, b)$-entry is $1$ if and only if $a$ and $b$ agree outside $s$.
For a word $u = u_1 \cdots u_\ell$ over the atomic graphs, let $\boldsymbol{F}_u = (F_u, \alpha_u, \beta_u) \coloneqq \boldsymbol{I} \cdot u_1 \cdots u_\ell$ and $F^\circ_u \coloneqq (F_u)^\circ$.

\begin{lemma}\label{lem:words-patterns}
    Let $u$ be a word of length $\ell$ over the atomic graphs.
    \begin{enumerate}
        \item $(\boldsymbol{F}_u)_n = (u_1)_n \cdots (u_\ell)_n$, $\boldsymbol{1}^T (\boldsymbol{F}_u)_n \boldsymbol{1} = \hom_{F_u,n}$, and $\tr((\boldsymbol{F}_u)_n) = \hom_{F^\circ_u,n}$.
        \item $F_u$ and $F^\circ_u$ have at most $2(c+1) + \ell$ vertices, $\pw(F_u) \leq 2c+1$, and $\pw(F^\circ_u) \leq 4c+3$.
        \item For every bipartite multigraph $F$ with $\pw(F) \leq c$, there is a word $u$ such that $F_u$ is the disjoint union of $F$ and $2(c+1)$ isolated vertices.
    \end{enumerate}
\end{lemma}
\begin{proof}
    1.~This follows from the above, as $\boldsymbol{I}_n$ is the identity matrix.

    2.~Every atomic graph $\boldsymbol{R}_s$ adds one vertex, and $\boldsymbol{E}_{s,t}$ adds none.
    For $i = 0, \dots, \ell$, let $X_i$ be the image of the out-labels of $\boldsymbol{F}_{u_1 \cdots u_i}$, viewed in $F_u$.
    A vertex that loses its out-label never regains it, and the edge added by $u_i = \boldsymbol{E}_{s,t}$ has both endpoints in $X_i$.
    Hence, the bags $X_0, \dots, X_\ell$ of size $\leq 2(c+1)$ form a path decomposition of $F_u$.
    Since $X_0$ is the image of $\alpha_u$ and every vertex of $F^\circ_u$ obtained by identification contains a vertex of $X_0$, adding $X_0$ to every bag yields a path decomposition of $F^\circ_u$.

    3.~Let $B_1, \dots, B_p$ be the bags of a path decomposition of $F$ of width $\leq c$.
    The vertices on either side of $F$ correspond to intervals of $[p]$, at most $c+1$ of which overlap in any point.
    Colour them greedily with the $c+1$ slots of their side, so that vertices sharing a bag receive distinct slots.
    Assign every edge of $F$ to a bag containing it.
    Let $u$ be the concatenation, for $i = 1, \dots, p$, of the atomic graphs $\boldsymbol{E}_{s,t}$ for the edges assigned to $B_i$, where $s$ and $t$ are the slots of their endpoints, followed by the atomic graphs $\boldsymbol{R}_s$ for the slots $s$ of the vertices in $B_i \setminus B_{i+1}$, where $B_{p+1} \coloneqq \emptyset$.
    Identify the vertices of $F$ with slot $s$, in the order of their intervals, with the successive vertices carrying the out-label $s$: the first with the vertex $s$ of $\boldsymbol{I}$, and each further one with the vertex added by the $\boldsymbol{R}_s$ after the last bag of its predecessor.
    As intervals with the same slot are disjoint, the out-label $s$ sits on the vertex of $B_i$ with slot $s$ while the atomic graphs for $B_i$ are composed, so every edge of $F$ is added between its endpoints.
    The vertex carrying the out-label $s$ at the end is isolated, which gives the $2(c+1)$ isolated vertices.
\end{proof}

\begin{lemma}\label{thm:pathwidth-log-depth}
    Let $c \geq 0$, $n \geq 2$, $k \coloneqq 12(c+1)$, and $\Delta \coloneqq 2\lceil (4c+6) \log_{5/4} n \rceil + 1$.
    For every bipartite multigraph $F$ with $\pw(F) \leq c$,
    the polynomial $\hom_{F,n}$ is a linear combination of homomorphism polynomials $\hom_{F',n}$ for patterns $F'$ with $\fw_\Delta(F') \leq k-1$.
\end{lemma}
\begin{proof}
    By \cref{lem:words-patterns,obs:no-isolated-vertices}, $\hom_{F,n} = n^{-2(c+1)} \boldsymbol{1}^T (\boldsymbol{F}_w)_n \boldsymbol{1}$ for some word $w$.
    By \cref{lem:words-patterns} and \cref{lem:trace-reduction} applied to the $m$ matrices of the atomic graphs with $N = n^{2(c+1)}$, the matrix $(\boldsymbol{F}_w)_n$ is a linear combination of matrices $(\boldsymbol{F}_u)_n$ with $|u| < N^2$ whose coefficients are polynomials in traces $\tr((\boldsymbol{F}_v)_n)$ with $|v| \leq N^2$.
    By \cref{lem:words-patterns}, $\boldsymbol{1}^T (\boldsymbol{F}_w)_n \boldsymbol{1}$ is thus a linear combination of products $\hom_{F_u,n} \prod_i \hom_{F^\circ_{v_i},n} = \hom_{F',n}$, where $F'$ is the disjoint union of $F_u$ and the $F^\circ_{v_i}$.
    Each of these components has pathwidth, and hence treewidth, $\leq 4c+3$ and at most $2(c+1) + N^2$ vertices, where $2(2(c+1) + N^2) \leq 3N^2 \leq n^{4c+6}$.
    By \cref{thm:bodlander}, each component has a tree decomposition with bags of size $\leq 12(c+1) = k$ and depth $\leq \Delta - 1$.
    Adding a root with an empty bag above all of them shows that $\fw_\Delta(F') \leq k-1$.
\end{proof}

\restatethm{cor:symVBP-symVAC1}{\corVBPAC}
\begin{proof}
    By \cref{thm:symv-to-hom,thm:symVBP-symVS}, every family in $\symVBP$ is, for all sufficiently large $n$, a linear combination of homomorphism polynomials for patterns of pathwidth $\leq c$ for some constant~$c$.
    By \cref{thm:pathwidth-log-depth}, each of them is a linear combination of homomorphism polynomials for patterns $F$ with $\fw_\Delta(F) \leq k-1$, where $k = 12(c+1)$ and $\Delta = O(c \log n)$.
    By \cref{thm:bounded-depth-characterisation,obs:support-orbit}, these linear combinations are computed by matrix-symmetric circuits of support depth $\leq \Delta$ and orbit size $\leq n^k$ for all $n \geq 2k+1$.
\end{proof}

\subsection{Separations of \texorpdfstring{$\symVAC^k$}{symVACk} and \texorpdfstring{$\symVP$}{symVP}}\label{app:symvac1-separation}

We prove \cref{thm:symVAC1-separation,thm:symVACk-hierarchy}.
The idea is to evaluate matrix-symmetric polynomials on weighted encodings of relational structures that witness the Weisfeiler--Leman iteration number lower bounds of \textcite{grohe_iteration_2025}.

\subsubsection{Set-up}
Let $\Sigma$ be a finite set of labels.
A \emph{$\Sigma$-labelled bipartite graph} $G$ consists of disjoint sets $A(G)$ and $B(G)$ and relations $E_\sigma(G) \subseteq A(G) \times B(G)$ for $\sigma \in \Sigma$.
We view it as a relational structure with universe $A(G) \cup B(G)$, unary relations $A$ and $B$, and binary relations $E_\sigma$.
A homomorphism $F \to G$ maps $A(F)$ to $A(G)$ and $B(F)$ to $B(G)$ and preserves every $E_\sigma$; we write $\hom(F, G)$ for their number.
Write $\mathsf{C}_d^{r}$ for the sentences of first-order logic with counting quantifiers with at most $d$ variables and quantifier rank at most $r$, and $\mathsf{C}_d \coloneqq \bigcup_r \mathsf{C}_d^{r}$.

For $K, q \geq 0$, let $\mathcal{T}^K_q$ denote the class of graphs $F$ that admit a rooted tree decomposition $(T, \beta)$ such that $|\beta(t)| \leq K$ for all $t \in V(T)$ and the bags along every root-leaf path of $T$ contain at most $q$ vertices in total \cite{adler_going_2026}.
We first translate between this notion and \cref{def:forestwidth}.

\begin{lemma}\label{lem:tkq-fw}
    Let $K, \Delta, q \geq 1$, and let $F$ be a graph.
    \begin{enumerate}
        \item If $\fw_\Delta(F) \leq K - 1$, then $F \in \mathcal{T}^K_{\Delta K}$.
        \item If $F \in \mathcal{T}^K_q$, then $\fw_q(F) \leq K - 1$.
    \end{enumerate}
\end{lemma}
\begin{proof}
    1.~A decomposition witnessing $\fw_\Delta(F) \leq K-1$ has an empty root bag and depth $\leq \Delta$, so every root-leaf path meets at most $\Delta$ non-empty bags of size $\leq K$.

    2.~Let $(T, \beta)$ be as in the definition of $\mathcal{T}^K_q$.
    As long as some node $t$ has a child $c$ with $\beta(c) \subseteq \beta(t)$, contract the edge $tc$ into $t$, keeping the bag $\beta(t)$; this preserves the properties of $(T, \beta)$.
    If the root bag is non-empty, add a new root with an empty bag.
    Now every node $t$ other than the root has a vertex $v_t \in \beta(t)$ that is not in the bag of its parent.
    Since the nodes whose bags contain $v_t$ form a subtree of $T$, the vertices $v_t$ for the non-root nodes $t$ on a root-leaf path are pairwise distinct.
    Hence, every root-leaf path has at most $q$ non-root nodes, i.e.\ the decomposition has depth $\leq q$.
\end{proof}

We use the following known correspondence.

\begin{theorem}[\cite{dawar_lovasz-type_2021,adler_going_2026}]\label{thm:hom-logic}
    Let $G$ and $H$ be $\Sigma$-labelled bipartite graphs and $K, q \geq 1$.
    Then $G$ and $H$ are $\mathsf{C}_K^{q}$-equivalent if and only if $\hom(F, G) = \hom(F, H)$ for all $\Sigma$-labelled bipartite graphs $F \in \mathcal{T}^K_q$.
\end{theorem}

\subsubsection{Translating structures into weighted graphs}

We translate the relational structures of \cite{grohe_iteration_2025} to weighted graphs on which we evaluate the homomorphism matrices.
First, we handle edge colours via interpolation.
Then, we reduce from general relational structures to edge-coloured graphs.

\paragraph{Edge colours via interpolation.}

Let $G$ be a $\Sigma$-labelled bipartite graph such that every pair $(a, b) \in A(G) \times B(G)$ lies in at most one relation $E_\sigma(G)$, and let $N \geq |A(G)|, |B(G)|$.
For $\omega \in \mathbb{K}^\Sigma$, let $M^\omega_G \in \mathbb{K}^{N \times N}$ be the matrix whose rows and columns are indexed by $A(G)$ and $B(G)$, padded by zero rows and columns, with entry $\omega_\sigma$ at $(a, b) \in E_\sigma(G)$ and $0$ at all other positions.
Matrix-symmetric polynomials do not depend on the choice of the indexing.
Let $F$ be a bipartite multigraph with bipartition $A \uplus B$ and without isolated vertices.
For a map $\lambda \colon E(F) \to \Sigma$, let $F^\lambda$ be the $\Sigma$-labelled bipartite graph with $A(F^\lambda) = A$, $B(F^\lambda) = B$, and $(a, b) \in E_\sigma(F^\lambda)$ if some edge $e = ab$ has $\lambda(e) = \sigma$.
Expanding the definition yields
\begin{equation}\label{eq:weighted-hom}
    \hom_{F,N}(M^\omega_G) = \sum_{\lambda \colon E(F) \to \Sigma} \Bigl(\prod_{e \in E(F)} \omega_{\lambda(e)}\Bigr) \hom(F^\lambda, G).
\end{equation}
Here, we use that a homomorphism $F^\lambda \to G$ maps all parallel edges of $F$ to the same pair, which lies in at most one relation.

\begin{lemma}\label{lem:multiplicity}
    Let $G$ and $H$ be $\Sigma$-labelled bipartite graphs as above with $|A(G)| = |A(H)|$ and $|B(G)| = |B(H)|$, and let $N \geq |A(G)|, |B(G)|$.
    Let $L$ be a $\Sigma$-labelled bipartite graph with $\hom(L, G) \neq \hom(L, H)$.
    Then there are a bipartite multigraph $F$ whose underlying simple graph is a subgraph of the Gaifman graph of $L$, and $\omega \in \mathbb{K}^\Sigma$ such that $\hom_{F,N}(M^\omega_G) \neq \hom_{F,N}(M^\omega_H)$.
\end{lemma}
\begin{proof}
    Deleting an isolated vertex of $L$ divides both $\hom(L, G)$ and $\hom(L, H)$ by the same number $|A(G)|$ or $|B(G)|$, which is non-zero since $\hom(L, G) \neq \hom(L, H)$, so we may suppose that $L$ has no isolated vertices.
    Let $F_0$ be the bipartite multigraph with an edge $e = ab$ labelled $\lambda(e) \coloneqq \sigma$ for every $\sigma \in \Sigma$ and $(a,b) \in E_\sigma(L)$, so that $L = F_0^\lambda$.
    Let $e_1, \dots, e_s$ be the edges of $F_0$, and let $F$ be obtained by replacing $e_i$ by $2^{i-1}$ parallel edges.
    By \eqref{eq:weighted-hom}, $\hom_{F,N}(M^\omega_{D})$ is a polynomial in $\omega$ for $D \in \{G, H\}$.
    Labellings of $F$ that are not constant on the parallel copies of some $e_i$ contribute $0$, since every pair lies in at most one relation of $D$.
    The remaining labellings correspond to labellings $\mu$ of $F_0$, contribute $\prod_i \omega_{\mu(e_i)}^{2^{i-1}} \hom(F_0^\mu, D)$, and are determined by their monomial via the binary expansions of the exponents.
    So the coefficient of $\prod_i \omega_{\lambda(e_i)}^{2^{i-1}}$ is $\hom(L, D)$.
    Hence, $\hom_{F,N}(M^\omega_G) - \hom_{F,N}(M^\omega_H)$ is a non-zero polynomial in $\omega$, and some $\omega \in \mathbb{K}^\Sigma$ does not annihilate it.
\end{proof}

\begin{lemma}\label{lem:hom-transfer}
    Let $G$, $H$, and $N$ be as in \cref{lem:multiplicity}, and let $K, q \geq 1$.
    If $G$ and $H$ are $\mathsf{C}_K^{q}$-equivalent, then $p(M^\omega_G) = p(M^\omega_H)$ for all $\omega \in \mathbb{K}^\Sigma$ and all linear combinations $p$ of $\hom_{F,N}$ for bipartite multigraphs $F \in \mathcal{T}^K_q$.
\end{lemma}
\begin{proof}
    Isolated vertices of $F$ contribute the same factor $N$ on both sides by \cref{obs:no-isolated-vertices}, and deleting them preserves membership in $\mathcal{T}^K_q$.
    For $F$ without isolated vertices and $\mu \colon E(F) \to \Sigma$, the Gaifman graph of $F^\mu$ is the underlying simple graph of $F$, so $F^\mu \in \mathcal{T}^K_q$ and $\hom(F^\mu, G) = \hom(F^\mu, H)$ by \cref{thm:hom-logic}.
    The claim follows by \eqref{eq:weighted-hom}.
\end{proof}

\paragraph{Incidence graphs.}
Let $\mathfrak{D}$ be a relational structure with universe $U$ over a finite vocabulary $\tau$ of arity $\leq m$, and let $\tau_s \subseteq \tau$ consist of the relation symbols of arity $s$.
The incidence graph $\mathrm{Inc}(\mathfrak{D})$ is the $\Sigma$-labelled bipartite graph with $\Sigma \coloneqq \{(s, J, S) \mid s \in [m],\ \emptyset \neq J \subseteq [s],\ S \subseteq \tau_s\}$,
$A(\mathrm{Inc}(\mathfrak{D})) \coloneqq U$, and $B(\mathrm{Inc}(\mathfrak{D})) \coloneqq \bigcup_{s \in [m]} U^s$.
For $u \in U$ and $\bar t \in U^s$ with $J \coloneqq \{j \in [s] \mid t_j = u\} \neq \emptyset$, the pair $(u, \bar t)$ lies in $E_{(s, J, S)}$ for $S \coloneqq \{R \in \tau_s \mid \bar t \in R^{\mathfrak{D}}\}$, and in no other relation.
Every pair lies in at most one relation.
If $\mathfrak{A}$ and $\mathfrak{B}$ both have $n$ elements, then the sides of $\mathrm{Inc}(\mathfrak{A})$ and $\mathrm{Inc}(\mathfrak{B})$ have the same sizes $n$ and $\sum_{s \in [m]} n^s < (n+1)^m$.

\begin{lemma}\label{lem:inc-logic}
    Let $\mathfrak{A}$ and $\mathfrak{B}$ be structures over a finite vocabulary of arity $\leq m$, and let $K, q \geq 1$.
    \begin{enumerate}
        \item If $\mathrm{Inc}(\mathfrak{A})$ and $\mathrm{Inc}(\mathfrak{B})$ are $\mathsf{C}_{K+1}^{q+1}$-equivalent, then $\mathfrak{A}$ and $\mathfrak{B}$ are $\mathsf{C}_K^{q}$-equivalent.
        \item If $\mathfrak{A}$ and $\mathfrak{B}$ are $\mathsf{C}_{mK}^{mq}$-equivalent, then $\mathrm{Inc}(\mathfrak{A})$ and $\mathrm{Inc}(\mathfrak{B})$ are $\mathsf{C}_K^{q}$-equivalent.
    \end{enumerate}
\end{lemma}
\begin{proof}
    1.~We translate every $\mathsf{C}_K^{q}$-sentence into a $\mathsf{C}_{K+1}^{q+1}$-sentence by relativising every quantifier $\exists^{\geq \nu} x$ to $A(x)$ and replacing every atom $R(z_1, \dots, z_s)$ by $\exists y \bigwedge_{j \in [s]} \bigvee_{(s, J, S) \in \Sigma,\, j \in J,\, R \in S} E_{(s,J,S)}(z_j, y)$, for a fresh variable $y$.
    The result has at most $K+1$ variables and quantifier rank at most $q+1$, since only atoms are replaced, and it holds in $\mathrm{Inc}(\mathfrak{D})$ if and only if the original sentence holds in $\mathfrak{D}$.

    2.~We translate every $\mathsf{C}_K^{q}$-sentence over $\mathrm{Inc}(\mathfrak{D})$ into a $\mathsf{C}_{mK}^{mq}$-sentence over $\mathfrak{D}$ via a $m$-dimensional interpretation.
    Every vertex of $\mathrm{Inc}(\mathfrak{D})$ is represented by a \emph{type} and a $m$-tuple over $U$:
    a vertex $u \in U$ by the type $\mathrm{el}$ and $(u, \dots, u)$,
    and a vertex $\bar t \in U^s$ by the type $(s, S)$, where $S \coloneqq \{R \in \tau_s \mid \bar t \in R^{\mathfrak{D}}\}$, and the $m$-tuple obtained from $\bar t$ by repeating its last entry.
    Distinct vertices of the same type have distinct representing tuples.
    For every type $\pi$, there is a quantifier-free formula $\chi_{\pi}(\bar y)$ over $\tau$ that holds exactly for the representing tuples of type $\pi$, in both structures.
    Translate a formula with at most $K$ variables by assigning to each variable $y$ a block $\bar y$ of $m$ variables and by a case distinction over the finitely many types of its free variables.
    Atoms translate into quantifier-free formulas:
    $A(y)$ and $B(y)$ become true or false depending on the type of $y$;
    $y = y'$ becomes $\bar y = \bar y'$ if $y$ and $y'$ have the same type and false otherwise;
    and $E_{(s,J,S)}(x, y)$ becomes $\bigwedge_{j \in J} x_1 = y_j \wedge \bigwedge_{j \in [s] \setminus J} x_1 \neq y_j$ if $x$ has type $\mathrm{el}$ and $y$ has type $(s, S)$, and false otherwise.
    A quantifier $\exists^{\geq \nu} y$ becomes the disjunction over all $(\nu_{\pi})_{\pi}$ with $\sum_{\pi} \nu_{\pi} = \nu$ of the conjunction over all types $\pi$ of $\exists^{\geq \nu_{\pi}} \bar y\, (\chi_{\pi}(\bar y) \wedge \dots)$, where $\dots$ is the translation for type $\pi$ of $y$.
    Counting quantifiers over $m$-tuples are expressible in counting logic with the $m$ quantified variables and quantifier rank $m$.
    The translation therefore has at most $mK$ variables and quantifier rank at most $mq$, and it holds in $\mathfrak{D}$ if and only if the original sentence holds in $\mathrm{Inc}(\mathfrak{D})$.
\end{proof}

We transfer pairs of structures to pairs of matrices.

\begin{lemma}\label{lem:transfer}
    Let $m, K_0, q_0 \geq 1$ and $n, N \geq 1$ with $N \geq (n+1)^m$.
    Let $\mathfrak{A}$ and $\mathfrak{B}$ be relational structures with $n$ elements over a finite vocabulary of arity $\leq m$ that are not $\mathsf{C}_{K_0}^{q_0}$-equivalent.
    Then there are matrices $M, M' \in \mathbb{K}^{N \times N}$ and a bipartite multigraph $F$ with $\fw_{q_0+1}(F) \leq K_0$ such that
    \begin{enumerate}
        \item $\hom_{F,N}(M) \neq \hom_{F,N}(M')$, and
        \item $\hom_{F',N}(M) = \hom_{F',N}(M')$ for every bipartite multigraph $F'$ with $\fw_\Delta(F') \leq K-1$ for some $K, \Delta \geq 1$ such that $\mathfrak{A}$ and $\mathfrak{B}$ are $\mathsf{C}_{mK}^{mK\Delta}$-equivalent.
    \end{enumerate}
\end{lemma}
\begin{proof}
    By \cref{lem:inc-logic} (contrapositively), $\mathrm{Inc}(\mathfrak{A})$ and $\mathrm{Inc}(\mathfrak{B})$ are not $\mathsf{C}_{K_0+1}^{q_0+1}$-equivalent.
    By \cref{thm:hom-logic}, some $\Sigma$-labelled bipartite graph $L \in \mathcal{T}^{K_0+1}_{q_0+1}$ satisfies $\hom(L, \mathrm{Inc}(\mathfrak{A})) \neq \hom(L, \mathrm{Inc}(\mathfrak{B}))$.
    Let $F$ and $\omega$ be as in \cref{lem:multiplicity} for $L$ and $N$, and let $M \coloneqq M^\omega_{\mathrm{Inc}(\mathfrak{A})}$ and $M' \coloneqq M^\omega_{\mathrm{Inc}(\mathfrak{B})}$.
    Then item~1 holds.
    Since $\mathcal{T}^{K_0+1}_{q_0+1}$ is closed under subgraphs, $F \in \mathcal{T}^{K_0+1}_{q_0+1}$, so $\fw_{q_0+1}(F) \leq K_0$ by \cref{lem:tkq-fw}.

    For item~2, let $F'$ be a bipartite multigraph and $K, \Delta \geq 1$ such that $\fw_\Delta(F') \leq K-1$ and $\mathfrak{A}$ and $\mathfrak{B}$ are $\mathsf{C}_{mK}^{mK\Delta}$-equivalent.
    Then $F' \in \mathcal{T}^K_{\Delta K}$ by \cref{lem:tkq-fw}.
    By \cref{lem:inc-logic}, $\mathrm{Inc}(\mathfrak{A})$ and $\mathrm{Inc}(\mathfrak{B})$ are $\mathsf{C}_K^{K\Delta}$-equivalent.
    By \cref{lem:hom-transfer}, $\hom_{F',N}(M) = \hom_{F',N}(M')$.
\end{proof}

\subsubsection{The separation}
By \cite[Theorem~5.2]{cai_optimal_1992}, two $d$-tuples receive the same colour after $r$ rounds of $d$-WL if, and only if, they satisfy the same $\mathsf{C}_{d+1}^r$-formulas.
Hence, if $d$-WL does not distinguish two structures after $r$ rounds, then they are $\mathsf{C}_{d+1}^r$-equivalent.
If $k$-WL distinguishes two structures, then they are not $\mathsf{C}_{k+1}$-equivalent, since counting the tuples of a given colour requires only $k$ further quantifiers.
Thus, the structures of \cref{thm:gln} are not $\mathsf{C}_{k+1}$-equivalent but $\mathsf{C}_{d+1}^{\lfloor n^{\varepsilon k} \rfloor}$-equivalent.
We show the following stronger version of \cref{thm:symVAC1-separation}.

\begin{theorem}\label{thm:symVAC1-separation-strong}
    There are $\eta, \theta > 0$, an infinite set $\mathcal{N} \subseteq \mathbb{N}$, and a family $(p_N) \in \symVP$ such that the following holds.
    For every sufficiently large $N \in \mathcal{N}$, every matrix-symmetric circuit of orbit size $\leq 2^{N^\eta}$ computing $p_N$ has support depth $\geq N^\theta$.
    In particular, $(p_N) \notin \symVAC$.
\end{theorem}

\begin{proof}
    Let $k \coloneqq k_0$, $\alpha$, and $\varepsilon$ be as in \cref{thm:gln}, and let $\eta \coloneqq \min\{\varepsilon, 1/(8k)\}/3$ and $\theta \coloneqq \varepsilon/3$.
    For $n \geq \alpha k^{14}$, let $d(n) \coloneqq \lfloor (n/(\alpha k^6))^{1/8} \rfloor \geq k$, let $\mathfrak{A}_n$ and $\mathfrak{B}_n$ be as in \cref{thm:gln} for $d(n)$, $k$, and $n$, and let $N(n) \coloneqq (n+1)^k$.
    Let $\mathcal{N} \coloneqq \{N(n) \mid n \geq \alpha k^{14}\}$.
    For $N = N(n) \in \mathcal{N}$, let $M_N$, $M'_N$, and $F_N$ be as in \cref{lem:transfer} for $m \coloneqq k$, $K_0 \coloneqq k+1$, and some $q_0$, and let $p_N \coloneqq \hom_{F_N,N}$.
    For $N \notin \mathcal{N}$, let $p_N \coloneqq 0$.

    \emph{Upper bound.}
    By \cref{lem:transfer}, $\fw_{q_0+1}(F_N) \leq k+1$.
    By \cref{thm:converse}, $p_N$ is computed by matrix-symmetric circuits of orbit size $\leq N^{k+2}$ for $N \geq 2k+5$.
    Hence, $(p_N) \in \symVP$.

    \emph{Lower bound.}
    Let $N = N(n) \in \mathcal{N}$ be sufficiently large, and let $C$ be a matrix-symmetric circuit of orbit size $\leq f \coloneqq 2^{N^\eta}$ and support depth $\leq \Delta$ computing $p_N$, where $\Delta \geq 1$.
    By \cref{lem:reduction}, we may suppose that $C$ is reduced.
    Let $\kappa \coloneqq \kappa_f(N)$.
    As $f < \binom{N}{\lfloor N/2 \rfloor}$, we have $\kappa \leq N/2$, so $2^{\kappa-1} \leq \binom{N}{\kappa-1} \leq f$ by minimality of $\kappa$.
    Hence $\kappa - 1 \leq N^\eta$, and in particular $3\kappa \leq N+2$.
    Let $K \coloneqq \max\{\kappa-1, 2\} \leq N^\eta$.
    By \cref{thm:support,thm:bounded-depth-characterisation}, $p_N$ is a linear combination of $\hom_{F,N}$ for bipartite multigraphs $F$ with $\fw_\Delta(F) \leq K-1$.
    Since $n + 1 = N^{1/k}$ and $\eta < 1/(8k)$, we have $kK \leq kN^\eta \leq d(n)$.
    If $kK\Delta \leq \lfloor n^{\varepsilon k} \rfloor$, then $\mathfrak{A}_n$ and $\mathfrak{B}_n$ are $\mathsf{C}_{kK}^{kK\Delta}$-equivalent, and \cref{lem:transfer} yields $p_N(M_N) = p_N(M'_N)$, contradicting $\hom_{F_N,N}(M_N) \neq \hom_{F_N,N}(M'_N)$.
    Hence, $\Delta > \lfloor n^{\varepsilon k} \rfloor/(kK) \geq (N^{1/k}-1)^{\varepsilon k}/(2kN^\eta) \geq N^\theta$, as $\eta + \theta < \varepsilon$.
    Applying this to $\Delta \coloneqq \max\{\Delta', 1\}$, where $\Delta'$ is the support depth of $C$, yields $\Delta' \geq N^\theta$.
    Since every family in $\symVAC$ is computed by circuits of polynomial orbit size and polylogarithmic support depth, $(p_N) \notin \symVAC$.
\end{proof}

\cref{thm:symVAC1-separation} follows from \cref{thm:symVAC1-separation-strong}.

For \cref{thm:symVACk-hierarchy}, we use a version of \cref{thm:gln} with a tunable number of rounds, at the price of constantly many variables.

\begin{lemma}[\protect{cf.\ \cite[Lemmas~5.1, 5.2, and~5.15]{grohe_iteration_2025}}]\label{lem:gln-tunable}
    There are constants $\ell_0 \geq 3$ and $\xi \geq 1$ such that for all $h \geq \ell_0$ and $r \geq 1$ there are relational structures $\mathfrak{A}$ and $\mathfrak{B}$ with the same number $n \leq 2\xi h^2r$ of elements over a finite vocabulary of arity $\leq \ell_0$ that are $\mathsf{C}_h^r$-equivalent but not $\mathsf{C}_{\ell_0}^{\xi hr}$-equivalent.
\end{lemma}
\begin{proof}
    The round bound is not stated in \cite{grohe_iteration_2025}, so we inspect the construction in the proof of \cite[Lemma~5.2]{grohe_iteration_2025}.
    Let $R_0 \geq 9$ be the constant of \cite[Corollary~5.12]{grohe_iteration_2025}, let $\ell_0 \coloneqq R_0 + 1$, and let $\delta$ be the constant of \cite[Lemma~5.2]{grohe_iteration_2025}.
    Let $h \geq \ell_0$ and $r \geq 1$, and let $\Lambda \coloneqq 2(h+1)(r+1)$.
    For $\ell_{\mathrm{hi}} \coloneqq h$, the proof of \cite[Lemma~5.2]{grohe_iteration_2025} constructs a layered graph $Z = (V, W, E)$ with layers $V_0, \dots, V_\Lambda$ and $W_1, \dots, W_\Lambda$ such that, for all $i \in [\Lambda]$, the graph $Z[V_i \cup W_i]$ is a perfect matching, every $w \in W_i$ has $\ell_0$ neighbours, and these lie in $V_{i-1} \cup V_i$.
    It considers the set of XOR-constraints
    \[
        \mathcal{C} \coloneqq \{(N_Z(w), 0) \mid w \in W\} \cup \{(\{v\}, 0) \mid v \in V_0\} \cup \{(\{v_\Lambda\}, 1)\}
    \]
    of arity $\leq \ell_0$ for some $v_\Lambda \in V_\Lambda$, and shows that $|V| \leq \delta h^2 r$ and that Falsifier does not win the $r$-round $h$-pebble game on $\mathcal{C}$.
    Let $\mathfrak{A} \coloneqq \mathfrak{A}(\mathcal{C})$ and $\mathfrak{B} \coloneqq \mathfrak{B}(\mathcal{C})$ be the structures of \cite[Section~5.1]{grohe_iteration_2025}.
    They have $2|V| \leq 2\delta h^2 r$ elements and arity $\leq \ell_0$, and they are $\mathsf{C}_h^r$-equivalent by \cite[Lemma~5.1]{grohe_iteration_2025}.

    We count the rounds of the strategy of Falsifier in the $\ell_0$-pebble game from the proof of \cite[Lemma~5.15]{grohe_iteration_2025}.
    In the first round, Falsifier pebbles $v_\Lambda$.
    If Verifier answers $0$, then $(\{v_\Lambda\}, 1)$ is violated.
    Otherwise, suppose that the current position assigns $1$ to some $v_i \in V_i$ with $i \geq 1$, and let $w_i \in W_i$ be the matching partner of $v_i$.
    Then $N_Z(w_i) \cap V_i = \{v_i\}$.
    Falsifier keeps the pebbles on the current pebbled vertices in $N_Z(w_i)$, including $v_i$, and pebbles the remaining vertices of $N_Z(w_i)$ one per round.
    This takes at most $\ell_0 - 1$ rounds and at most $\ell_0$ pebbles.
    Unless $(N_Z(w_i), 0)$ is violated, some $v_{i-1} \in N_Z(w_i) \cap V_{i-1}$ is assigned $1$.
    Once some $v_0 \in V_0$ is assigned $1$, the constraint $(\{v_0\}, 0)$ is violated.
    Hence, Falsifier wins within $1 + \Lambda(\ell_0 - 1) \leq 1 + 8(\ell_0 - 1)hr \leq 8\ell_0 hr$ rounds, using $(h+1)(r+1) \leq 4hr$.
    By \cite[Lemma~5.1]{grohe_iteration_2025}, $\mathfrak{A}$ and $\mathfrak{B}$ are not $\mathsf{C}_{\ell_0}^{8\ell_0 hr}$-equivalent.
    The claim follows with $\xi \coloneqq \max\{\delta, 8\ell_0\}$.
\end{proof}

\begin{proof}[Proof of \cref{thm:symVACk-hierarchy}]
    Let $\ell_0$ and $\xi$ be as in \cref{lem:gln-tunable}, let $k \geq 1$, and write $\log = \log_2$.
    For $N \geq 2$, let $h(N) \coloneqq \lfloor \sqrt{\log N} \rfloor$ and $r(N) \coloneqq h(N) \lceil \log^k N \rceil$.
    Let $N_0 \geq 2\ell_0 + 3$ be such that $h(N) \geq \ell_0$ and $(2\xi h(N)^2 r(N) + 1)^{\ell_0} \leq N$ for all $N \geq N_0$; it exists since the left-hand side is polylogarithmic in $N$.
    For $N \geq N_0$, write $h \coloneqq h(N)$ and $r \coloneqq r(N)$, let $\mathfrak{A}_N$ and $\mathfrak{B}_N$ be as in \cref{lem:gln-tunable} for $h$ and $r$, and let $M_N$, $M'_N$, and $F_N$ be as in \cref{lem:transfer} for $m \coloneqq \ell_0$, $K_0 \coloneqq \ell_0$, and $q_0 \coloneqq \xi hr$.

    \emph{The family.}
    Let $p_N \coloneqq \hom_{F_N, N}$ for $N \geq N_0$ and $p_N \coloneqq 0$ otherwise.
    By \cref{lem:transfer}, $\fw_{\xi hr+1}(F_N) \leq \ell_0$.
    By \cref{thm:converse}, $p_N$ is computed by a matrix-symmetric circuit of orbit size $\leq N^{\ell_0+1}$ and support depth $\leq \xi hr + 1 = \xi h^2 \lceil \log^k N \rceil + 1 \leq 3\xi \log^{k+1} N$.
    Hence, $(p_N) \in \symVAC^{k+1}$.

    \emph{The lower bound.}
    Suppose that $(p_N) \in \symVAC^k$, witnessed by a constant $c \geq 1$.
    Let $N$ be sufficiently large, and let $C$ be a matrix-symmetric circuit of orbit size $\leq f \coloneqq N^c$ and support depth $\leq \Delta \coloneqq \lceil c \log^k N \rceil$ computing $p_N$.
    By \cref{lem:reduction}, we may suppose that $C$ is reduced.
    Let $\kappa \coloneqq \kappa_f(N)$.
    Since $\binom{N}{\lfloor c \rfloor + 1} > N^c$, we have $\kappa \leq c + 1$ and hence $3\kappa \leq N+2$.
    Let $K \coloneqq \max\{\kappa - 1, 2\} \leq c + 1$.
    By \cref{thm:support,thm:bounded-depth-characterisation}, $p_N$ is a linear combination of $\hom_{F,N}$ for bipartite multigraphs $F$ with $\fw_\Delta(F) \leq K-1$.
    Since $h \to \infty$, we have $\ell_0 K \leq h$ and $\ell_0 K \Delta \leq \ell_0 (c+1)^2 \lceil \log^k N \rceil \leq r$.
    Hence, $\mathfrak{A}_N$ and $\mathfrak{B}_N$ are $\mathsf{C}_{\ell_0 K}^{\ell_0 K \Delta}$-equivalent, and \cref{lem:transfer} yields $p_N(M_N) = p_N(M'_N)$, contradicting $\hom_{F_N,N}(M_N) \neq \hom_{F_N,N}(M'_N)$.
    Hence, $(p_N) \notin \symVAC^k$.
\end{proof}

\begin{remark}\label{rem:hierarchy-black-box}
    Without inspecting the proofs of \cite{grohe_iteration_2025}, a weaker hierarchy follows from \cref{thm:gln} and \cite[Theorem~1.1]{grohe_iteration_2025}:
    with $k_0$ and $\varepsilon$ as in \cref{thm:gln}, $\symVAC^j \subsetneq \symVAC^{\lfloor j/\varepsilon \rfloor + k_0}$ for every $j \geq 1$.
    In particular, the hierarchy $(\symVAC^j)_{j}$ does not collapse.

    Indeed, let $k \coloneqq k_0$ and $g \coloneqq \lfloor j/(\varepsilon k) \rfloor + 1$, so that $g\varepsilon k > j$ and $gk \leq j/\varepsilon + k$.
    For large $N$, let $n \coloneqq \lfloor \log^g N \rfloor$ and $d \coloneqq \lfloor (n/(\alpha k^6))^{1/8} \rfloor$, and let $\mathfrak{A}_N$ and $\mathfrak{B}_N$ be as in \cref{thm:gln} for $d$, $k$, and $n$.
    By \cite[Theorem~1.1]{grohe_iteration_2025}, $k$-WL stabilises on $\mathfrak{A}_N$ and $\mathfrak{B}_N$ after $O(kn^{k-1}\log n)$ rounds.
    Hence, $k$-WL distinguishes them after this many rounds, and they are not $\mathsf{C}_{k+1}^{q_0}$-equivalent for $q_0 \coloneqq n^k$ by \cite[Corollary~2.2]{grohe_iteration_2025}, for large $N$.
    Let $M_N$, $M'_N$, and $F_N$ be as in \cref{lem:transfer} for $m \coloneqq k$, $K_0 \coloneqq k+1$, and $q_0$, and let $p_N \coloneqq \hom_{F_N,N}$.
    As in the proof of \cref{thm:symVACk-hierarchy}, \cref{thm:converse} yields circuits of orbit size $\leq N^{k+2}$ and support depth $\leq n^k + 1 = O(\log^{gk} N)$, so $(p_N) \in \symVAC^{gk}$.
    Conversely, for circuits of orbit size $\leq N^c$ and support depth $\leq \lceil c \log^j N \rceil$, the argument of the lower bound in the proof of \cref{thm:symVACk-hierarchy} applies, since $\mathfrak{A}_N$ and $\mathfrak{B}_N$ are $\mathsf{C}_{d+1}^{\lfloor n^{\varepsilon k} \rfloor}$-equivalent, $d \to \infty$, and $n^{\varepsilon k} = \Omega(\log^{g \varepsilon k} N)$ with $g\varepsilon k > j$.
    Hence, $(p_N) \notin \symVAC^j$.

    The gap between $j$ and $\lfloor j/\varepsilon \rfloor + k_0$ arises since the upper bound of \cite[Theorem~1.1]{grohe_iteration_2025} on the number of rounds is polynomially larger than the lower bound $n^{\varepsilon k}$.
    \Cref{lem:gln-tunable} closes this gap.
\end{remark}

\end{document}